\documentclass[fleqn,10pt]{article}
\usepackage{latexsym, graphicx, epsfig, amsmath, amssymb,amsfonts}
\usepackage{natbib,amsthm,version}
\usepackage{amsbsy,bm,multirow,enumerate}
\usepackage[titletoc,page]{appendix}
\usepackage[mathscr]{eucal}
\usepackage{mathtools}
\usepackage{color}
\usepackage[utf8]{inputenc}
\usepackage[english]{babel}
\usepackage{amsthm}
\usepackage{enumerate}
\usepackage[hidelinks]{hyperref}
\usepackage{subfigure}
\usepackage[lined,boxed]{algorithm2e}

\DeclareMathOperator{\sech}{sech}

\newtheorem{theorem}{Theorem}[section]

\newtheorem{lemma}[theorem]{Lemma}
\newtheorem{remark}{Remark}[section]

\theoremstyle{definition}

\title{
  A Method for Representing Periodic Functions and
  Enforcing Exactly  Periodic Boundary  Conditions
  with Deep Neural Networks
} 
\author{
  Suchuan Dong\thanks{Author of correspondence.
    Email: sdong@purdue.edu}, \ Naxian Ni
  \\
  Center for Computational and Applied Mathematics \\
  Department of Mathematics \\
  Purdue University, USA 
 } 

\date{(July 14, 2020)}
\begin{document}
\maketitle



\begin{abstract}

  We present a simple and effective method for representing periodic functions
  and enforcing exactly the periodic boundary conditions for solving
  differential equations with
  deep neural networks (DNN). The method stems from some simple
  properties about function compositions involving periodic functions.
  It essentially composes a DNN-represented arbitrary function with
  a set of independent  periodic functions with adjustable
  (training) parameters.
  We distinguish two types of periodic conditions: those imposing
  the periodicity requirement on the function and all its derivatives
  (to infinite order), and those imposing periodicity on the function
  and its derivatives up to a finite order $k$ ($k\geqslant 0$).
  The former will be referred
  to as $C^{\infty}$ periodic conditions, and the latter $C^{k}$ periodic
  conditions. We define operations that constitute a $C^{\infty}$
  periodic layer and a $C^k$ periodic layer (for any $k\geqslant 0$).
  A deep neural network
  with a $C^{\infty}$ (or $C^k$) periodic layer incorporated as the
  second layer automatically and exactly satisfies the $C^{\infty}$
  (or $C^k$) periodic conditions.
  We present extensive numerical experiments on ordinary
  and partial differential equations with $C^{\infty}$ and $C^k$
  periodic boundary conditions to verify and demonstrate that the proposed
  method indeed enforces exactly, to the machine accuracy,
  the periodicity for the DNN solution and its derivatives.

\end{abstract}


\vspace{0.05cm}

\section{Introduction}
\label{sec:intro}

%

Deep neural networks (DNN) have emerged in the past few years
as a promising alternative to the classical numerical methods
(such as finite difference and finite element)
for solving ordinary and partial differential equations (PDE).
DNN-based solvers transform the PDE solution problem into an
optimization problem.
They typically represent the unknown field function
in terms of a deep neural network, thanks to the universal approximation
property of DNNs~\cite{HornikSW1989,HornikSW1990,Cotter1990,Li1996}.
Then these methods compute the solution
by minimizing a loss function that consist of
 residual norms of the governing equation and also possibly of the boundary
and initial conditions
in strong or weak forms
(see e.g.~\cite{LagarisLF1998,LagarisLP2000,RuddF2015,SirignanoS2018,EY2018,RaissiPK2019,ZangBYZ2020,Samaniegoetal2020}, among others).
DNN solutions are smooth analytical functions (depending on
the activation function used) and, once the network is trained,
can be evaluated for the function value and its derivatives exactly
at any point inside the domain.


Boundary (and initial) conditions play a critical role in the solution
of PDEs and make the problem well
posed~\cite{Gresho1991,SaniG1994,DongKC2014,Dong2015clesobc,NiYD2019}.
To enforce the boundary conditions (BC) in DNN solvers,
an often-used approach is the penalty method, by incorporating
a penalty term consisting of the residual norm of the boundary conditions
into the loss function.
The penalty method enforces the boundary conditions only approximately,
and the penalty coefficient strongly influences the DNN training and
convergence~\cite{ChenDW2020}. Choosing an appropriate or near-optimal
penalty coefficient is largely an art, usually conducted by trial and error.

Enforcing the boundary conditions exactly with deep neural networks, if feasible,
would be highly desirable. In this case, the DNN is constructed such that
the boundary conditions are automatically and exactly satisfied.
The constrained optimization problem for the PDE solution
will then become less constrained or unconstrained,
which will greatly facilitate the DNN training.
Enforcing exactly the boundary conditions with deep neural networks is,
however, highly non-trivial.
For Dirichlet and Neumann type boundary conditions, several researchers
have investigated the problem and promising techniques
are available (see e.g.~\cite{LagarisLF1998,LagarisLP2000,McFallM2009,BergN2018}).
In~\cite{LagarisLF1998} the unknown solution is decomposed into two components.
One component satisfies the Dirichlet/Neumann
boundary conditions and has no
training parameters, while the other component vanishes on the boundary
and is represented by a neural network in the domain. 
In~\cite{LagarisLP2000} the authors decompose the unknown solution into
a function represented by a deep neural network plus a linear combination of radial
basis functions. The combination coefficients of the radial basis functions
are determined by solving a linear system at every iteration of the DNN
evaluation to satisfy the Dirichlet or Neumann boundary conditions.
This process is understandably computationally expensive~\cite{LagarisLP2000}.
In~\cite{McFallM2009} the authors rewrite the solution into
two parts, similar to~\cite{LagarisLF1998}, with one part satisfying
the Dirichlet/Neumann boundary
conditions and the other part vanishing on the boundary but otherwise unconstrained.
In order to deal with complex domain boundaries, the authors of~\cite{McFallM2009} introduce
a multiplicative length factor in front of the unconstrained part, which
heuristically represents the distance of a point to the domain boundary.
These ideas are further developed by~\cite{BergN2018},
where the Dirichlet boundary condition is considered.
In~\cite{BergN2018} 
the Dirichlet boundary data extension and the distance function to
the boundary are both represented by low-capacity deep neural networks
and pre-trained.
This simplifies the implementation of the DNN solver. However,
the enforcement of the boundary condition becomes only approximate.
Similar ideas for the Dirichlet/Neumann type boundary conditions
have also appeared in more recent works, see e.g.~\cite{RaoSL2020}
for the DNN simulation of elastodynamic problems.


Periodic boundary conditions are widely encountered in computational science
of various areas, especially when the physical domain involved in
is infinite or homogeneous along
one or more directions~\cite{DongKER2006,Dong2008jfm}.
In such cases, usually only one  cell
will be computed in numerical simulations, and periodic boundary conditions
are imposed on the cell boundaries. 
With classical numerical methods,
another often-used technique for this type of problems is to express the
unknown field function in terms of Fourier expansions, leading to
what is known as the Fourier spectral or
pseudo-spectral method~\cite{CanutoHQZ1988,Dong2007,DongZ2011}.
While both are referred to as periodic conditions,
the periodicity requirements imposed by the numerical method, when Fourier expansions
are used and when they are not, are different.
With the use of Fourier expansions,
the method seeks a smooth periodic function as the solution
to the governing equations, which
automatically satisfies the periodicity 
for the solution value and {\em all} its derivatives (to infinite order) on
the cell boundaries.
On the other hand, when Fourier expansions are not used in the
method, the periodicity needs to be imposed explicitly on
the cell boundaries, and this can only be imposed for
the solution value and its derivatives up to a certain finite order.
We will distinguish these two types of periodic boundary conditions in this work.
We refer to the former as the $C^{\infty}$ periodic conditions,
and the latter as the $C^k$ periodic conditions, where $k\geqslant 0$
denotes the highest derivative the periodic condition imposes on.

The penalty method has been used to enforce the periodic boundary
conditions (for up to the first derivative) with DNN-based PDE solvers
in some recent studies; see e.g.~\cite{ChenDW2020,NgomM2020}.
In \cite{ChenDW2020} a penalty term representing the residual norms of
the periodic conditions for the function and its first derivative
is included in the loss function.
In \cite{NgomM2020} a sinusoidal activation function is employed
with a shallow neural network (one hidden layer) to mimic the Fourier decompositions
of the function, which is termed
the Fourier neural network by some
researchers (see e.g.~\cite{GallantW1988,Silvscu1999,Liu2013,ZhumekenovUTACK2019}),
and then the periodicity condition for the solution is imposed by
a penalty term in the loss function.
The penalty method can only impose the periodic boundary conditions
approximately.
It becomes more difficult, and perhaps impractical, to
enforce $C^k$ periodic conditions using the penalty for moderate or large $k$ values. 
In particular, it is practically impossible
to impose the $C^{\infty}$ periodic conditions with the penalty method.

How to enforce exactly the $C^{\infty}$ and $C^k$ (for any $k\geqslant 0$)
periodic conditions with deep neural networks is the focus of the current work.
This problem seems to have barely been investigated before.
To the best of the authors' knowledge, the only work close in theme to the current effort
is perhaps~\cite{GokuzumNK2019},
in which the authors enforce the periodicity condition for the solution value only,  by
adopting a similar idea to~\cite{LagarisLF1998} and
constructing a trial function with two parts. On part enforces the periodicity
for the solution value, and the other part vanishes on the boundary and
is represented by a DNN~\cite{GokuzumNK2019}.


In the current paper we present a method for enforcing exactly the $C^{\infty}$
and $C^k$ (for any $k\geqslant 0$) periodic boundary conditions with deep
neural networks. The DNN resulting from the current method, by design, automatically
and exactly satisfies the $C^{\infty}$ or $C^k$ (for any prescribed $k$)
periodic conditions. This method is based on some simple properties about
function compositions 
involving periodic functions (Lemmas \ref{lem:lem_1} and \ref{lem:lem_2}
in Section \ref{sec:method}),
and leverages the universal approximation power of deep neural networks.
It essentially composes a DNN-represented arbitrary function with
a set of independent known periodic functions with adjustable parameters.
We consider the feed-forward neural network architecture~\cite{GoodfellowBC2016}
in this work,
and define the operations that constitute
a $C^{\infty}$ periodic layer and a $C^k$
periodic layer. To enforce the $C^{\infty}$ periodic conditions,
one only needs to set the second
layer of the DNN (i.e.~the first hidden layer) as a $C^{\infty}$ periodic layer.
To enforce the $C^k$ periodic conditions, one
only needs to set the second layer of the DNN as
a $C^k$ periodic layer.
The $C^{\infty}$ periodic layer constructs a set of independent
$C^{\infty}$ periodic functions
with a user-prescribed period, based on sinusoidal functions, affine mappings and
nonlinear activation functions (such as ``tanh'' and ``sigmoid'').
The $C^k$ periodic layer constructs a set of independent
$C^k$ periodic functions, based on
the generalized Hermite interpolation polynomials, affine mappings
and nonlinear activation
functions. The output of the overall DNN,
with the $C^{\infty}$ (or $C^k$) periodic layer incorporated
therein, automatically and exactly satisfies
the $C^{\infty}$ (or $C^k$) periodic conditions.

The operations involved in the $C^{\infty}$ and $C^k$ periodic layers can be
implemented as user-defined Tesorflow/Keras (www.tensorflow.org and keras.io) layers,
and incorporated into a DNN using Keras in a straightforward way.
We present a number of numerical experiments with  ordinary
and partial differential equations to verify 
and demonstrate that the proposed method indeed enforces exactly, to the machine accuracy,
the periodic boundary conditions as expected.
All the numerical examples in the current paper are implemented and conducted
based on Tensorflow, Keras and Python.


The contributions of this paper consist of two aspects:
(i) the method for representing smooth periodic functions and exactly
enforcing $C^{\infty}$ periodic boundary conditions with deep neural networks;
(ii) the method for exactly enforcing the $C^k$ (for any $k\geqslant 0$)
periodic boundary conditions with deep neural networks.

The rest of this paper is structured as follows.
In Section \ref{sec:method} we define the operations that
constitute the $C^{\infty}$ periodic layer
and the $C^k$ periodic layer in one and higher dimensions, and establish
that a deep neural network with these layers incorporated
as the second layer exactly
satisfies the $C^{\infty}$ or $C^k$ periodic boundary conditions
for a given domain.
In Section \ref{sec:tests} we present extensive numerical experiments on
periodic function approximations, and on solving
the Helmholtz equations in one and two dimensions,
the diffusion equation, and the wave equation, with
$C^{\infty}$ and $C^k$ periodic boundary conditions.
We demonstrate numerically that the proposed
method enforces exactly the periodic
boundary conditions, to the machine accuracy,
for the DNN solution and its corresponding higher derivatives.
Section \ref{sec:summary} then concludes the presentation
with some closing remarks.


\section{Enforcing Exact Periodic Conditions with DNN}
\label{sec:method}

\subsection{$C^{\infty}$ and $C^k$ Periodic Conditions}

Consider a smooth periodic function $f(x)$ with period $L$
defined on the real axis,
\begin{equation}\label{equ:peri_func}
  f(x+L)=f(x), \quad \forall x\in (-\infty,\infty).
\end{equation}
Now restrict $f(x)$ to a finite interval $[a, b]$, where $b-a=L$.
Then $f$ satisfies the following relations on
the boundaries:
\begin{equation}\label{equ:pbc_inf}
  f(a)=f(b), \ \
  f'(a)=f'(b), \ \
  f''(a)=f''(b), \ \ \dots, \ \
  f^{(m)}(a)= f^{(m)}(b), \ \ \dots
\end{equation}
We refer to the conditions in \eqref{equ:pbc_inf} as
the $C^{\infty}$ periodic conditions.
Hereafter we will refer to a smooth function $f(x)$ satisfying
these conditions as a $C^{\infty}$ periodic function on $[a,b]$,
or simply a periodic function.

In practice, the function may not be smooth and
the conditions in~\eqref{equ:pbc_inf}
may only be required  for
the derivatives up to a finite order $k$ ($k\geqslant 0$), i.e.
\begin{equation}\label{equ:pbc_ck}
  f^{(l)}(a) = f^{(l)}(b),
  \quad 0\leqslant l\leqslant k,
\end{equation}
where $f^{(0)}(x)=f(x)$ by convention.
We refer to the $(k+1)$ conditions in~\eqref{equ:pbc_ck}
as the $C^k$ periodic conditions.
With a slight abuse of notation,
we will refer to a function $f(x)$ satisfying
the conditions \eqref{equ:pbc_ck} as a $C^k$ periodic function
on $[a,b]$.

Our goal here is to devise a method for representing $C^{\infty}$
and $C^k$ periodic functions with
deep neural networks  such that,
by design, the output of the DNN automatically and exactly
satisfies the $C^{\infty}$ or $C^k$ periodic conditions.
Such neural networks will be referred to as
$C^{\infty}$ or $C^k$ periodic deep neural networks.
When solving a boundary value problem
or initial/boundary value problem together with the $C^{\infty}$ or $C^k$
periodic boundary conditions, one can use the method developed
herein to construct periodic DNNs as the trial functions
that automatically take into account
the periodic boundary
conditions.

\subsection{Enforcing Exact $C^{\infty}$ Periodic Conditions with DNN}
\label{sec:pbc}

We present a method below for representing $C^{\infty}$ periodic functions
and enforcing exactly the $C^{\infty}$ periodic conditions with DNN.
The method is based on the following property about function compositions
involving periodic functions.
\begin{lemma}\label{lem:lem_1}
  Let $v(x)$  be a given smooth periodic function
  with period $L$ on the real axis, i.e.~$v(x+L)=v(x)$ for all $x\in (-\infty,\infty)$,
  and f(x) denote
  an arbitrary smooth function. Define
  $u(x) = f(v) = f(v(x))$. 
  Then
  \begin{subequations}\label{equ:p_cond}
  \begin{align}
    &
    u(x+L) = u(x), \quad \forall x\in (-\infty,\infty);
    \\
    &
    u^{(l)}(a) = u^{(l)}(b), \quad l= 0, 1, 2, \dots
  \end{align}
  \end{subequations}
  where $a$ and $b$ denote two real numbers with $b-a=L$.
  
\end{lemma}
\noindent This lemma can be proven by straightforward verifications.


We seek a DNN representation for an arbitrary smooth
periodic function with a prescribed period $L$,
such that the $C^{\infty}$ periodic conditions are automatically satisfied.
In light of Lemma \ref{lem:lem_1}, our basic idea for 
the representation is to compose an arbitrary DNN-presented function,
together with a set of independent known periodic functions with period $L$
and adjustable (training) parameters.
Let us first use a single known periodic function with prescribed
period $L$ for illustration.
We consider the sinusoidal functions,
\begin{equation}\label{equ:def_p}
  p(x) = A\cos(\omega x + \phi) + c, \quad
  \text{with} \ \omega = \frac{2\pi}{L},
\end{equation}
where the constants $A$, $c$ and $\phi$ are scalar adjustable (training) parameters.
Here $\omega$ is a fixed constant as given above and ensures
that $p(x)$ has a period $L$.
Let $\sigma(\cdot)$ denote a nonlinear activation function (such as
``tanh'' or ``sigmoid'').
We define
\begin{equation}
  v(x) = \sigma(p(x)) = \sigma(A\cos(\omega x + \phi) + c).
\end{equation}
This step is crucial. The nonlinear function $\sigma(\cdot)$
will generate higher-frequency components in the output.
So while $p(x)$ has a single frequency $\omega$,
$v(x)$ contains not only the frequency $\omega$,
but components with other and higher frequencies, all with a common period $L$.
Finally, we consider an arbitrary function $f(x)$ represented by a DNN,
and define
\begin{equation}
  u(x) = f_{\text{dnn}}(v(x)),
\end{equation}
where $f_{\text{dnn}}$ denotes the DNN-presented arbitrary function.
By Lemma \ref{lem:lem_1}, this $u(x)$ satisfies
the $C^{\infty}$ periodic conditions exactly.

%

\begin{figure}
  \centerline{
    \includegraphics[height=2.0in]{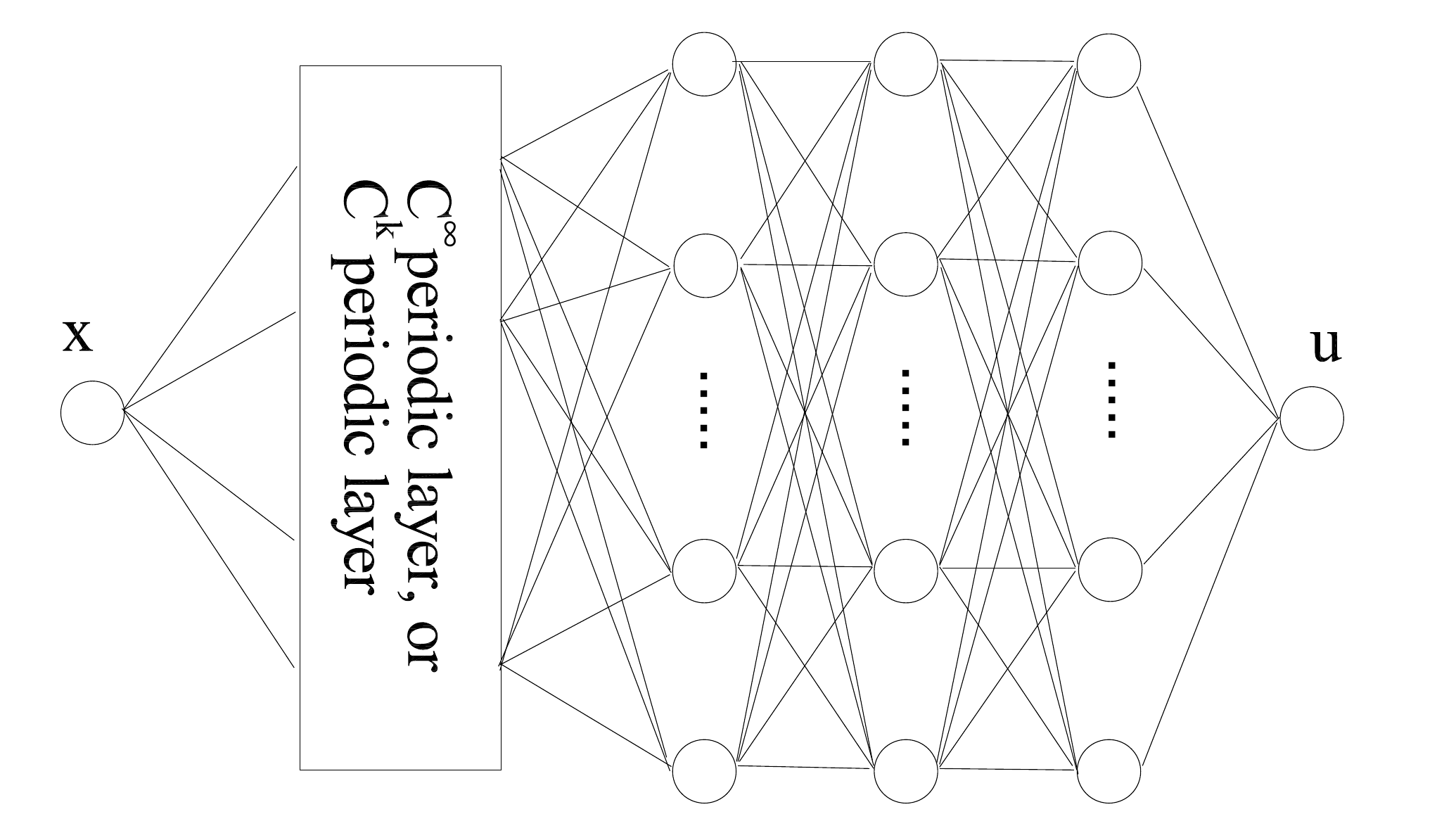}
  }
  \caption{
    Sketch of a feed-forward deep neural network
    with a $C^{\infty}$  or $C^k$ periodic layer incorporated
    as the second layer.
  }
  \label{fig:dnn}
\end{figure}

\begin{figure}
  \centerline{
    \includegraphics[height=2.5in]{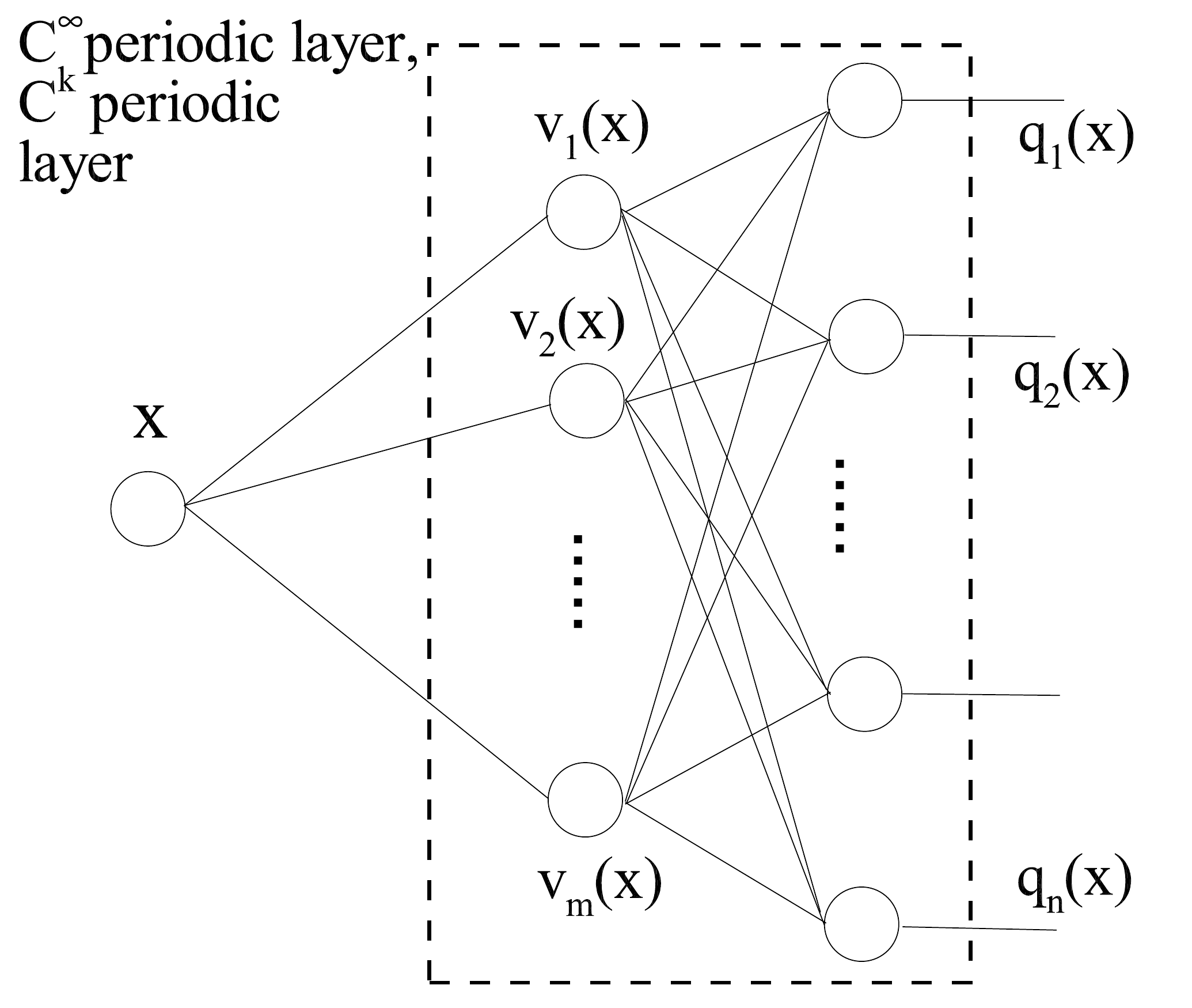}(a)
    \qquad
    \includegraphics[height=2.5in]{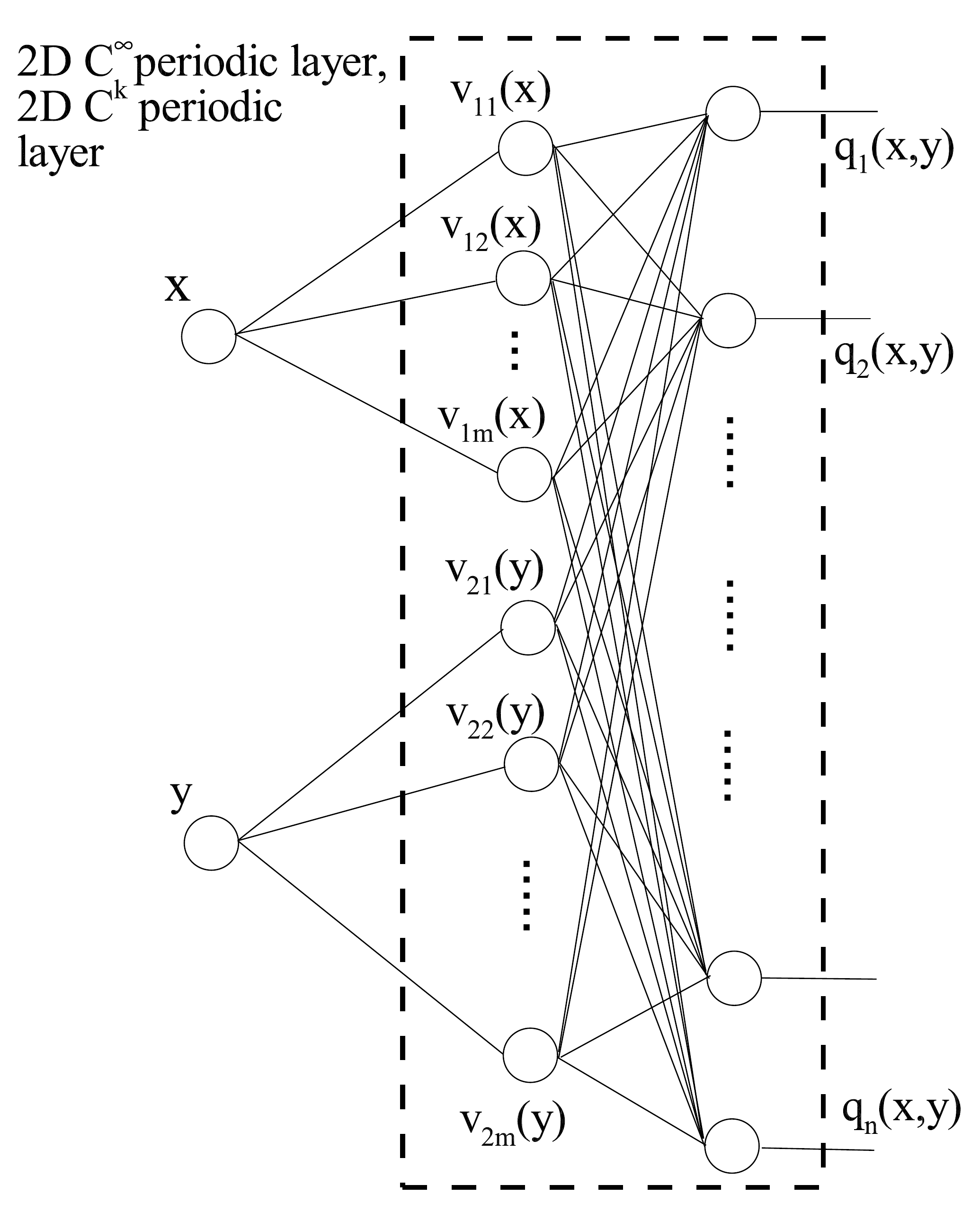}(b)
  }
  \caption{
    Sketch illustrating the internal structures of
    (a) 1D, and (b) 2D $C^{\infty}$ or $C^k$ periodic layers.
  }
  \label{fig:peri_lay}
\end{figure}


In practice, we would like to compose the DNN-represented arbitrary function $f_{dnn}(\cdot)$
with a set of independent periodic functions $v(x)$ as defined above
with adjustable parameters.
This leads to the idea of a periodic layer with multiple nodes (neurons)
within the layer.
We consider
the feed-forward deep neural network architecture~\cite{GoodfellowBC2016}
in the current work. Figure \ref{fig:dnn} illustrates the idea of
the current DNN with a sketch.
Let $x$ denote
the input layer to the network, and $u(x)$ denote the output layer
of the network.
We use the second layer (behind the input $x$) to implement the
set of independent known periodic functions (copies of $v(x)$) with period $L$
and adjustable parameters,
so that the output of the network $u(x)$ satisfies exactly
the $C^{\infty}$ periodic conditions~\eqref{equ:p_cond}.
We refer to such a layer as a $C^{\infty}$ periodic layer, or simply
a periodic layer.

The operations within the $C^{\infty}$ periodic layer are defined as follows
(see Figure \ref{fig:peri_lay}(a)).
Let $\mathcal{L}_p(m,n)$ denote the $C^{\infty}$ periodic layer,
where $n$ denotes the number of
nodes in the output of this layer
and $m$ denotes the size of the set of independent periodic
functions $v(x)$.
Here both $m$ and $n$ are hyper-parameters of the $C^{\infty}$ periodic
layer $\mathcal{L}_p(m,n)$. The operations within $\mathcal{L}_p(m,n)$ are defined by:
\begin{subequations}\label{equ:p_layer}
\begin{align}
  &
  v_i(x) = \sigma(A_i\cos(\omega x + \phi_i) + c_i),
  \quad 1\leqslant i \leqslant m;
  \label{equ:def_zi} \\
  &
  q_j(x) = \sigma\left(\sum_{i=1}^{m} v_i(x) W_{ij} + B_j \right),
  \quad 1\leqslant j \leqslant n.
  \label{equ:def_qi}
\end{align}
\end{subequations}
In these equations $q_j(x)$ ($1\leqslant j\leqslant n$)
are the output of this layer, and the fixed constant $\omega$
is given in \eqref{equ:def_p} for a prescribed period $L$.
$\sigma(\cdot)$ is the
nonlinear activation function, and 
it is used twice in this layer.
The training parameters of $\mathcal{L}_p(m,n)$ are the constants
$A_i$, $\phi_i$, $c_i$,
$W_{ij}$ and $B_j$, with $1\leqslant i\leqslant m$
and $1\leqslant j\leqslant n$.




In the current work, we have employed
Tensorflow (www.tensorflow.org) and Keras (keras.io)
to implement the operations of the $C^{\infty}$ periodic layer
as described above. The $C^{\infty}$ periodic layer is implemented
as a user-defined Tensorflow/Keras layer, and
it can be used in the same way as the built-in core Keras layers.


\begin{remark}\label{rem:rem_1}
  The operations of  the $C^{\infty}$ periodic layer defined by \eqref{equ:p_layer}
  can be extended to two, three and higher dimensions (2D/3D) in a straightforward way.
  Here we outline the idea with two dimensions only (see Figure \ref{fig:peri_lay}(b)).
  Let $(x_1,x_2)$ denote the coordinates in two dimensions,
  and $u(x_1,x_2)$ denote a smooth periodic function to be approximated,
  with properties
  \begin{equation}
    u(x_1+L_1,x_2) = u(x_1,x_2), \quad
    u(x_1,x_2+L_2) = u(x_1,x_2),
    \quad \forall x_1,x_2 \in (-\infty,\infty),
  \end{equation}
  where $L_1$ and $L_2$ are the periods in the $x_1$ and $x_2$ directions,
  respectively.
  Equivalently, we can write the periodicity conditions in terms of
  a single periodic cell $[a_1,b_1]\times[a_2,b_2]$,
  \begin{equation}
    \left\{
    \begin{split}
      &
      \frac{\partial^{\alpha}}{\partial x_1^{\alpha}}u(a_1,x_2)
      =  \frac{\partial^{\alpha}}{\partial x_1^{\alpha}}u(b_1,x_2),
      \quad \forall x_2\in [a_2,b_2], \\
      &
      \frac{\partial^{\alpha}}{\partial x_2^{\alpha}}u(x_1,a_2)
      =  \frac{\partial^{\alpha}}{\partial x_2^{\alpha}}u(x_1,b_2),
      \quad \forall x_1\in [a_1,b_1],
      \quad \alpha = 0, 1, 2, \dots
    \end{split}
    \right.
  \end{equation}
  where $a_1$, $b_1$, $a_2$ and $b_2$ are
  given constants satisfying $b_1-a_1=L_1$ and $b_2-a_2=L_2$.
  In this case, we define the 2D periodic layer, $\mathcal{L}^{2D}_p(m,n)$,
  with the following operations:
  \begin{subequations}\label{equ:2d_pbc}
    \begin{align}
      &
      v_{1i}(x_1) = \sigma\left( A_{1i}\cos(\omega_1 x_1 + \phi_{1i}) + c_{1i} \right),
      \quad 1\leqslant i\leqslant m;
      \label{equ:v1}\\
      &
      v_{2i}(x_2) = \sigma\left( A_{2i}\cos(\omega_2 x_2 + \phi_{2i}) + c_{2i} \right),
      \quad 1\leqslant i\leqslant m;
      \label{equ:v2}\\
      &
      q_j(x_1,x_2) = \sigma\left( \sum_{i=1}^{m}v_{1i}(x_1)W_{ij}^{(1)}
      +\sum_{i=1}^{m}v_{2i}(x_2)W_{ij}^{(2)} + B_j
      \right),
      \quad 1\leqslant j\leqslant n.
      \label{equ:qout}
    \end{align}
  \end{subequations}
  In these equations, $m$ and $n$ are  hyper-parameters of the layer $\mathcal{L}_p^{2D}$,
  $q_j(x_1,x_2)$ ($1\leqslant j\leqslant n$) denote the output of this layer,
  and  the constants $\omega_1$ and $\omega_2$ are defined by
  \begin{equation}
    \omega_1 = \frac{2\pi}{L_1}, \ \
    \omega_2 = \frac{2\pi}{L_2},
  \end{equation}
  with prescribed periods $(L_1,L_2)$.
  The training parameters of $\mathcal{L}^{2D}_p(m,n)$ consist of the constants:
  \begin{equation*}
    A_{1i}, \ A_{2i}, \ \phi_{1i}, \ \phi_{2i}, \ c_{1i}, \ c_{2i}, \
    W_{ij}^{(1)}, \ W_{ij}^{(2)}, \ B_j,
    \quad 1\leqslant i\leqslant m, \
    1\leqslant j\leqslant n.
  \end{equation*}
  By composing an arbitrary DNN-represented function with the 2D $C^{\infty}$
  periodic layer defined above, we attain an overall DNN whose
  output automatically
  and exactly satisfies the 2D $C^{\infty}$ periodic conditions.
  The $C^{\infty}$ periodic layer for three and higher dimensions
  can be defined in a similar way.

\end{remark}

\begin{remark}\label{lem:lem_a}
  In two or higher dimensions, if the $C^{\infty}$ periodic conditions are
  imposed only in some (not all) directions, the $C^{\infty}$ periodic layer
  as defined above can be modified in a simple way to accommodate the situation.
  For example, consider the 2D $C^{\infty}$ periodic layer defined
  in \eqref{equ:2d_pbc} and suppose that the $C^{\infty}$ periodic conditions
  are imposed only in the $x_1$ direction with period $L_1$, but not
  in the $x_2$ direction. In this case, we can retain the equations \eqref{equ:v1}
  and \eqref{equ:qout}, and replace \eqref{equ:v2} by the following equation
  \begin{equation}
    v_{2i}(x_2) = \sigma\left( A_{2i}x_2 + c_{2i} \right),
      \quad 1\leqslant i\leqslant m,
      \label{equ:v2_a}
  \end{equation}
  where the constants $A_{2i}$ and $c_{2i}$ are the training parameters.
  The modified 2D periodic layer consisting of equations \eqref{equ:v1}, \eqref{equ:v2_a}
  and \eqref{equ:qout}, when composed with a DNN-represented arbitrary
  function, will give rise to an overall DNN that automatically and exactly satisfies
  the $C^{\infty}$ periodic conditions in the $x_1$ direction.
  
\end{remark}

\subsection{Enforcing Exact $C^k$ Periodic Conditions with DNN}
\label{sec:ck_pbc}

We present in this subsection a method for representing $C^k$
periodic functions and enforcing exactly the $C^k$ periodic conditions
(for any $k\geqslant 0$) with DNN.
The method is based on the following simple property about function
compositions involving $C^k$ periodic functions:
\begin{lemma}\label{lem:lem_2}
  Let $v(x)$ ($x\in[a,b]$) denote a given function
  with continuous derivatives up to the order $k$ and satisfying 
  the following property,
  \begin{equation} \label{equ:v_rela}
    v^{(l)}(a) = v^{(l)}(b), \quad 0\leqslant l\leqslant k.
  \end{equation}
  Let $f(x)$ denote an arbitrary function defined on the
  real axis with continuous derivatives up to the order $k$.
  Define $u(x) = f(v) = f(v(x))$ ($x\in[a,b]$). Then
  \begin{equation} \label{equ:u_rela}
    u^{(l)}(a) = u^{(l)}(b), \quad 0\leqslant l\leqslant k.
  \end{equation}

\end{lemma}
\begin{proof}
  By induction one can show that
  $u^{(m)}(x) = g(v, v', v'', \dots, v^{(m)})$ for $0\leqslant m\leqslant k$.
  In other words,
  $u^{(m)}(x)$ depends on $x$ only through $v$ and its derivatives.
  Equation \eqref{equ:u_rela} follows immediately from this
  relation and the conditions \eqref{equ:v_rela}.
\end{proof}

We seek a DNN representation for an arbitrary $C^k$ periodic
function on $[a,b]$, such that the $C^k$ periodic conditions
are automatically and exactly satisfied.
In light of Lemma \ref{lem:lem_2}, our basic idea for the
representation is to compose an arbitrary function represented by
a DNN, together with a set of independent known $C^k$ periodic functions
with adjustable (training) parameters.
To construct a $C^k$ periodic function $v(x)$ on $[a,b]$ in Lemma \ref{lem:lem_2}, i.e.~satisfying
the conditions~\eqref{equ:v_rela}, we note that
these conditions are reminiscent of the Hermite interpolation conditions.
So the Hermite interpolation polynomial of degree at most $(2k+1)$
can be used to construct $v(x)$.
Once $v(x)$ is obtained, we compose an arbitrary DNN-represented function
$f$  with $v(x)$, and the resultant
function satisfies the $C^k$ periodic conditions exactly.

Let us now use a single $C^k$ periodic function $v(x)$ ($x\in[a,b]$)
to illustrate the idea in some detail.
Let $s_i$ ($0\leqslant i\leqslant k$) denote $(k+1)$ adjustable (training) parameters.
Let $h(x)$ denote the unique Hermite
interpolation polynomial of degree at most $(2k+1)$ that satisfies
the following $(2k+2)$  interpolation conditions:
\begin{equation}\label{equ:int_cond}
  \left\{
  \begin{split}
    & h(a) = s_0, \quad h(b) = s_0; \\
    & h'(a) = s_1, \quad h'(b) = s_1; \\
    & \cdots \\
    & h^{(k)}(a) = s_k, \quad  h^{(k)}(b) = s_k.
  \end{split}
  \right.
\end{equation}
The Newton form for
$h(x)$ can be computed based on the divided differences, and
the explicit Lagrange form for $h(x)$ is available
in e.g.~\cite{Spitzbart1960,Traub1964}.
We then define
\begin{equation}\label{equ:def_v}
    \left\{
    \begin{split}
      &
      p(x) = h(x) + (r_0 + r_1x)(x-a)^{k+1}(x-b)^{k+1}, \\
      &
      v(x) = \sigma(p(x)) = \sigma\left( h(x) + (r_0 + r_1x)(x-a)^{k+1}(x-b)^{k+1} \right),
    \end{split}
    \right.
\end{equation}
where $r_0$ and $r_1$ are two additional adjustable parameters, and
$\sigma(\cdot)$ is a nonlinear activation function (e.g.~``tanh'' or ``sigmoid'').
It is straightforward to verify that the $v(x)$ (and also $p(x)$) given
in \eqref{equ:def_v} satisfies the conditions \eqref{equ:v_rela}.
The explicit forms for $p(x)$ in \eqref{equ:def_v} corresponding to $k=0$, $1$ and $2$
are given by:
\begin{equation*}
  \begin{split}
    & p(x) = s_0 + (r_0+r_1x)(x-a)(x-b), \quad \text{for} \ k=0; \\
    & p(x) = s_0 + s_1(x-a)(b-x)(a+b-2x)
    + (r_0+r_1x)(x-a)^2(x-b)^2,
    \quad \text{for} \ k=1; \\
    & p(x) = \xi^3(x-a)^3\left[
      s_0 + (-3s_0\xi + s_1)(x-b)
      + \left(6s_0\xi^2 - 3s_1\xi + \frac{s_2}{2}\right)(x-b)^2
      \right] \\
    &\qquad\quad
    + \xi^3(b-x)^3\left[
      s_0 + (3s_0\xi+s_1)(x-a)
      + \left(6s_0\xi^2 + 3s_1\xi + \frac{s_2}{2}\right)(x-a)^2
      \right] \\
    &\qquad\quad
    + (r_0+r_1x)(x-a)^3(x-b)^3,
    \quad \text{for} \ k=2,
  \end{split}
\end{equation*}
where $\xi = \frac{1}{b-a}$.
%
%
Let $f_{\text{dnn}}(x)$ denote an arbitrary function represented
by a deep neural network.
With $v(x)$ given by \eqref{equ:def_v}, we finally define
\begin{equation}
  u(x) = f_{\text{dnn}}(v(x)).
\end{equation}
Then by Lemma \ref{lem:lem_2}
$u(x)$ satisfies exactly the $C^k$ periodic conditions~\eqref{equ:u_rela}.

In practice, we would like to compose the DNN-represented arbitrary function
$f_{dnn}(\cdot)$ with a set of independent $C^k$ periodic functions $v(x)$
with adjustable parameters.
This leads to the idea of a $C^k$ periodic layer with multiple nodes (neurons)
within the layer.
Consider again the feed-forward neural network architecture,
and let $x$ denote the input and $u(x)$ denote the output of
the network.
Analogous to the $C^{\infty}$ periodic layer in Section \ref{sec:pbc},
we define a $C^k$ periodic layer below, and use it as
the second layer (behind the input $x$)
of the network to implement the set of independent $C^k$ periodic functions
(copies of $v(x)$) and enforce the $C^k$ periodic conditions.
Figure \ref{fig:dnn} sketches the DNN with a $C^k$ periodic layer incorporated
as the second layer.

The operations within the $C^k$ periodic layer are defined as follows
(see Figure \ref{fig:peri_lay}(a)).
Let $\mathcal{L}_{C^k}(m,n)$ denote the $C^k$ periodic layer, where
$n$ denotes the number of nodes in the output of
this layer and $m$ denotes the size of the set of
independent $C^k$ periodic functions $v(x)$.
Both $m$ and $n$ are  hyper-parameters of this layer.
Given the input $x$, we compute the output
$q_j(x)$ ($1\leqslant j\leqslant n$) of the $C^k$ periodic
layer $\mathcal{L}_{C^k}(m,n)$ by:
\begin{subequations}\label{equ:ckp_layer}
\begin{align}
  &
  v_i(x) = \sigma\left(h_i(x) + (r_{0i}+r_{1i}x)(x-a)^{k+1}(x-b)^{k+1} \right),
  \quad 1\leqslant i\leqslant m;
  \label{equ:def_vi} \\
  &
  q_j(x) = \sigma\left(\sum_{i=1}^{m}v_i(x) W_{ij} + B_j \right),
  \quad 1\leqslant j\leqslant n.
  \label{equ:qj_ck}
\end{align}
\end{subequations}
In these equations $\sigma(\cdot)$ is the nonlinear activation function, and
$h_i(x)$ ($1\leqslant i\leqslant m$) are the Hermite interpolation polynomials
of degree at most $(2k+1)$ satisfying the conditions
\begin{equation}
  h_i^{(l)}(a) = s_{li}, \ \ h_i^{(l)}(b) = s_{li},
  \quad 0\leqslant l\leqslant k, \ 1\leqslant i\leqslant m.
\end{equation}
The constant parameters involved in these equations,
$r_{0i}$, $r_{1i}$, $W_{ij}$, $B_j$, $s_{li}$, for $1\leqslant i\leqslant m$,
$1\leqslant j\leqslant n$ and $0\leqslant l\leqslant k$,
are the training parameters of
the $C^k$ periodic layer $\mathcal{L}_{C^k}(m,n)$.
By incorporating 
the $C^k$ periodic layer defined by \eqref{equ:ckp_layer} as
the second layer, the resultant DNN automatically and exactly satisfies
the $C^k$ periodic conditions with its output.


\begin{remark}\label{rem:rem_2}
  The operations of the $C^k$ periodic layer $\mathcal{L}_{C^k}(m,n)$
  defined by \eqref{equ:ckp_layer}
  can be extended to two, three and higher dimensions in a straightforward
  fashion. Here we use two dimensions only to illustrate the idea
  (see Figure \ref{fig:peri_lay}(b)).
  Let $x_1$ and $x_2$ ($x_1\in[a_1,b_1]$, $x_2\in[a_2,b_2]$)
  denote the coordinates in two dimensions,
  and $u(x_1,x_2)$ denote a 2D $C^k$ periodic function, satisfying
  the $C^k$ periodic conditions:
  \begin{equation}\label{equ:ckbc_2d}
    \left\{
    \begin{split}
      &
    \frac{\partial^{\alpha}}{\partial x_1^{\alpha}} u(a_1,x_2)
    = \frac{\partial^{\alpha}}{\partial x_1^{\alpha}}u(b_1,x_2),
    \quad \forall x_2\in[a_2,b_2],    \\
    &
    \frac{\partial^{\alpha}}{\partial x_2^{\alpha}}u(x_1,a_2)
    = \frac{\partial^{\alpha}}{\partial x_2^{\alpha}} u(x_1,b_2),
    \quad \forall x_1\in[a_1,b_1],
    \quad \alpha = 0, 1, \dots, k.
    \end{split}
    \right.
  \end{equation}
  In this case,
  we define the 2D $C^k$ periodic layer $\mathcal{L}_{C^k}^{2D}(m,n)$ with
  the following operations:
  \begin{subequations}\label{equ:2d_ckpbc}
    \begin{align}
      &
      v_{1i}(x_1) = \sigma\left(h_{1i}(x_1)
      + \left(r_{0i}^{(1)}+r_{1i}^{(1)}x_1\right)(x_1-a_1)^{k+1}(x_1-b_1)^{k+1}  \right),
      \quad 1\leqslant i\leqslant m;
      \label{equ:v1_ck}
      \\
      &
      v_{2i}(x_2) = \sigma\left(h_{2i}(x_2)
      + \left(r_{0i}^{(2)}+r_{1i}^{(2)}x_2\right)(x_2-a_2)^{k+1}(x_2-b_2)^{k+1}  \right),
      \quad 1\leqslant i\leqslant m;
      \label{equ:v2_ck}
      \\
      &
      q_j(x_1,x_2) = \sigma\left(
      \sum_{i=1}^{m}v_{1i}(x_1)W_{ij}^{(1)} + \sum_{i=1}^{m}v_{2i}(x_2)W_{ij}^{(2)} + B_j
      \right),
      \quad 1\leqslant j\leqslant n.
      \label{equ:q_ck}
    \end{align}
  \end{subequations}
  In the above equations, $m$ and $n$ are the hyper-parameters of this layer,
  $q_j(x_1,x_2)$ ($1\leqslant j\leqslant n$)
  denote the output of this layer, and $h_{1i}(x_1)$ and $h_{2i}(x_2)$
  are the Hermite interpolation polynomials of degree at most
  $(2k+1)$ satisfying the conditions:
  \begin{equation}
    \left\{
    \begin{split}
      &
      h_{1i}^{(l)}(a_1) = s_{li}^{(1)}, \quad
      h_{1i}^{(l)}(b_1) = s_{li}^{(1)}, \quad 0\leqslant l\leqslant k,
      \ \ 1\leqslant i\leqslant m;\\
      &
      h_{2i}^{(l)}(a_2) = s_{li}^{(2)}, \quad
      h_{2i}^{(l)}(b_2) = s_{li}^{(2)}, \quad 0\leqslant l\leqslant k,
      \ \ 1\leqslant i\leqslant m.
    \end{split}
    \right.
  \end{equation}
  The constant parameters involved in the above equations,
  \begin{equation*}
    s_{li}^{(1)}, \ s_{li}^{(2)}, \ r_{0i}^{(1)}, \ r_{0i}^{(2)}, \
    r_{1i}^{(1)}, \ r_{1i}^{(2)}, \ W_{ij}^{(1)}, \ W_{ij}^{(2)}, \ B_j,
    \quad 0\leqslant l\leqslant k, \
    1\leqslant i\leqslant m, \
    1\leqslant j\leqslant n,
  \end{equation*}
  are the training parameters of the layer $\mathcal{L}_{C^k}^{2D}(m,n)$.
  By using the 2D $C^k$ periodic layer $\mathcal{L}_{C^k}^{2D}(m,n)$
  as the second layer of a DNN and with $(x_1,x_2)$ as the input,
  the resultant DNN will automatically and exactly satisfy the
  2D $C^k$ periodic boundary conditions~\eqref{equ:ckbc_2d}.
  The $C^k$ periodic layer for three and higher dimensions
  can be defined in a similar way.
  
\end{remark}

\begin{remark}\label{lem:lem_b}
  In two and higher dimensions, if the $C^k$ periodic conditions
  are only imposed in some (not all) directions, the $C^k$ periodic
  layer as defined above can be modified in a simple way to accommodate
  the situation. The modification is similar to what is
  discussed in Remark \ref{lem:lem_a} for the modified $C^{\infty}$ periodic layer.
  For illustration, let us consider the 2D $C^k$ periodic layer defined by
  \eqref{equ:2d_ckpbc}, and suppose that the $C^k$ periodic conditions
  are imposed only in the $x_1$ direction, not in the $x_2$ direction.
  In this case, we can retain the equations \eqref{equ:v1_ck} and \eqref{equ:q_ck},
  and replace equation \eqref{equ:v2_ck} by the following equation for $v_{2i}(x_2)$,
  \begin{equation}\label{equ:v2_ck_b}
    v_{2i}(x_2) = \sigma\left(r_{0i}^{(2)} + r_{1i}^{(2)}x_2 \right),
  \end{equation}
  where the constants $r_{0i}^{(2)}$ and $r_{1i}^{(2)}$ are training parameters.
  The modified 2D $C^k$ periodic layer consisting of equations
  \eqref{equ:v1_ck}, \eqref{equ:v2_ck_b} and \eqref{equ:q_ck},
  when used as the second layer of a DNN,
  will impose exactly the $C^k$ periodic conditions in the $x_1$ direction
  in the output of this DNN.

\end{remark}



\section{Numerical Examples}
\label{sec:tests}

We present several numerical examples in what follows
to demonstrate the effectiveness of the method presented
in the previous section.
We consider the approximation of periodic functions, and
the solution of the
Helmholtz equation,
the unsteady diffusion equation and the wave equation,
together with periodic boundary conditions, using deep
neural networks.
We employ a variant of the deep Galerkin method~\cite{SirignanoS2018}
for solving the differential equations with DNN and also for enforcing
the initial conditions with unsteady problems.
The periodic boundary conditions (BC) are dealt with based on
the method from Section \ref{sec:method}.
Note that with the current method 
the periodic boundary conditions are satisfied automatically
by the DNN. Therefore there is no need to account for the
periodic boundary conditions in the loss function.
The application codes for all the tests reported here are implemented
using Tensorflow/Keras
and Python, with either Adam~\cite{KingmaB2014}
or L-BFGS~\cite{NocedalW2006} as the optimizer.
We employ the hyperbolic tangent (``tanh'') as the nonlinear
activation function in all the tests.

\subsection{Approximation of Periodic Functions}
\label{sec:peri_func}

Let us look into the DNN approximation of  periodic functions
using the method developed in Section \ref{sec:method}.
We employ three different functions to illustrate the
performance characteristics of the method:
a $C^{\infty}$ periodic function, a $C^0$ periodic function,
and a non-periodic function.
Note that in the case of the non-periodic function, we are
essentially seeking a periodic function approximation of
the non-periodic function.

Consider first the  function
\begin{equation}\label{equ:u1}
  u_1(x) = \sin(2\pi x+0.25\pi) + \cos(9\pi x-0.1\pi)
  -2\sin(7\pi x + 0.33\pi)
\end{equation}
on the domain $\Omega = \{ x | 0\leqslant x\leqslant 2\}$.
This is a $C^{\infty}$ periodic function on this domain.
We would like to approximate $u_1(x)$ using a $C^{\infty}$
periodic DNN, and using a $C^k$ periodic DNN with $k=0$ and $k=1$.


To approximate $u_1(x)$, we employ a feed-forward 
neural network~\cite{GoodfellowBC2016} as illustrated in Figure \ref{fig:dnn}.
The input to the network is $x$, and the output is the
approximation $u(x)$. We use $3$ hidden layers in between,
each with a width of $30$ neurons.
The hyperbolic tangent (``tanh'') function is used
as the activation function for all the hidden layers,
and no activation is applied on the output layer.
As discussed in Section \ref{sec:method}, we set the second layer of
the network (i.e.~the first hidden layer) as a $C^{\infty}$ periodic layer
for the $C^{\infty}$ periodic approximation, and as a $C^k$ periodic layer
for the $C^k$ periodic approximation.
More specifically, we employ a $\mathcal{L}_p(m,n)$ with $m=11$ and $n=30$
for the $C^{\infty}$ periodic layer,
and a $\mathcal{L}_{C^k}(m,n)$ with
$m=11$ and $n=30$ for the $C^k$-periodic layer with $k=0$ and $1$.
In other words, a set of $11$ independent periodic functions $v(x)$ has been
used within
the $C^{\infty}$ periodic layer and the $C^k$ periodic layer.
For the $C^{\infty}$ periodic layer $\mathcal{L}_p(m,n)$,
the constant $\omega$ in equation
\eqref{equ:def_zi} is set to, according to equation \eqref{equ:def_p},
\begin{equation}
  \omega = \frac{2\pi}{V_{\Omega}} = \frac{2\pi}{2} = \pi,
\end{equation}
where $V_{\Omega}=\int_{\Omega}dx=2$ is the size of the domain $\Omega$.

\begin{table}
  \centering
  \begin{tabular}{ll | ll}
    \hline
    parameter & value & parameter & value \\ \hline
    hidden layers & depth=3, width=30 & $N_e$ & 3 \\
    1st hidden layer & $C^{\infty}$ periodic layer $\mathcal{L}_p(11,30)$,
    & $Q$ & 30 ($C^{\infty}$ periodic DNN),  \\
    & or $C^k$ periodic layer $\mathcal{L}_{C^k}(11,30)$ &
    & or 40 ($C^k$ periodic DNN) \\
    activation & $\tanh$ & optimizer & Adam \\
    maximum epochs & 10000 & learning rate & $1e-3$ \\
    input data & $x_i^e$ ($0\leqslant e\leqslant N_e-1$, $0\leqslant i\leqslant Q-1$)
    & label data & $u_1(x_i^{e})$, or $u_2(x_i^e)$, or $u_3(x_i^e)$ \\
    $x_i^e$ & Gauss-Lobatto-Legendre quadrature & $\omega$ & $2\pi/V_{\Omega}$  \\
    \hline
  \end{tabular}
  \caption{Function approximation: DNN and simulation parameters.}
  \label{tab:peri_func}
\end{table}


We minimize the following loss function with this DNN,
\begin{equation}\label{equ:loss_u1}
  \begin{split}
  \text{Loss} &= \frac{1}{V_{\Omega}}\int_{\Omega}\left| u(x)-u_1(x) \right|^2 dx
  =\frac{1}{V_{\Omega}}\sum_{e=0}^{N_e-1}\int_{\Omega_e} \left| u(x)-u_1(x) \right|^2 dx \\
  &=\frac{1}{V_{\Omega}}\sum_{e=0}^{N_e-1}\sum_{i=0}^{Q-1}\left|u(x^e_i)-u_1(x^e_i)\right|^2 J^e w_i,
  \end{split}
\end{equation}
where 
$N_e$ is the number of elements (i.e.~sub-intervals) we have partitioned
the domain $\Omega$ into
in order to compute the integral, $Q$ is the number of
quadrature points within each element,
$\Omega_e$ denotes the sub-interval occupied by the
element $e$ ($0\leqslant e\leqslant N_e-1$),
$x_i^e$ ($0\leqslant i\leqslant Q-1$)
are the Gauss-Lobatto-Legendre quadrature points within
$\Omega_e$ for $0\leqslant e\leqslant N_e-1$, $J^e$ is the Jacobian of
the element $\Omega_e$ with respect to the standard element $[-1,1]$,
and $w_i$ ($0\leqslant i\leqslant Q-1$)
are the quadrature weights associated with
the Gauss-Lobatto-Legendre quadrature points.
In the numerical experiments we have employed $3$ elements ($N_e=3$)
to partition the domain $\Omega$, with $\Omega_0=[0,0.7]$,
$\Omega_1=[0.7,1.4]$ and $\Omega_2=[1.4,2]$.
We employ $30$ quadrature points ($Q=30$) within each element
for the $C^{\infty}$ periodic DNN, and $40$ quadrature
points ($Q=40$) within each element for the $C^k$ ($k=0,1$)
periodic DNN.
The input data to the network consist of all the quadrature points $x_i^e$
($0\leqslant i\leqslant Q-1$, $0\leqslant e\leqslant N_e-1$),
and the label data consist of $u_1(x_i^e)$.
The Adam optimizer has been used to train the network
for $10000$ epochs for each case, with the learning rate
fixed at the default value $10^{-3}$.
The options of ``early stopping'' and
``restore to best weight'' have been used in Keras during
the training process.
The main parameters for the DNN and the simulations are summarized in
Table \ref{tab:peri_func}.

\begin{figure}
  \centering
  \includegraphics[width=3.5in]{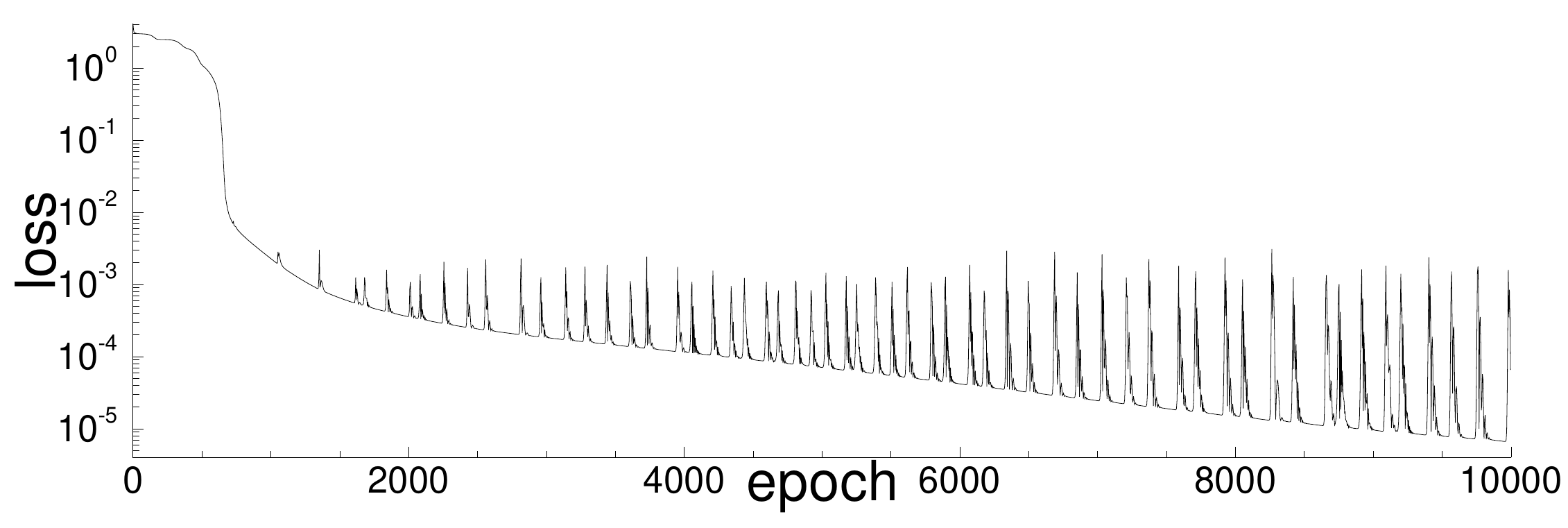}(a)
  \includegraphics[width=3.5in]{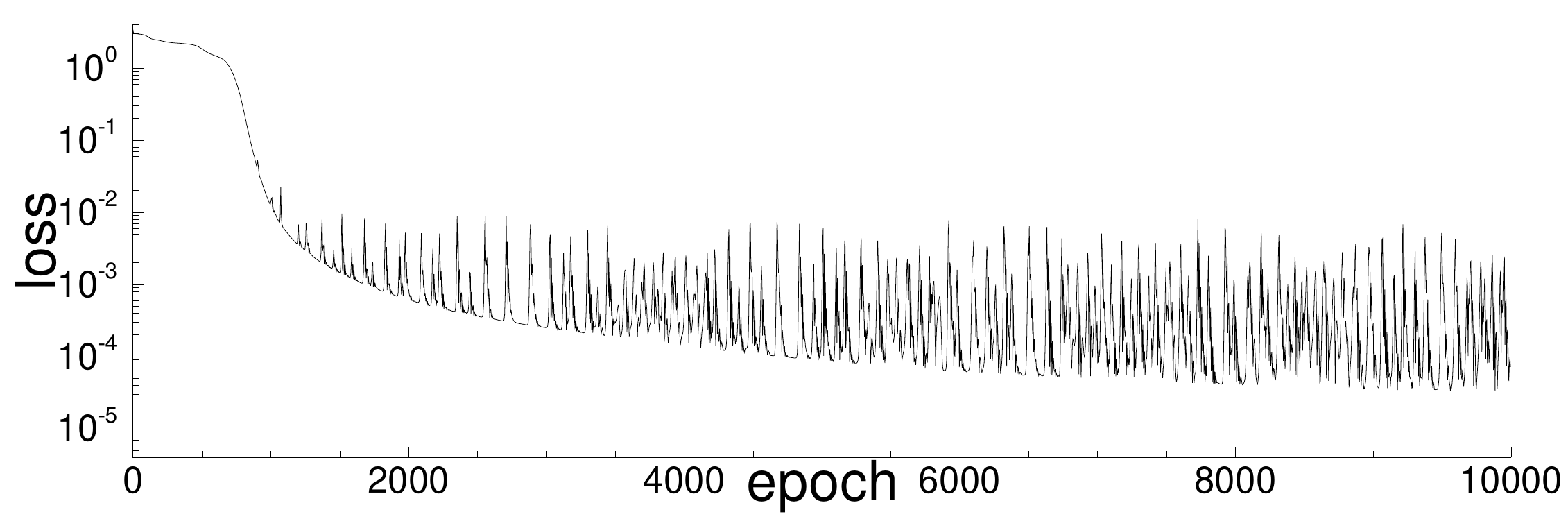}(b)
  \includegraphics[width=3.5in]{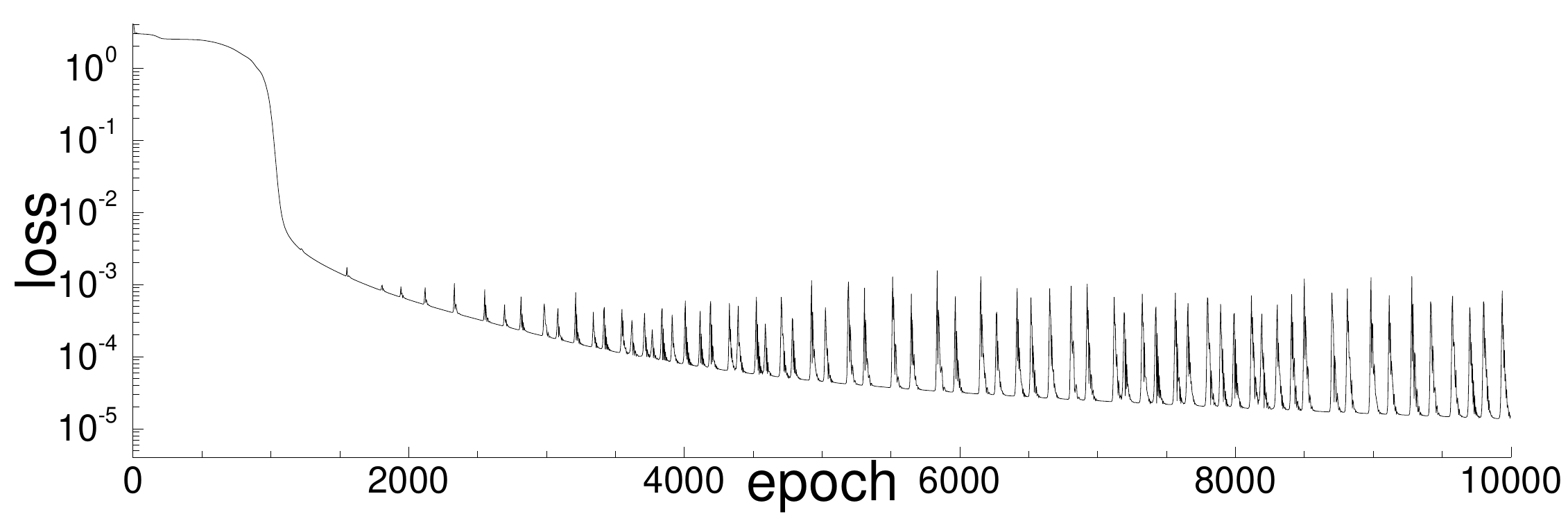}(c)
  \caption{
    Approximation of the periodic function $u_1(x)$: training
    loss histories corresponding to (a) $C^{\infty}$, (b) $C^0$, and (c) $C^1$
    periodic DNN approximations.
  }
  \label{fig:pf_loss}
\end{figure}


Figure \ref{fig:pf_loss} shows the training histories
of the loss function corresponding to
the $C^{\infty}$, $C^0$ and $C^1$ periodic DNN approximations.
The loss function decreases rather slowly as the training begins.
Then we observe a short stage when the loss function decreases sharply.
After that, the reduction in the loss function slows down again,
resulting in a long tail in the training history curve.
These characteristics seem to be common to the DNN training
for all the problems we have considered
in this work.
During the stage with slow reduction in the loss function (long tail),
we observe that the loss value fluctuates from time to time
during the training, resulting in a sequence of spikes in the training
history curves (Figure \ref{fig:pf_loss}).
This is likely due to the fixed learning rate employed here.
Reducing the learning rate gradually as the training progresses
will likely reduce the loss fluctuations.
In spite of the spikes, one can observe the trend of
decreasing loss as the training proceeds.
Since we have turned on the options of ``early stopping'' and
``restore to best weight'' in Keras, the spikes in the training
curves have no effect on the simulation results we have
obtained here.

\begin{figure}
  \centerline{
    \includegraphics[width=2in]{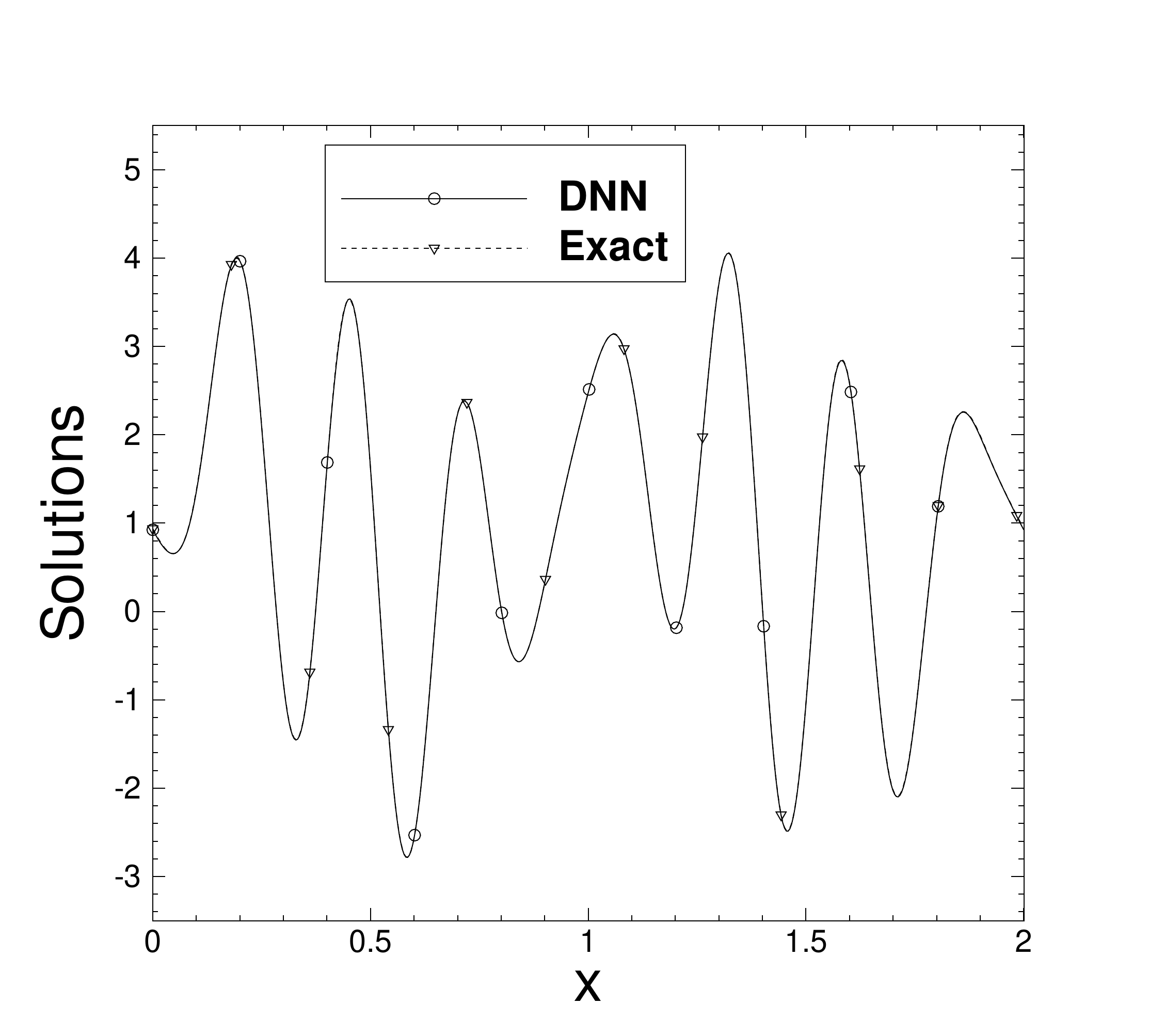}(a)
    \includegraphics[width=2in]{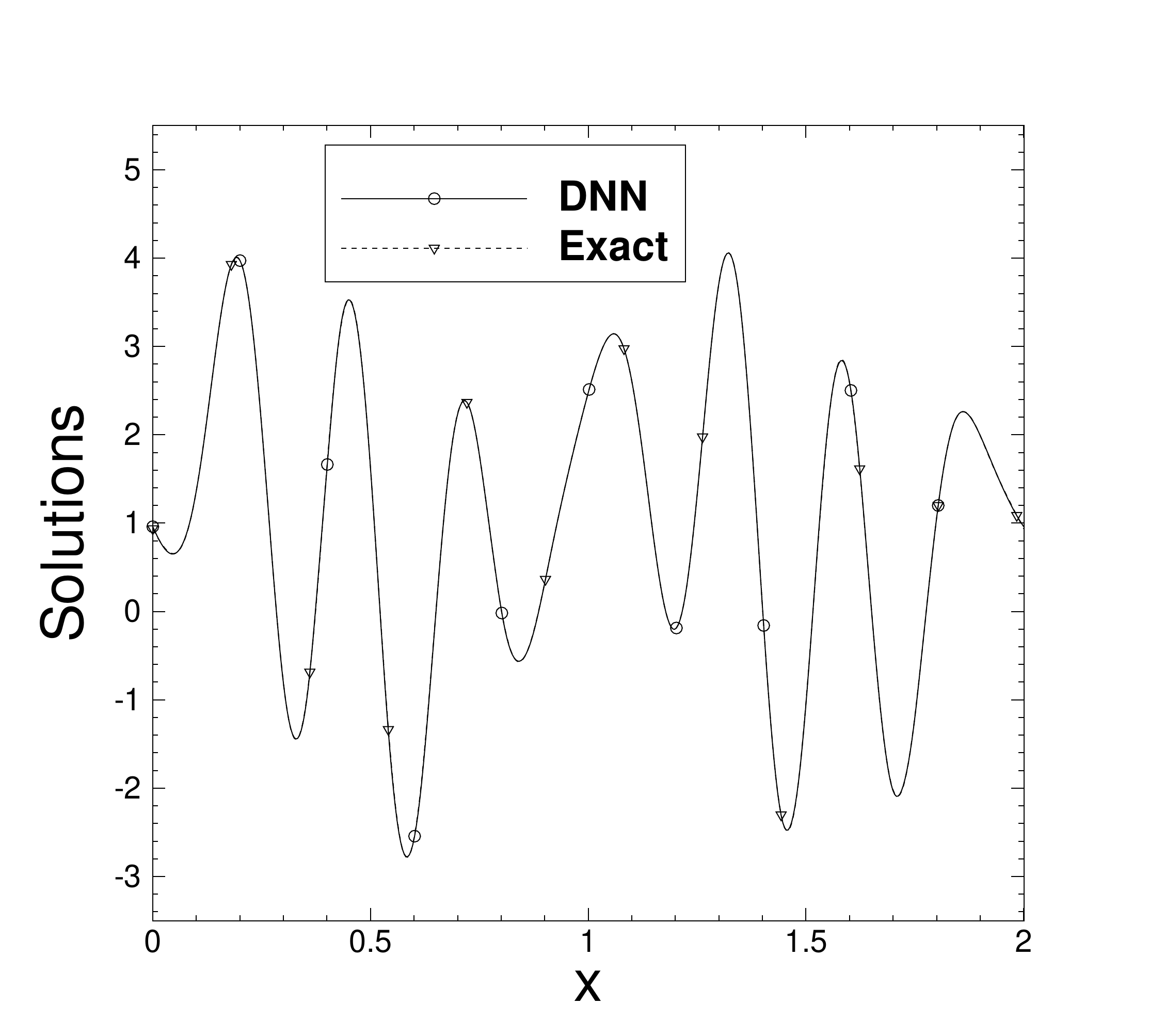}(b)
    \includegraphics[width=2in]{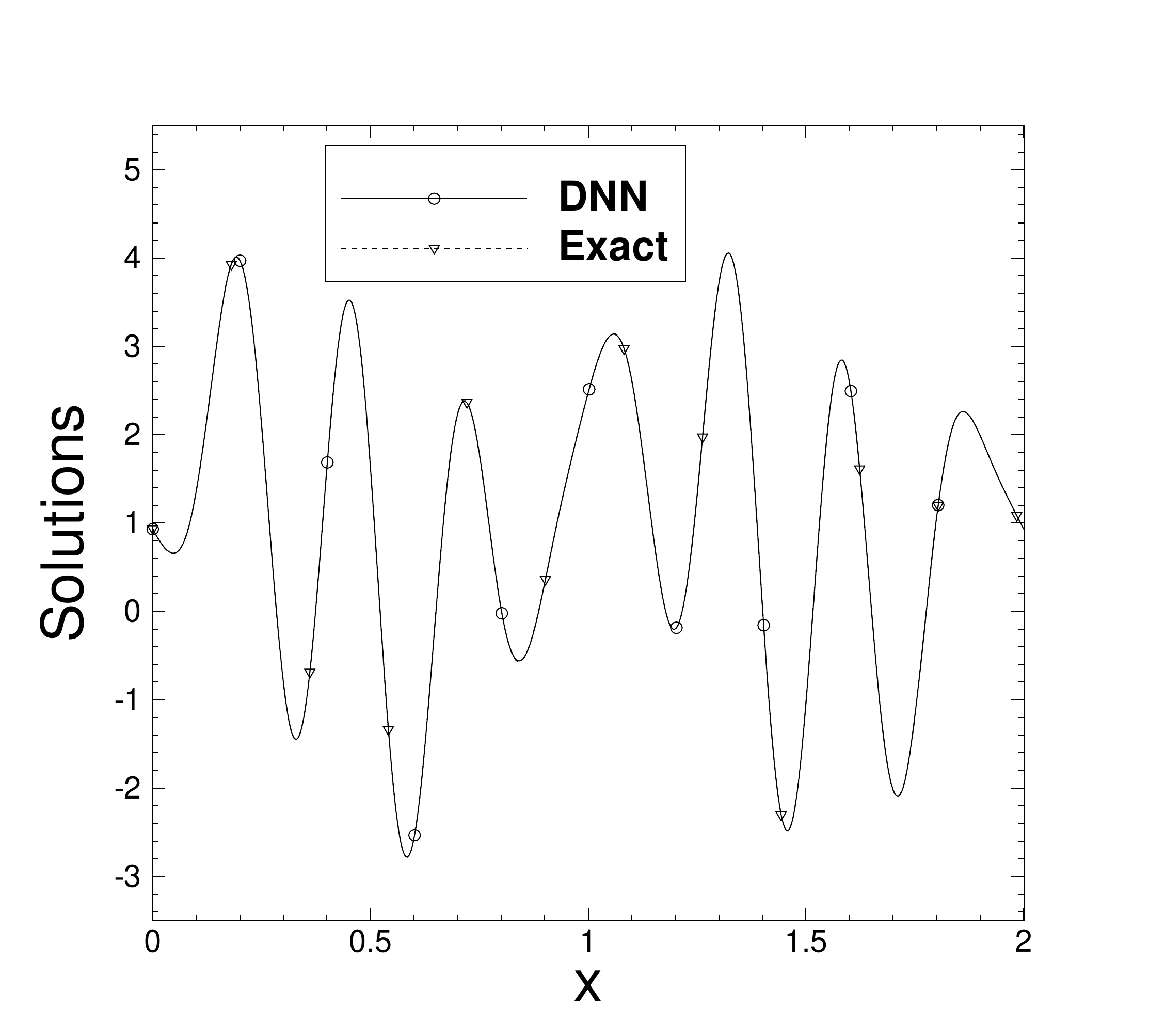}(c)
  }
  \centerline{
    \includegraphics[width=2in]{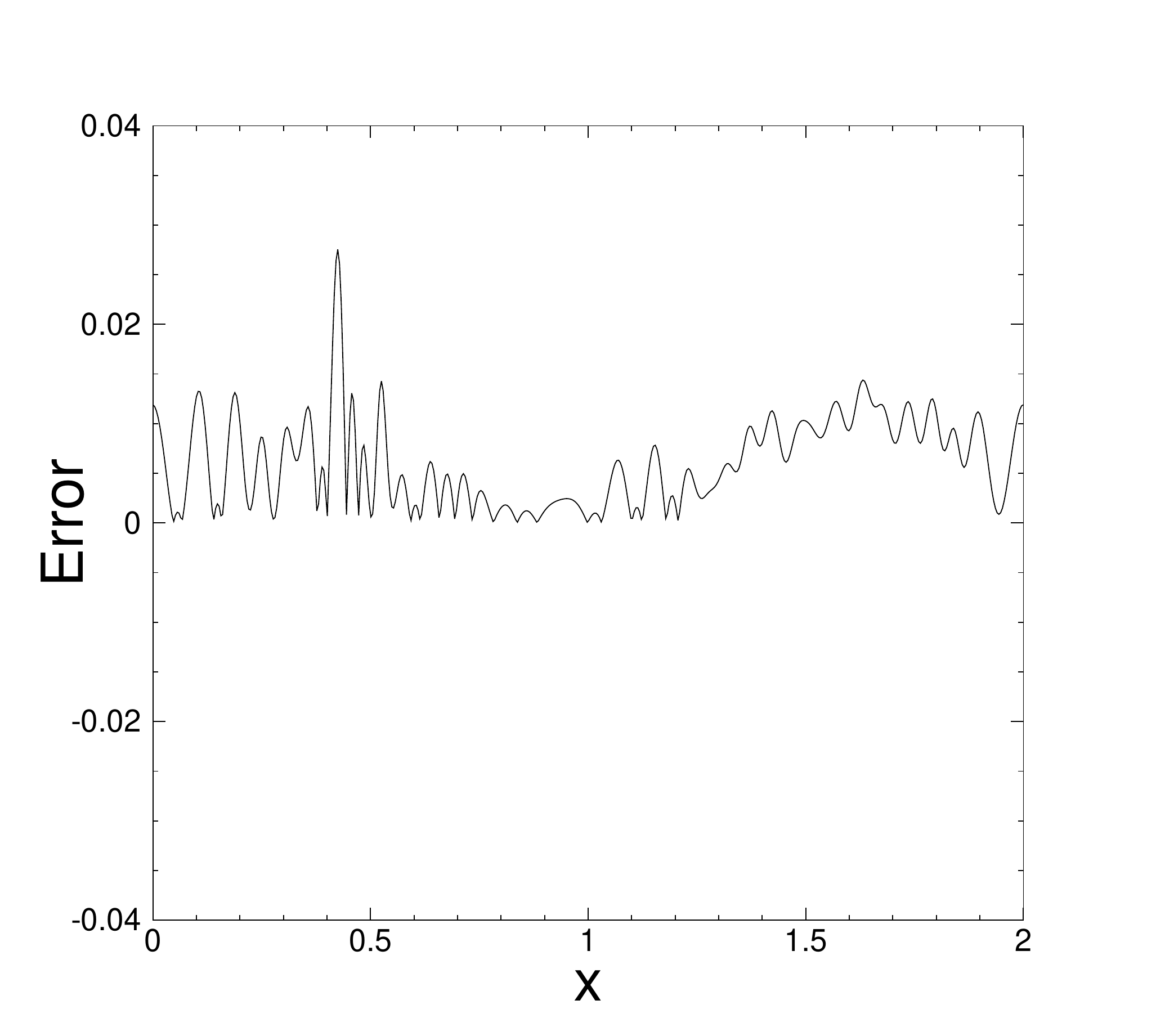}(d)
    \includegraphics[width=2in]{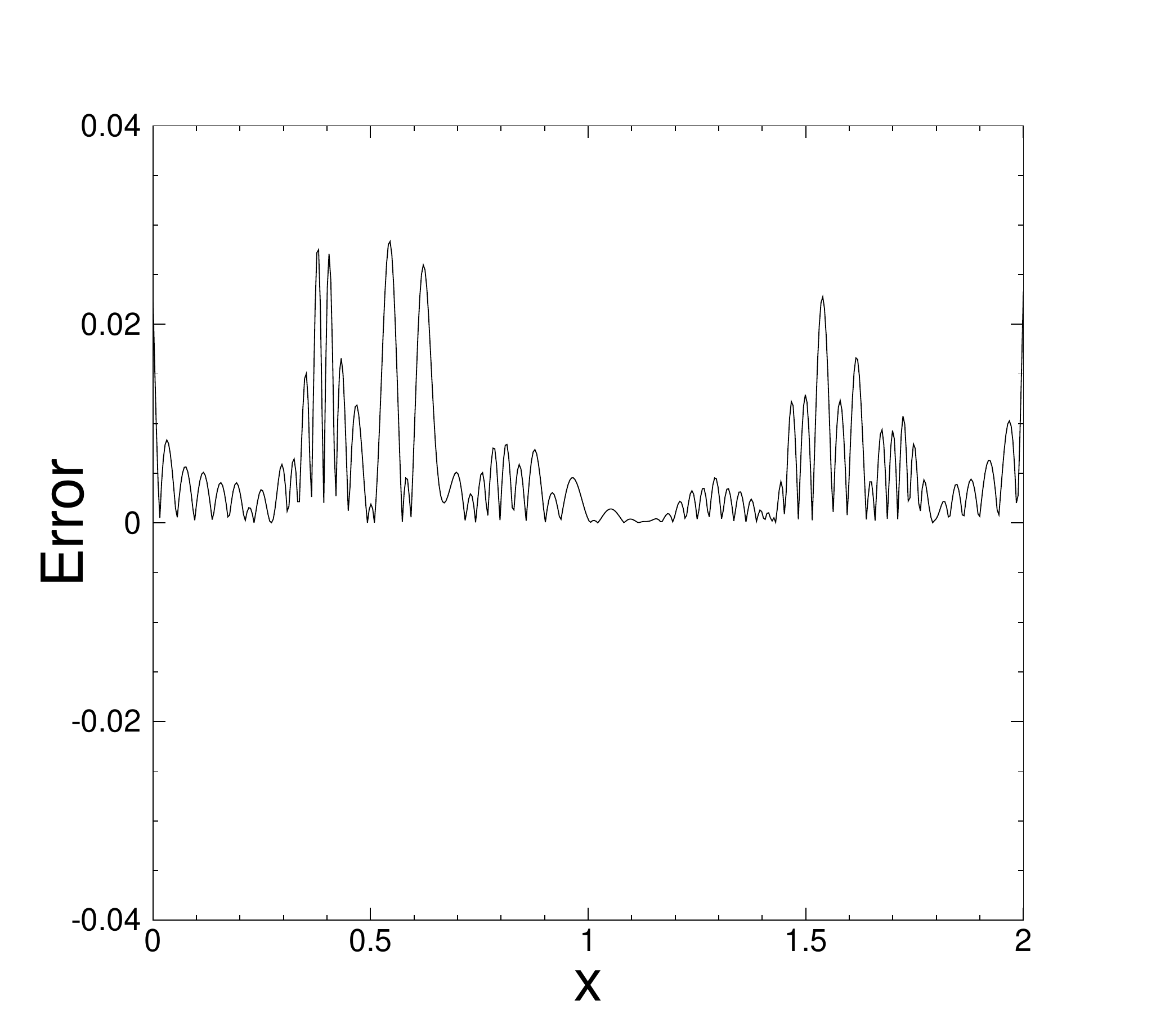}(e)
    \includegraphics[width=2in]{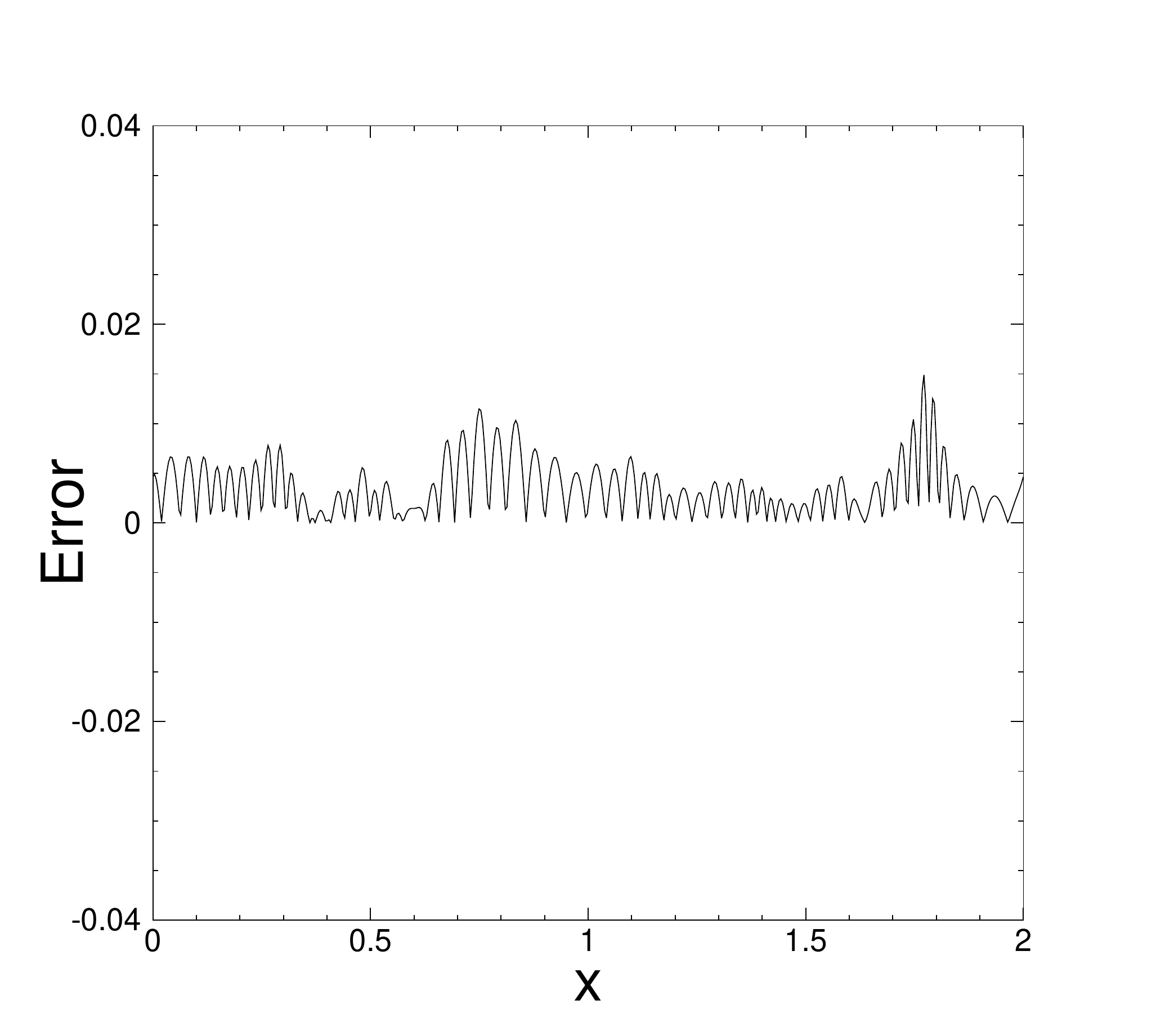}(f)
  }
  \caption{ Approximation of periodic function $u_1(x)$:
    comparison of DNN approximations (top row) and their errors
    (bottom row) with
    (a,d) $C^{\infty}$, (b,e) $C^0$, (c,f) $C^1$ periodic
    conditions. The exact function is shown in the plots
    (a,b,c) for reference.
  }
  \label{fig:pf_soln}
\end{figure}

\begin{table}[tp]
  \centering
  \begin{tabular}{lllll}
    \hline
     & $C^{\infty}$ periodic DNN  & $C^0$ periodic DNN & $C^1$ periodic DNN & Exact value \\
    \hline
    $u_1(0)$ & 9.2478271387350e-01 & 9.5998617007276e-01 &
    9.3202741230659e-01 & 9.3667924347381e-01 \\
    $u_1(2)$ & 9.2478271387350e-01 & 9.5998617007276e-01 &
    9.3202741230659e-01 & 9.3667924347381e-01 \\
    $u'_1(0)$ & -9.2010196979120e+00 & \fbox{-1.1103158994625e+01} &
    -9.3797238469930e+00 & -9.2086781968974e+00 \\
    $u'_1(2)$ & -9.2010196979120e+00 & \fbox{-7.1832211211812e+00} &
    -9.3797238469930e+00 & -9.2086781968972e+00 \\
    $u''_1(0)$ & 6.0998633006437e+01 & \fbox{8.4031868694133e+01} &
    \fbox{1.0357079019194e+02} & 4.4301828505852e+01 \\
    $u''_1(2)$ & 6.0998633006437e+01 & \fbox{8.6538176666730e+01} &
    \fbox{3.5624115949487e+01} &  4.4301828505854e+01 \\
    \hline
  \end{tabular}
  \caption{
    Approximations of the periodic function $u_1(x)$:
    Values of the function and its first and second derivatives at the left/right
    domain boundaries ($x=0, 2$)
    from the DNN approximations with $C^{\infty}$, $C^0$ and $C^1$
    periodic conditions and from the exact function.
    $14$ significant digits (double precision)
    are listed  to show
    that the current method enforces the periodic conditions exactly.
    The boxes highlight the values that mis-match on the
    boundaries.
  }
  \label{tab:pf_u1}
\end{table}

In Figure \ref{fig:pf_soln} we compare the DNN approximation results
of $u_1(x)$ (top row) obtained with $C^{\infty}$ (plot (a)),
$C^0$ (plot (b)), and $C^1$ (plot (c)) periodic conditions,
together with the exact function $u_1(x)$.
The distributions of the absolute error,
$|u(x)-u_1(x)|$,
corresponding to these approximations
are shown in Figures \ref{fig:pf_soln}(d,e,f), respectively.
It is observed that the DNN approximations computed with
all three methods agree well with the exact function $u_1(x)$,
and the approximation function curves overlap with
the exact function curve.

In Table \ref{tab:pf_u1} we list the values of
the function $u_1(x)$ and its derivatives $\frac{du_1}{dx}$
and $\frac{d^2u_1}{dx^2}$ on the domain
boundaries $x=0$ and $x=2$, obtained from
the  $C^{\infty}$, $C^0$ and $C^1$ periodic DNN approximations,
as well as from the exact $u_1(x)$ function
given in \eqref{equ:u1}.
The function derivatives have been computed based on auto-differentiation.
Once the DNN is trained, the derivatives computed in
this way are exact values corresponding to the given DNN representation.
We have shown $14$ significant digits (double precision) for the values in this table.
It is evident that the $C^{\infty}$ periodic DNN enforces
exactly, to the machine accuracy, the periodicity for the function as well as
its derivatives. On the other hand, the $C^0$ periodic DNN
enforces exactly the periodicity only for the function value,
and the $C^1$ periodic DNN enforces exactly the periodicity
only for the function value
and the first derivative.
These numerical results have verified the analyses about these methods 
in Section \ref{sec:method}.

We next consider the function
\begin{equation}\label{equ:u2}
  u_2(x) = \sin\frac{\pi x}{2}
\end{equation}
on the domain $\Omega = \{ x | 0\leqslant x\leqslant 2 \}$.
This is a $C^0$ periodic function on this domain,
with $u_2(0)=u_2(2)$ and $u_2'(0) \neq u_2'(2)$.
We would like to approximate $u_2(x)$ with $C^{\infty}$ and $C^k$ ($k=0$ and $1$)
periodic DNNs.

\begin{figure}
  \centerline{
    \includegraphics[width=2in]{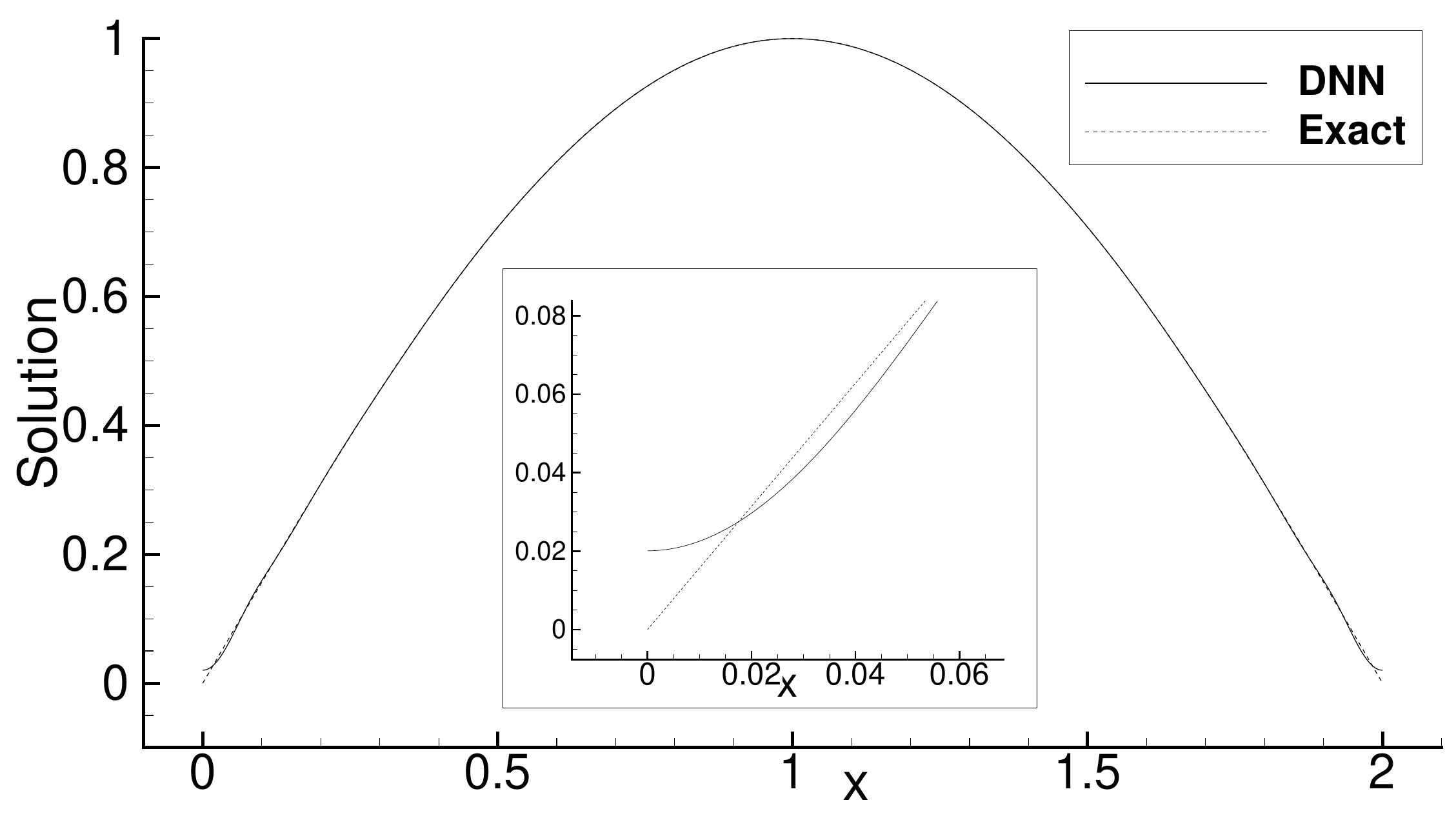}(a)
    \includegraphics[width=2in]{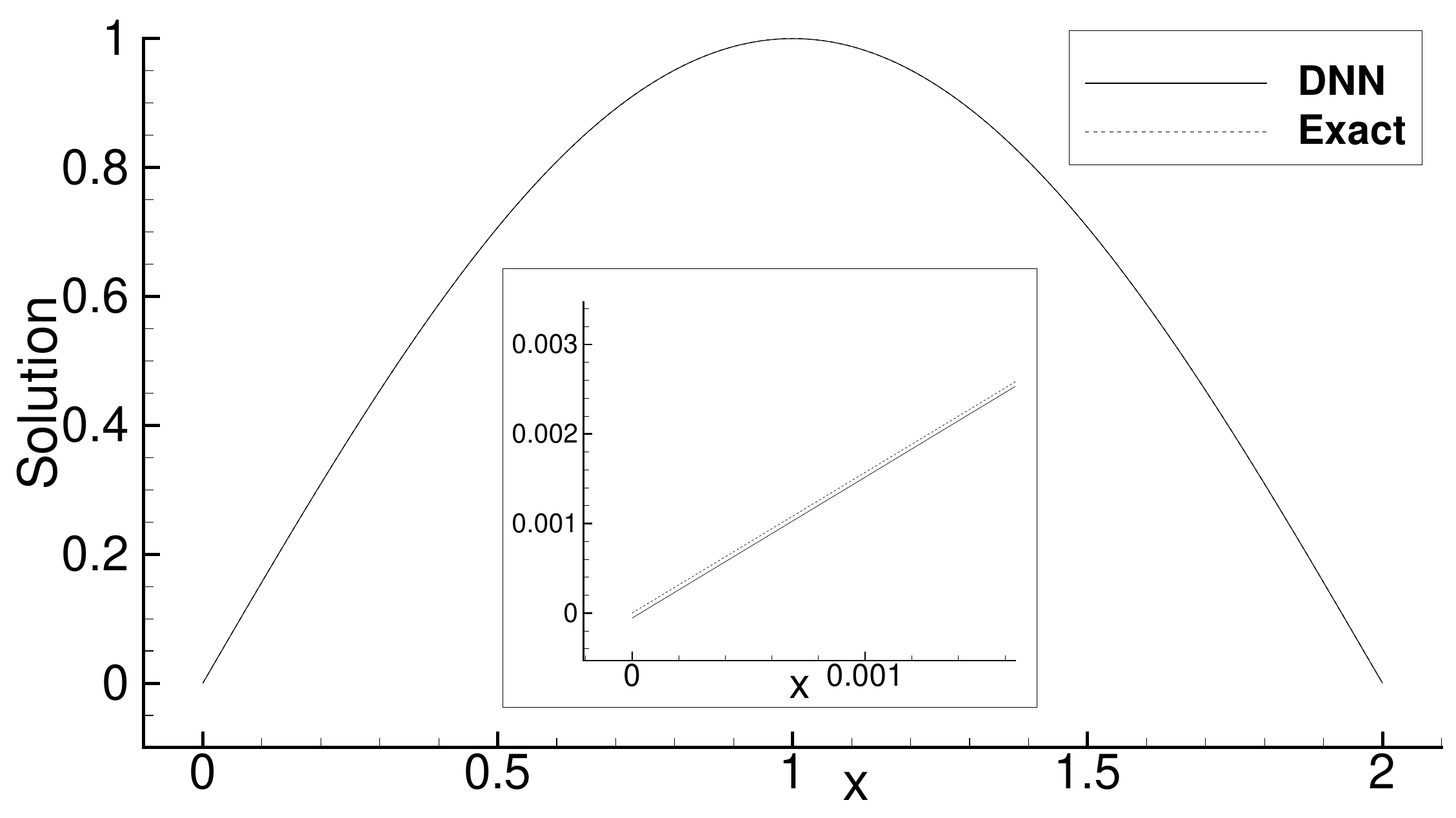}(b)
    \includegraphics[width=2in]{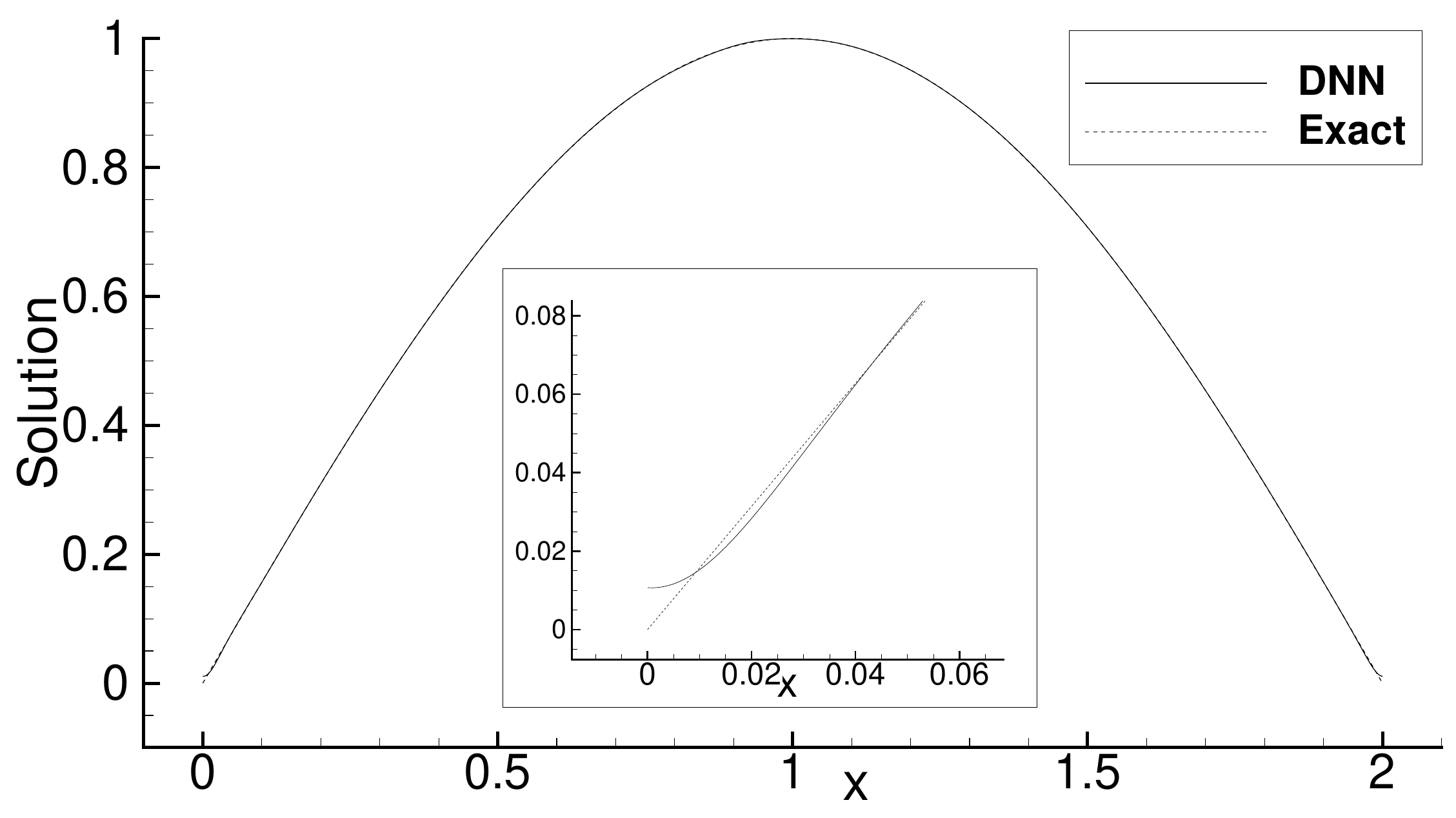}(c)
  }
  \centerline{
    \includegraphics[width=2in]{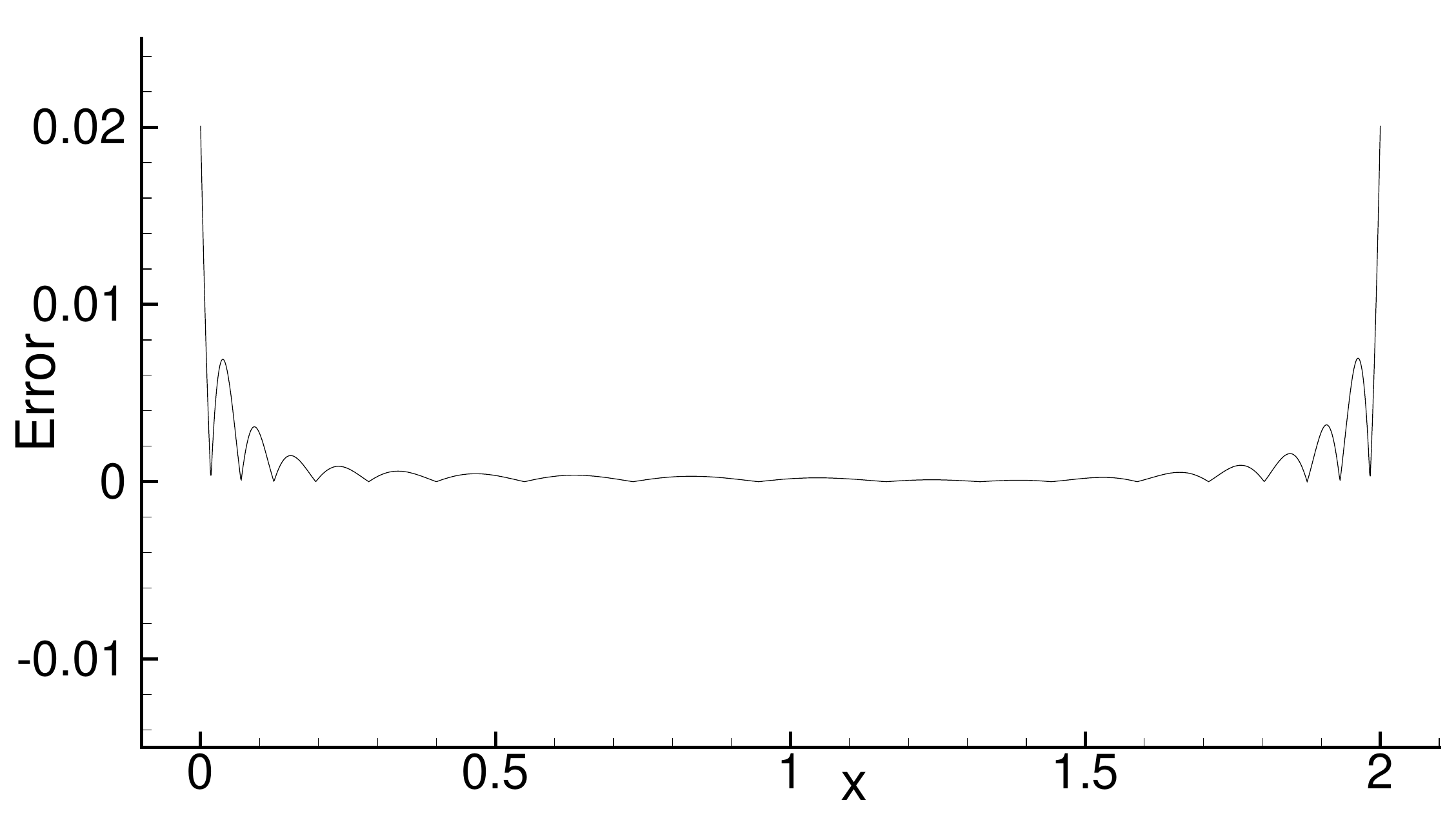}(d)
    \includegraphics[width=2in]{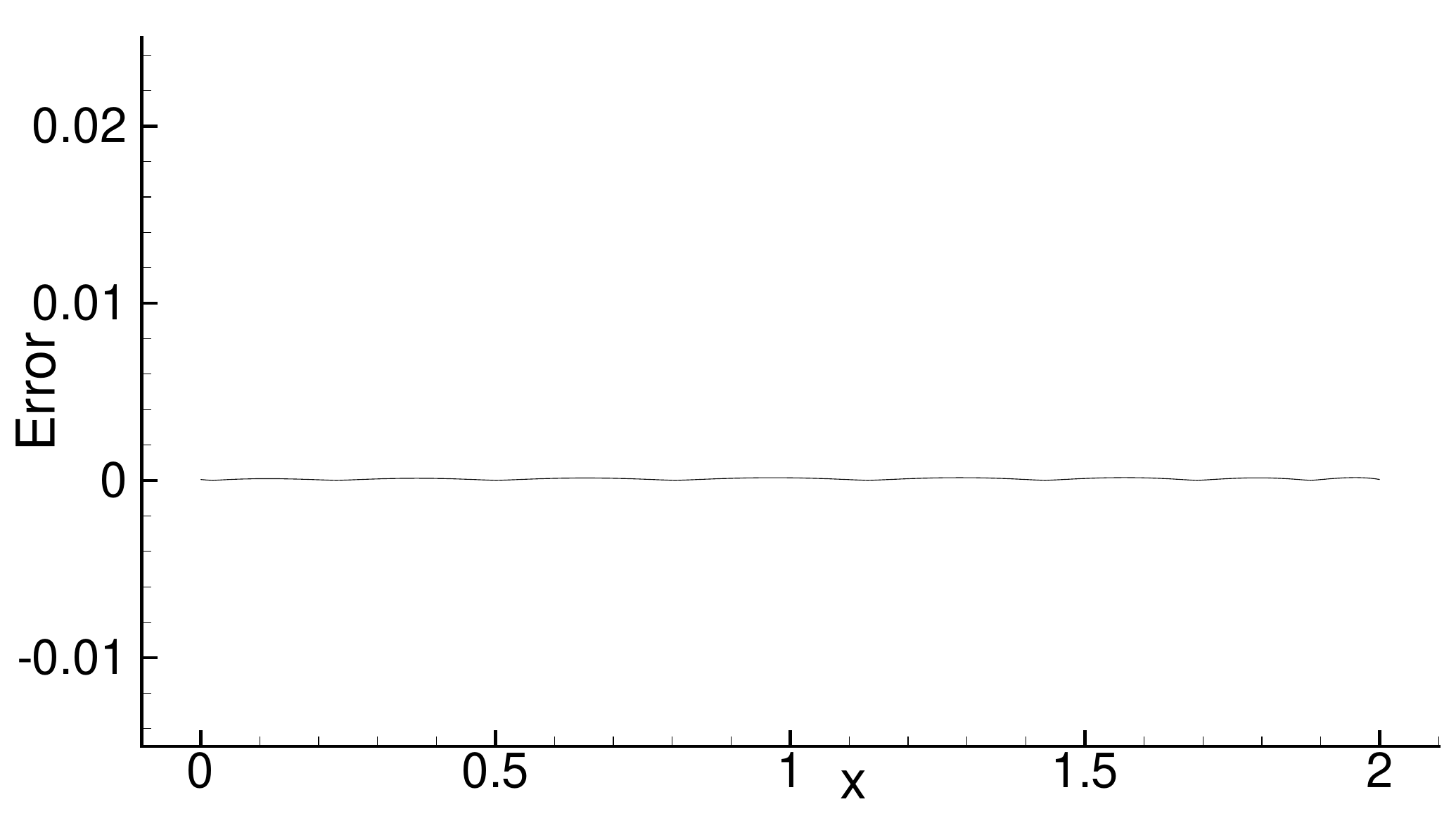}(e)
    \includegraphics[width=2in]{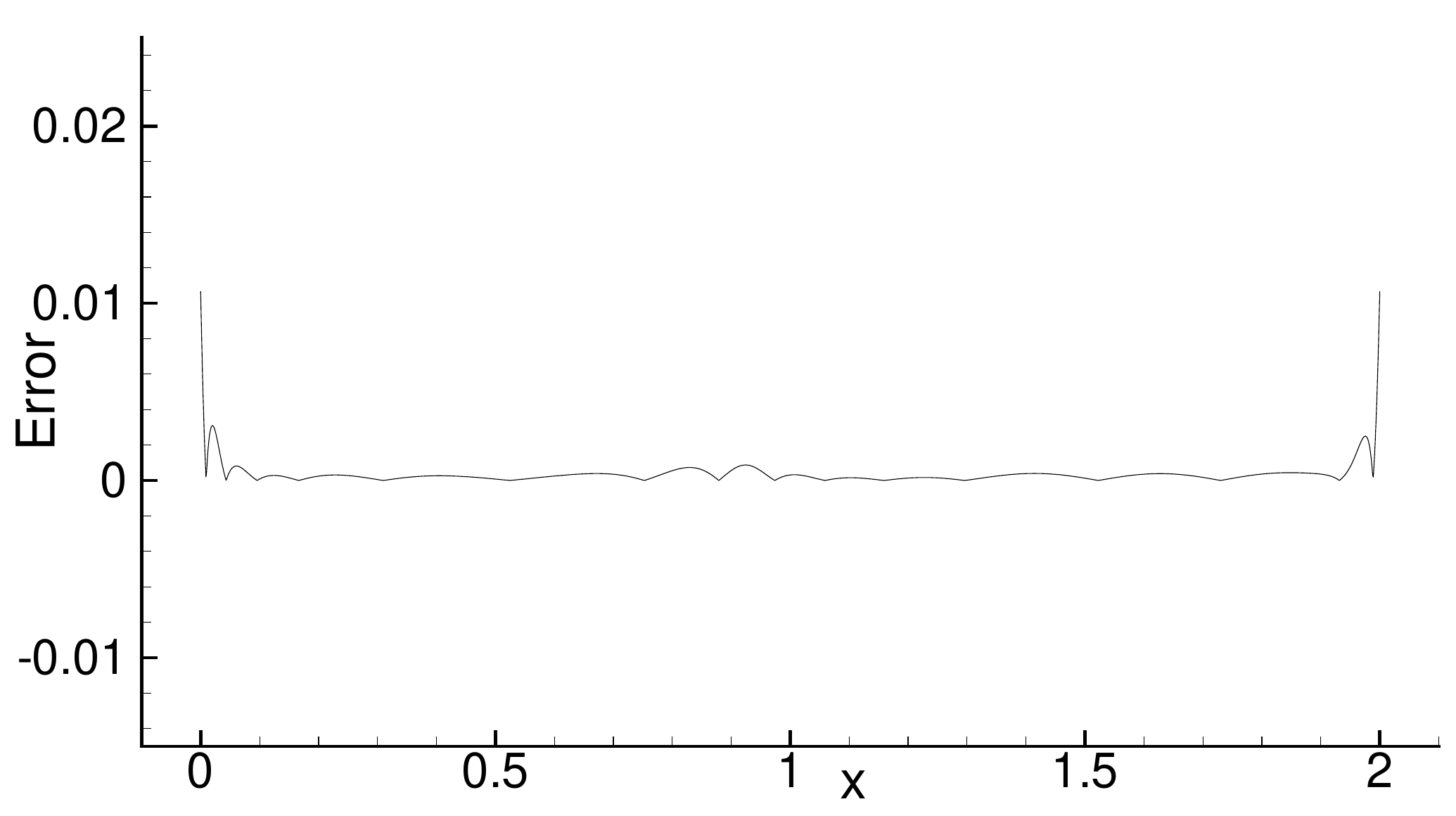}(f)
  }
  \caption{
    Approximation of $C^0$ periodic function $u_2(x)$:
    Comparison of approximation results (top row) and their errors (bottom row) obtained with
    (a,d) $C^{\infty}$, (b,e) $C^0$, and (c,f) $C^1$ periodic DNNs.
    The exact function is included for comparison.
    The insets are the magnified views near $x=0$.
  }
  \label{fig:c0_pf}
\end{figure}

\begin{table}
  \centering
  \begin{tabular}{lllll}
    \hline
     & $C^{\infty}$ periodic DNN & $C^0$ periodic DNN & $C^1$ periodic DNN & Exact value \\
    \hline
    $u_2(0)$ & 2.0077934937638e-02 & -5.6212180516792e-05 &
    1.0675453567003e-02 & 0  \\
    $u_2(2)$ & 2.0077934937638e-02 & -5.6212180516792e-05 &
    1.0675453567003e-02 & 0 \\
    $u'_2(0)$ & 2.0046381097937e-03 & \fbox{1.5738634304121e+00} &
    -8.9688048016955e-02 & 1.5707963267949e+00 \\
    $u'_2(2)$ & 2.0046381097902e-03 & \fbox{-1.5651374257149e+00} &
    -8.9688048016955e-02 & -1.5707963267949e+00 \\
    $u''_2(0)$ & 4.8956024103701e+01 & \fbox{-3.0860019634346e-02} &
    \fbox{1.1444431255354e+02} & 0 \\
    $u''_2(2)$ & 4.8956024103700e+01 & \fbox{1.7388612411815e-01} &
    \fbox{1.0788094853391e+02} & 0 \\
    \hline
  \end{tabular}
  \caption{
    Approximation of $C^0$ periodic function $u_2(x)$:
    Values of the function and its derivatives at the left/right
    domain boundaries ($x=0$ and $x=2$)
    from the  $C^{\infty}$, $C^0$ and $C^1$ periodic DNN approximations
    and from the exact function.
    $14$ significant digits 
    are listed  to demonstrate
    that the current method enforces the periodic conditions exactly.
  }
  \label{tab:c0_pf}
\end{table}

We employ the same DNN and simulation parameters to approximate $u_2(x)$
as for $u_1(x)$; see Table \ref{tab:peri_func}.
The loss function is given by \eqref{equ:loss_u1}, with
$u_1(x)$ replaced by $u_2(x)$.
Figure \ref{fig:c0_pf} is a comparison of the approximation results
and their errors obtained with $C^{\infty}$, $C^0$ and $C^1$
periodic DNNs.
The exact function $u_2(x)$ has also been included for comparison.
The $C^0$ periodic DNN produces results that are considerably more
accurate than the other two methods, as expected.
The $C^{\infty}$ and $C^1$ periodic DNN approximations produce
accurate results in the bulk of the domain, but exhibit larger errors
near/at the domain boundaries. The $C^{\infty}$ and $C^1$
periodic conditions appear to have the tendency of bending
the function curve near
the boundaries to achieve periodicity for the derivatives;
see the insets of Figures \ref{fig:c0_pf}(a,b,c).

To verify the periodicity of the DNN approximations on the boundaries,
we list in Table \ref{tab:c0_pf} the values
of the approximated $u_2(x)$ and its derivatives (up to order two)
on the domain boundaries $x=0$ and $2$ from different
approximations and from the exact function $u_2(x)$.
Again $14$ significant digits have been shown for each value.
It is observed that the current methods indeed enforce
exactly the periodicity for the approximation function and its
derivatives on the boundaries as expected.
In the $C^{\infty}$ periodic DNN approximation, the function 
and its derivatives (up to order 2 considered here) have
identical values on the two boundaries. In contrast,
with the $C^0$ periodic DNN approximation only the function value
is identical on the boundaries, and with the $C^1$ periodic DNN
approximation the function and the first derivative have
identical values on the two boundaries.

\begin{figure}
  \centerline{
    \includegraphics[width=2in]{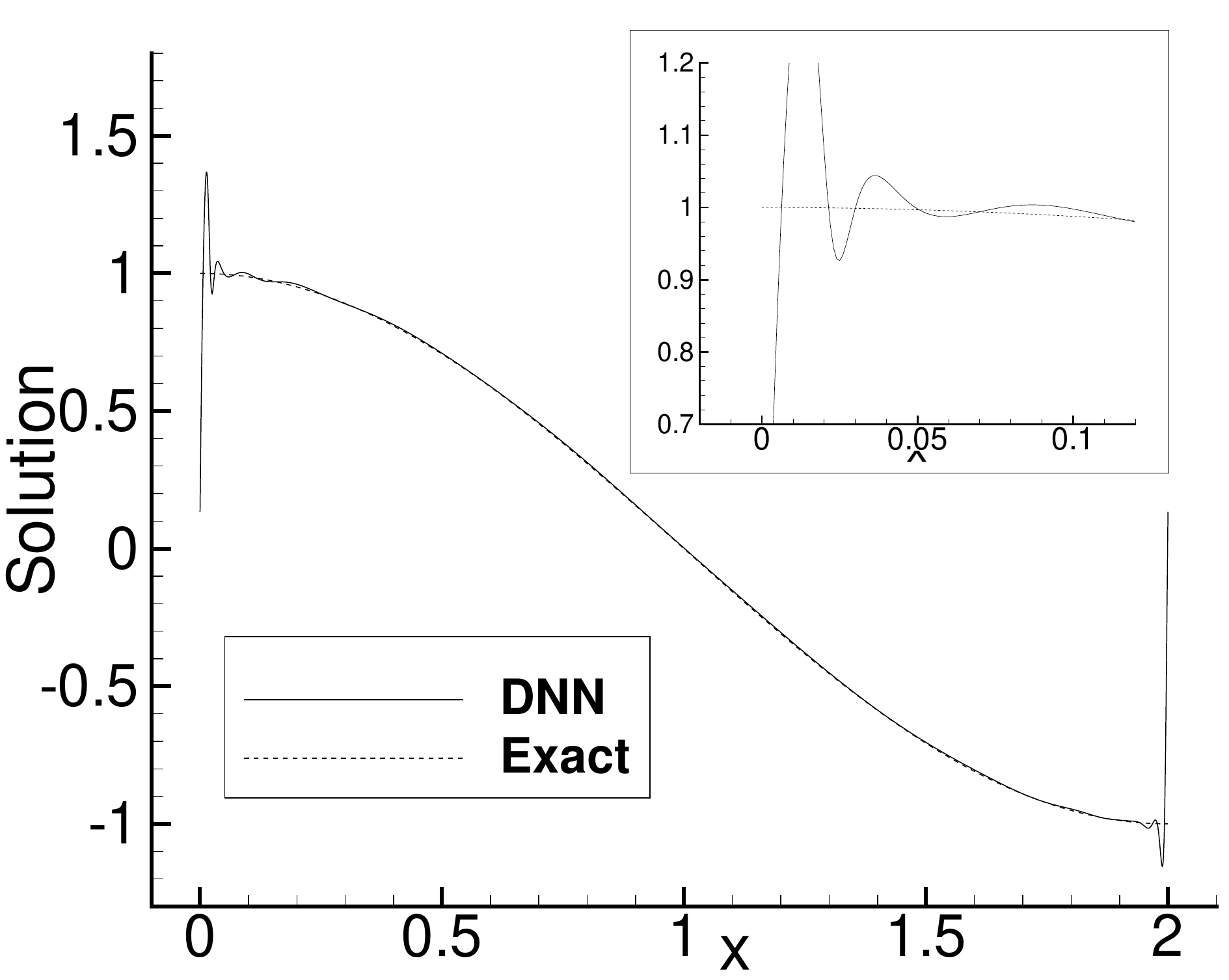}(a)
    \includegraphics[width=2in]{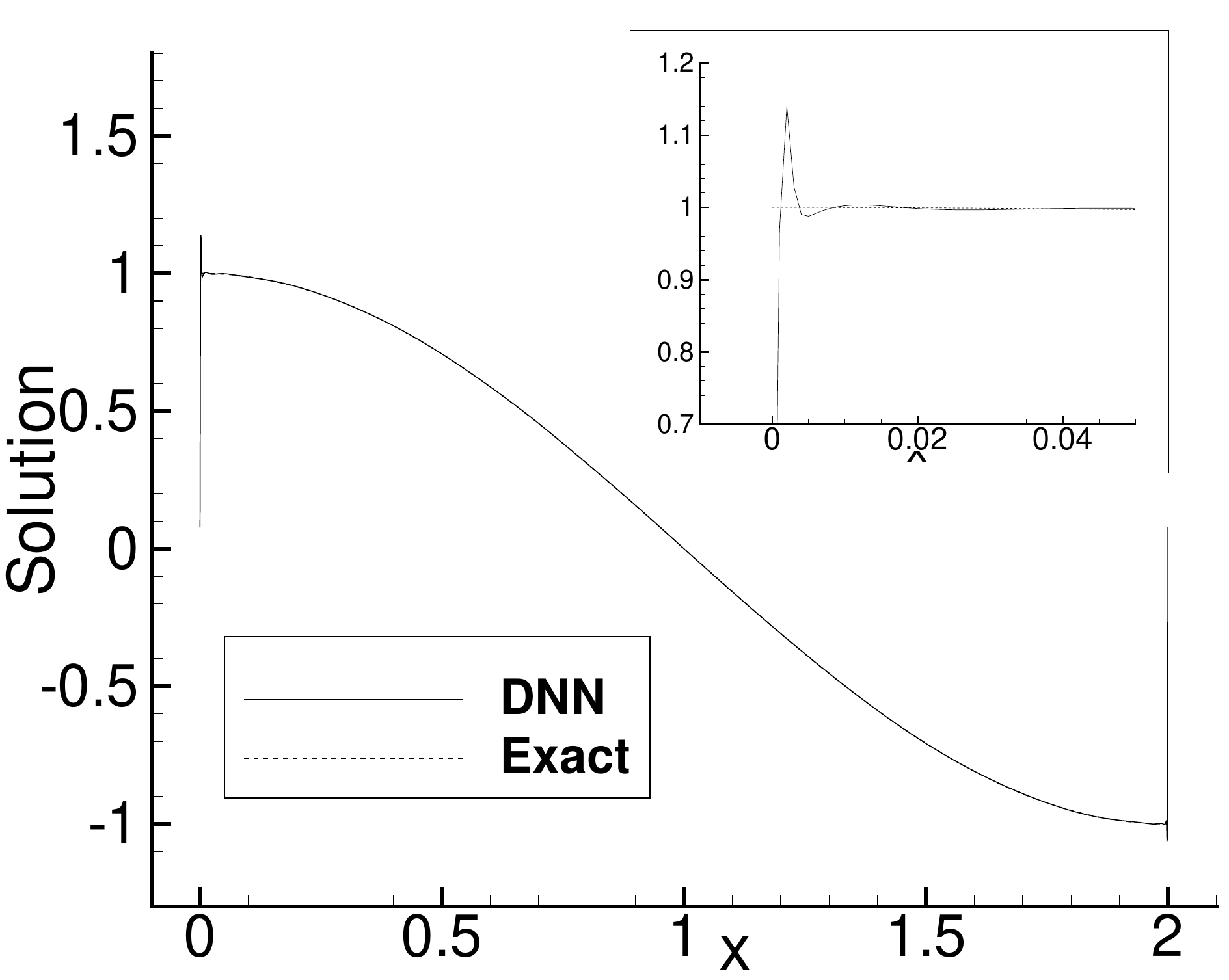}(b)
    \includegraphics[width=2in]{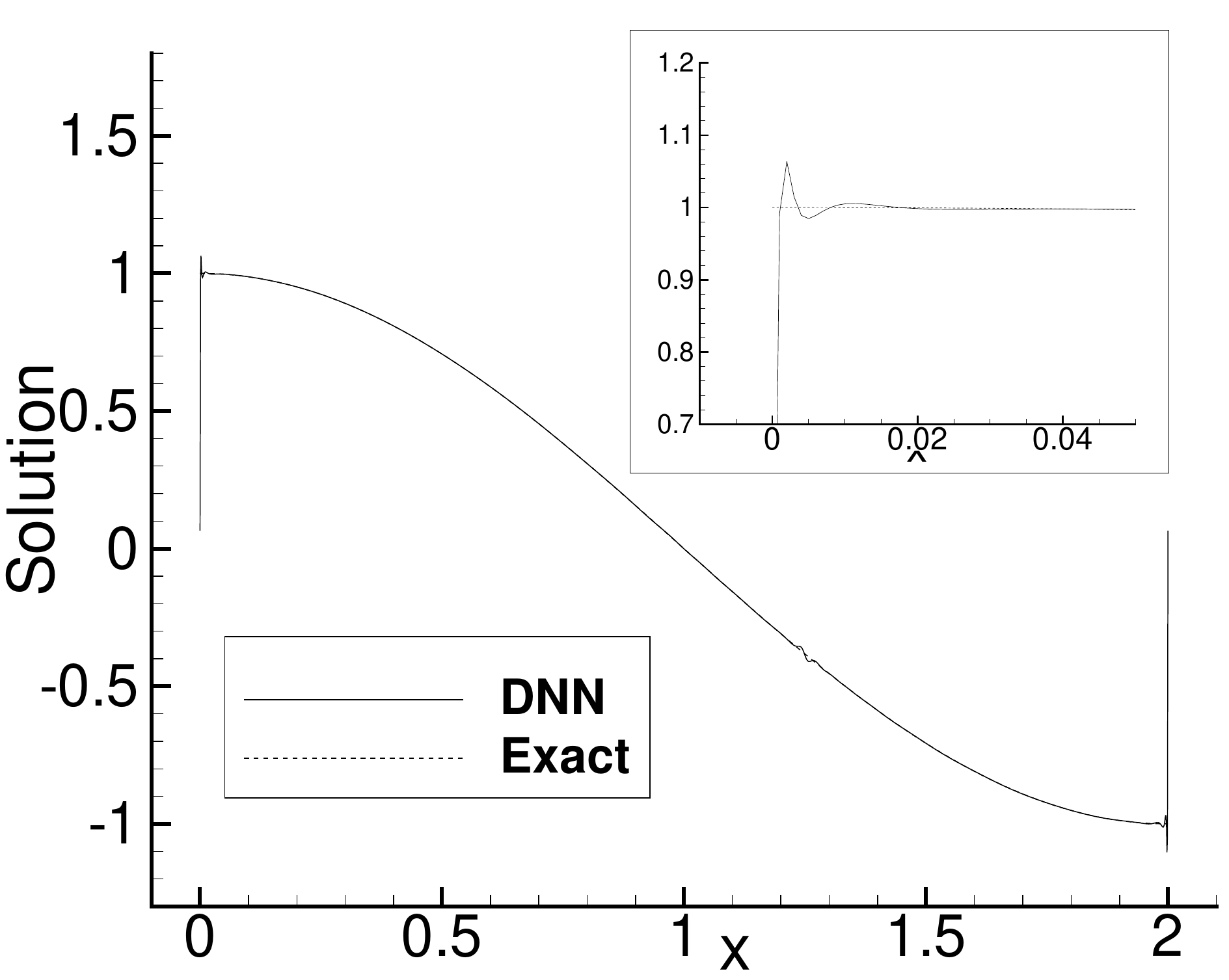}(c)
  }
  \centerline{
    \includegraphics[width=2in]{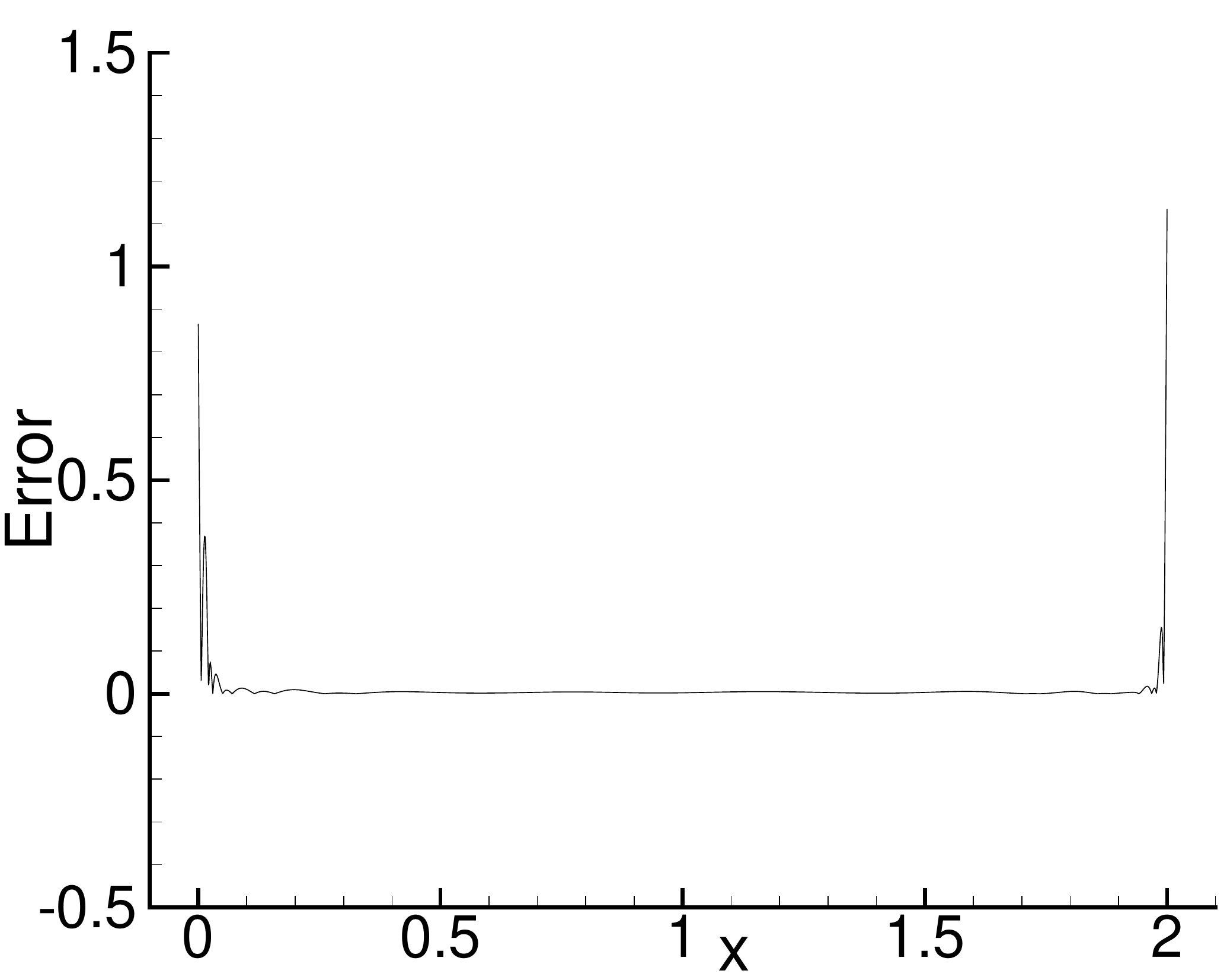}(d)
    \includegraphics[width=2in]{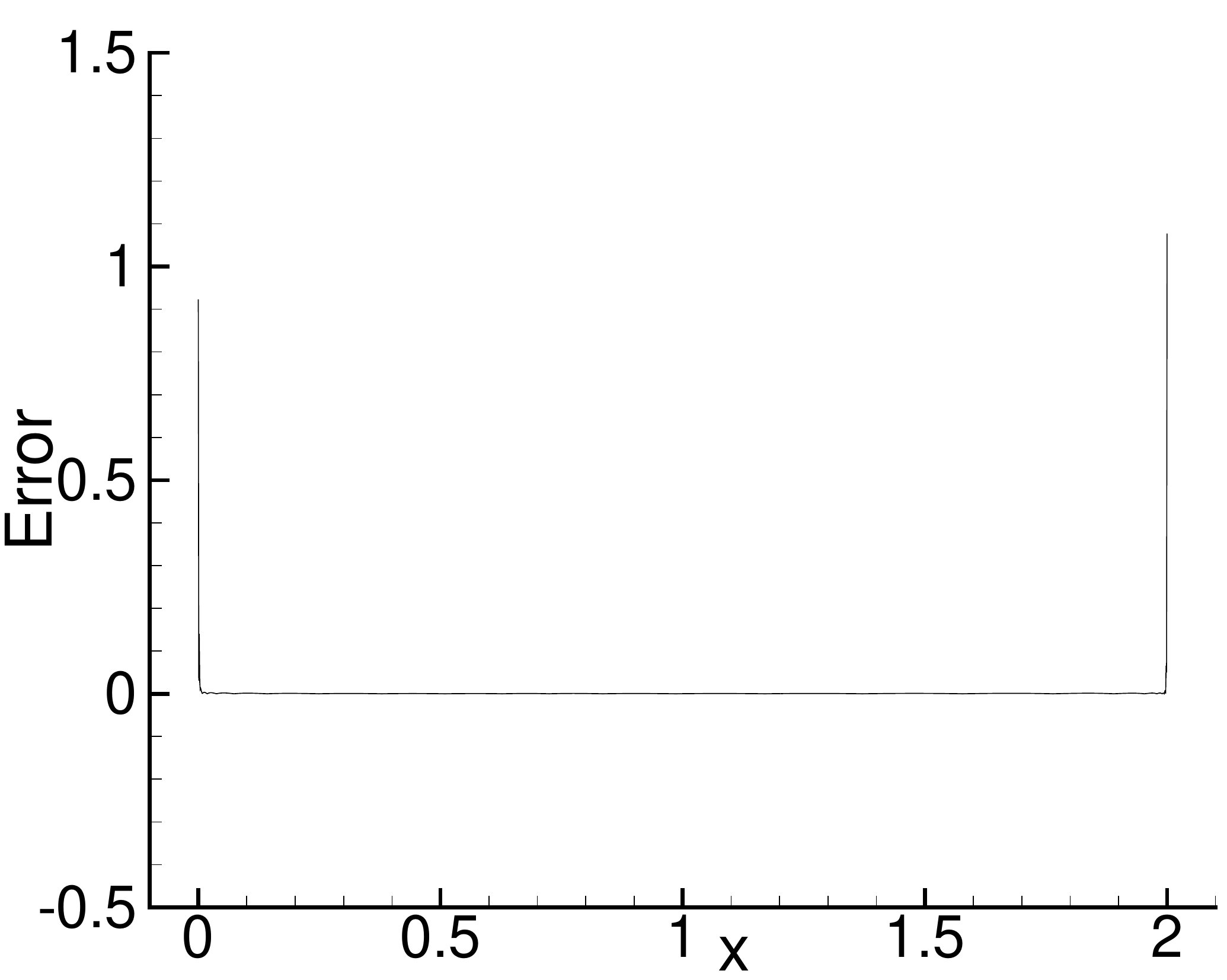}(e)
    \includegraphics[width=2in]{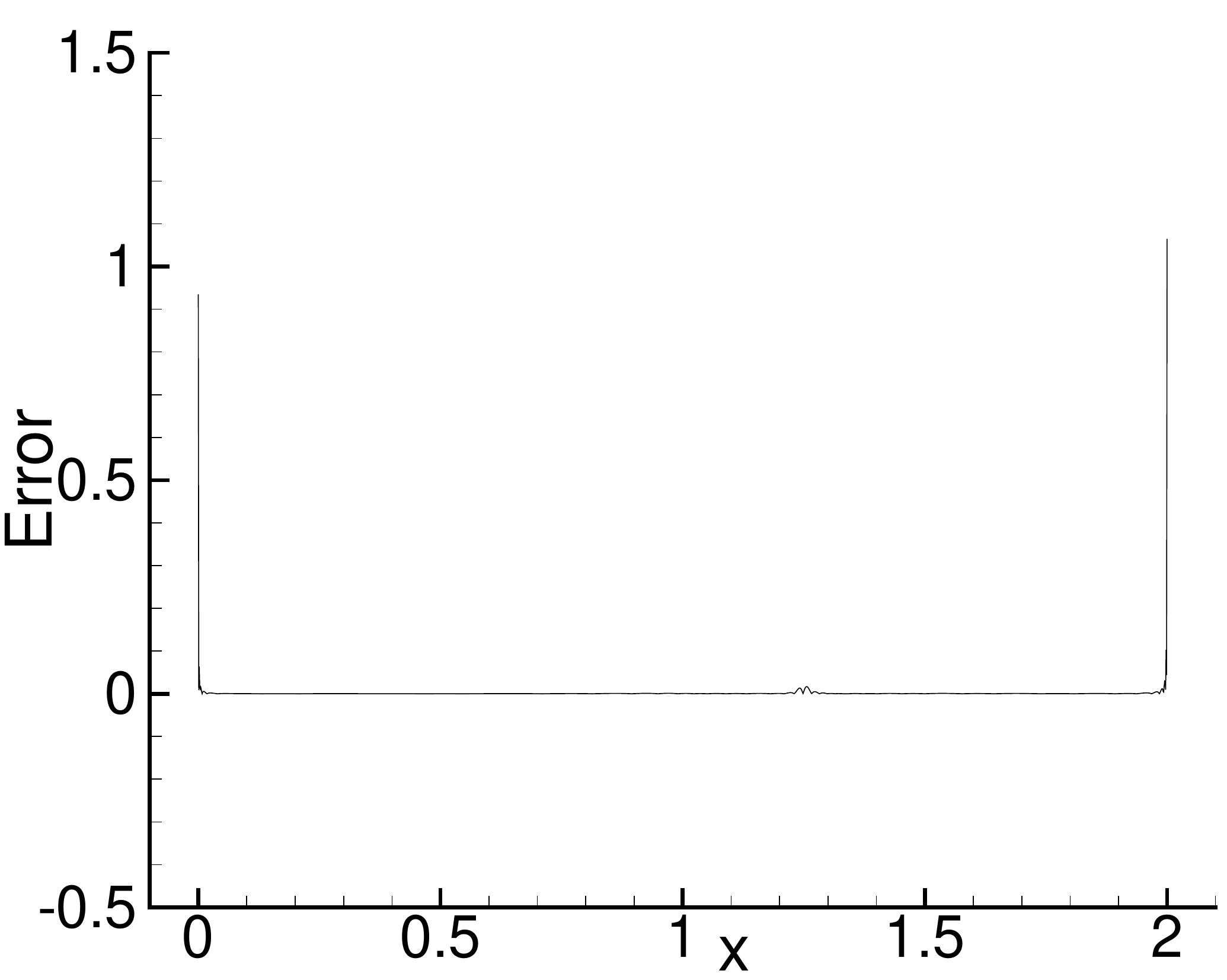}(f)
  }
  \caption{
    Approximation of non-periodic function $u_3(x)$: Approximation results (top row)
    and their errors (bottom row) obtained with
    (a,d) $C^{\infty}$, (b,e) $C^0$, (c,f) $C^1$, periodic DNNs.
    The exact profile of the function $u_3(x)$ is also included.
    The insets are magnified views near $x=0$.
  }
  \label{fig:non_pf}
\end{figure}

\begin{table}
  \centering
  \begin{tabular}{lllll}
    \hline
     & $C^{\infty}$ periodic DNN & $C^0$ periodic DNN & $C^1$ periodic DNN & Exact value \\
    \hline
    $u_3(0)$ & 1.3452550840065e-01 & 7.7219924514080e-02 &
    6.4853245159018e-02 & 1  \\
    $u_3(2)$ & 1.3452550840063e-01 & 7.7219924514080e-02 &
    6.4853245159018e-02 & -1 \\
    $u'_3(0)$ & 1.7647534129428e+02 & \fbox{7.0097116541954e+02} &
    1.5259262265715e+03 & 0 \\
    $u'_3(2)$ &  1.7647534129428e+02 &  \fbox{1.0015322821138e+03} &
    1.5259262265715e+03 & 0 \\
    $u''_3(0)$ & -9.2430741908226e+03 & \fbox{8.4509087153852e+05} &
    \fbox{-4.6596618576551e+05} & -2.4674011002723e+00 \\
    $u''_3(2)$ & -9.2430741908221e+03 & \fbox{-9.4100497028271e+05} &
    \fbox{-4.5176214142043e+05} & 2.4674011002723e+00 \\
    \hline
  \end{tabular}
  \caption{
    Approximation of the non-periodic function $u_3(x)$:
    Values of the function and its derivatives at the left/right
    domain boundaries ($x=0$ and $x=2$)
    from the approximations with $C^{\infty}$, $C^0$ and $C^1$
    periodic DNNs and from the exact function.
    $14$ significant digits 
    are listed
    to demonstrate
    that the current method enforces the periodic conditions exactly
    for the approximation function and its derivatives.
  }
  \label{tab:non_pf}
\end{table}

We finally consider a non-periodic function,
\begin{equation}
  u_3(x) = \cos\frac{\pi x}{2},
\end{equation}
on the domain $\Omega = \{x| 0\leqslant x\leqslant 2  \}$.
We would like to approximate this function using 
$C^{\infty}$ and $C^k$ ($k=0,1$) periodic DNNs.
The DNN and simulation parameter values employed here are the same as
those for $u_1(x)$ and $u_2(x)$ (see Table \ref{tab:peri_func}),
except for the number of quadrature points within each element.
Here for $u_3(x)$ we employ $Q=40$ with the $C^{\infty}$ periodic DNN,
and $Q=50$ with the $C^0$ and $C^1$ periodic DNNs.

Figure \ref{fig:non_pf} shows the approximation functions $u_3(x)$ and
their errors obtained with $C^{\infty}$, $C^0$ and $C^1$ periodic DNNs.
In the bulk of the domain the DNN approximations appear to be in
good agreement with the exact function $u_3(x)$ with all three methods.
In a region near the two boundaries, the periodic DNN approximations
exhibit large errors, and one can observe fluctuations in the
approximation functions (Gibbs phenomenon).
Table \ref{tab:non_pf} lists the values of the approximation function $u_3(x)$
and its derivatives on the two boundaries ($x=0,2$)
obtained with the $C^{\infty}$, $C^0$ and $C^1$ periodic DNNs
as well as the exact function $u_3(x)$.
The results again demonstrate that the current methods
enforce exactly the periodicity for the approximation function and its
derivatives.

\subsection{One-Dimensional Helmholtz Equation with Periodic BCs}

In this subsection we test the performance of the proposed method
with the one-dimensional (1D) Helmholtz equation,
\begin{equation}\label{equ:helm1d}
  \frac{d^2u}{dx^2} - \lambda u = f(x),
\end{equation}
on the domain $\Omega = \{x | a\leqslant x\leqslant b  \}$,
where $\lambda$ ($\lambda\geqslant 0$), $a$ and $b$ are given constants and
$f(x)$ is a prescribed source term. We impose periodic boundary conditions (BC)
on the domain boundaries, $x=a$ and $b$.

Specifically,
we consider two types of periodic boundary conditions.
The first type is the $C^1$
periodic condition,
\begin{equation}\label{equ:helm1d_c1}
  u(a) = u(b), \quad
  u'(a) = u'(b).
\end{equation}
The second type is the $C^{\infty}$ periodic condition,
\begin{equation}\label{equ:helm1d_cinf}
  u(a)=u(b), \quad
  u'(a)=u'(b), \quad
  u''(a) = u''(b), \quad \dots, \quad
  u^{(m)}(a) = u^{(m)}(b), \quad \dots
\end{equation}
Note that with the $C^{\infty}$ periodic condition \eqref{equ:helm1d_cinf},
we are effectively seeking a $C^{\infty}$ periodic function, with the period $L=b-a$,
on the infinite domain $x\in(-\infty, \infty)$ that solves
the equation \eqref{equ:helm1d}.
Since the Helmholtz equation is a second-order equation, imposing
the $C^0$ periodic condition only, i.e.~$u(a)=u(b)$, does not
lead to a unique solution to the problem.

For the numerical tests in this section we fix the problem parameters
to the following values:
\begin{equation}
  \lambda = 10, \quad
  a = 0, \quad
  b = 4, \quad
  L = b-a = 4.
\end{equation}
We choose the source term $f(x)$ such that the Helmholtz
equation \eqref{equ:helm1d}
has an analytic solution
\begin{equation}\label{equ:helm1d_anal}
  u(x) = \sin[3\pi(x+0.05)]\cos[2\pi(x+0.05)] + 2.
\end{equation}
This is a periodic function with $L=4$ as a period, and
it satisfies the boundary
conditions~\eqref{equ:helm1d_c1} and~\eqref{equ:helm1d_cinf}.



To simulate this problem, we employ a feed-forward neural network (Figure \ref{fig:dnn})
with $4$ hidden layers, with $20$ nodes in each layer, apart from
the input and output layers. The input to the network is the coordinate $x$ (1 node),
and the output of the network is the solution to the Helmholtz equation
$u$ (1 node). The second layer of the network (or the first hidden layer)
is set to be a $C^{\infty}$ periodic layer $\mathcal{L}_p(m,n)$ with $m=11$ and $n=20$,
in which we set the constant $\omega = \frac{2\pi}{L}=\frac{\pi}{2}$
in equation \eqref{equ:def_zi},
when the $C^{\infty}$ periodic boundary conditions in \eqref{equ:helm1d_cinf} are imposed.
When the $C^1$ periodic periodic boundary conditions in \eqref{equ:helm1d_c1} are imposed,
we set the second layer of the network to be a $C^1$ periodic layer
$\mathcal{L}_{C^1}(m,n)$ with $m=11$ and $n=20$, as detailed in Section \ref{sec:method}.

\begin{figure}
  \centerline{
    \includegraphics[width=3in]{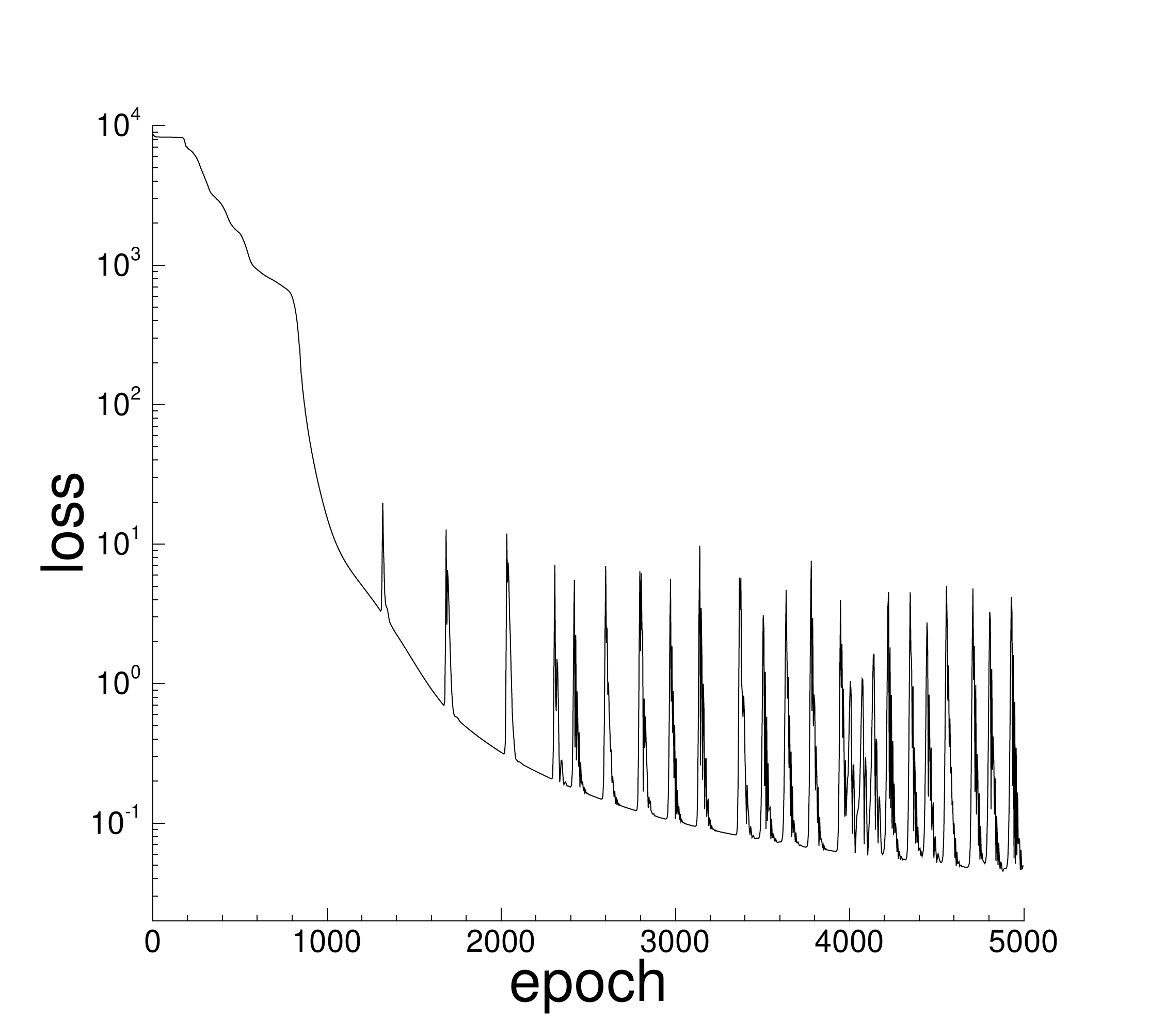}(a)
    \includegraphics[width=3in]{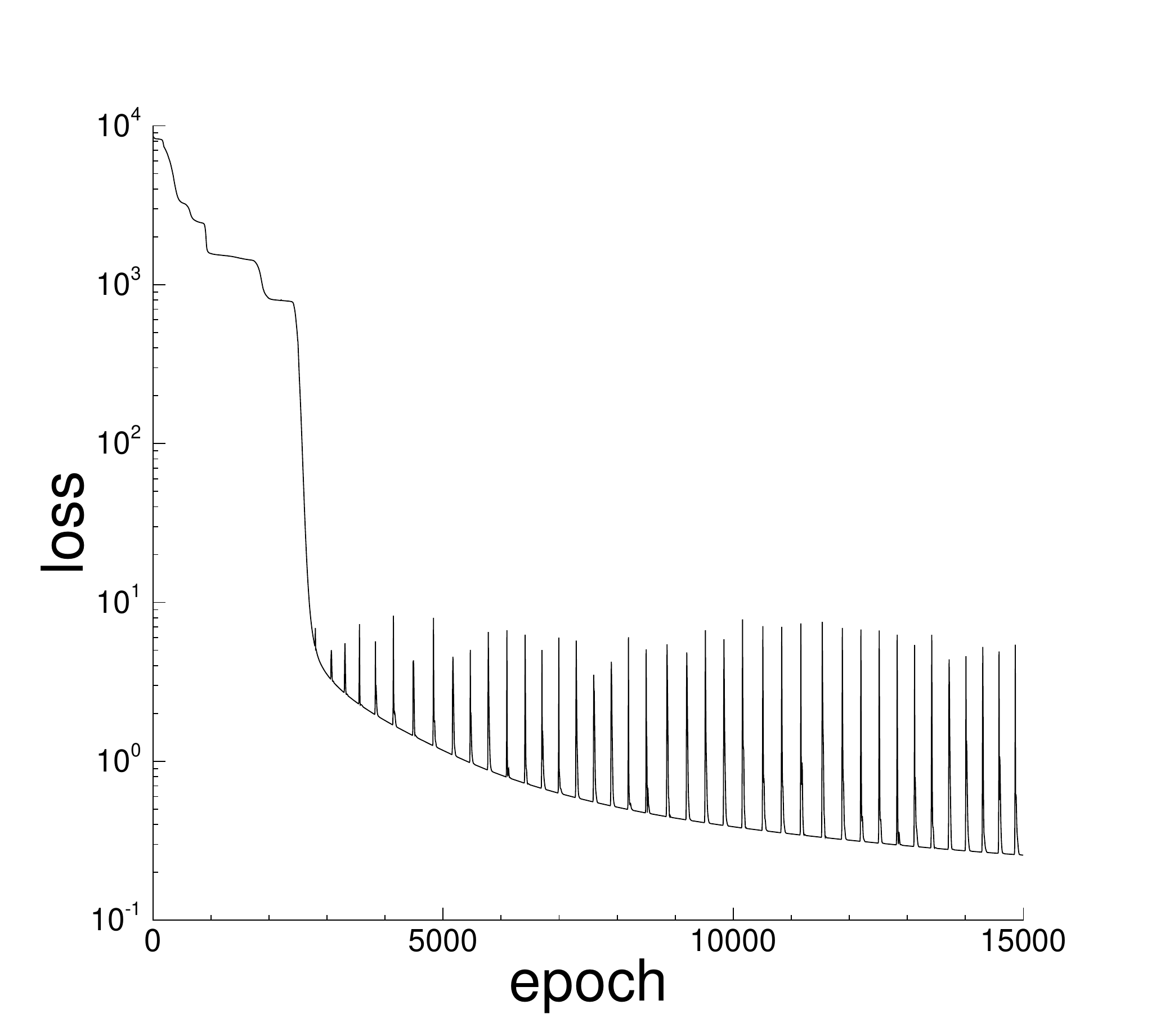}(b)
  }
  \caption{
    1D Helmholtz equation: training histories of
    the loss function with the (a) $C^{\infty}$
    and (b) $C^1$ periodic boundary conditions.
  }
  \label{fig:helm1d_loss}
\end{figure}


We minimize the following loss function with this DNN,
\begin{equation}\label{equ:loss_helm1d}
  \begin{split}
  \text{Loss} &= \frac{1}{L}\int_{\Omega}\left[
    \frac{d^2u}{dx^2} - \lambda u - f(x)
    \right]^2 dx
  =\frac{1}{L}\sum_{e=0}^{N_e-1}\int_{\Omega_e} \left[
    \frac{d^2u}{dx^2} - \lambda u - f(x)
    \right]^2 dx \\
  &= \frac{1}{L}\sum_{e=0}^{N_e-1}\sum_{i=0}^{Q-1}\left[
    \left.\frac{d^2u}{dx^2}\right|_{x_i^e} - \lambda u(x_i^e) - f(x_i^e)
    \right]^2 J^e w_i.
  \end{split}
\end{equation}
In this equation, $N_e$ is the number of elements (sub-intervals) we
have partitioned the domain $\Omega$ into in order to compute the integral,
$\Omega_e$ ($0\leqslant e\leqslant N_e-1$) denotes the region of element $e$,
$Q$ is the number of quadrature points within each element,
$J^e$ is the Jacobian of $\Omega_e$ with respect to the standard element $[-1,1]$,
$x_i^e$ ($0\leqslant i\leqslant Q-1$)
are the Gauss-Lobatto-Legendre quadrature points within element $e$
for $0\leqslant e\leqslant N_e-1$, and
$w_i$ ($0\leqslant i\leqslant Q-1$) are the weights associated with
the Gauss-Lobatto-Legendre quadrature.
The input data to the network consist of $x_i^e$ ($0\leqslant i\leqslant Q-1$,
$0\leqslant e\leqslant N_e-1$), and the label data consist of
$f(x_i^e)$ ($0\leqslant i\leqslant Q-1$,
$0\leqslant e\leqslant N_e-1$).
In the expression \eqref{equ:loss_helm1d},
$u(x_i^e)$ can be obtained from the output of the DNN, and
$\left.\frac{d^2u}{dx^2}\right|_{x_i^e}$ can be computed by
auto-differentiation.

\begin{figure}
  \centerline{
    \includegraphics[width=3in]{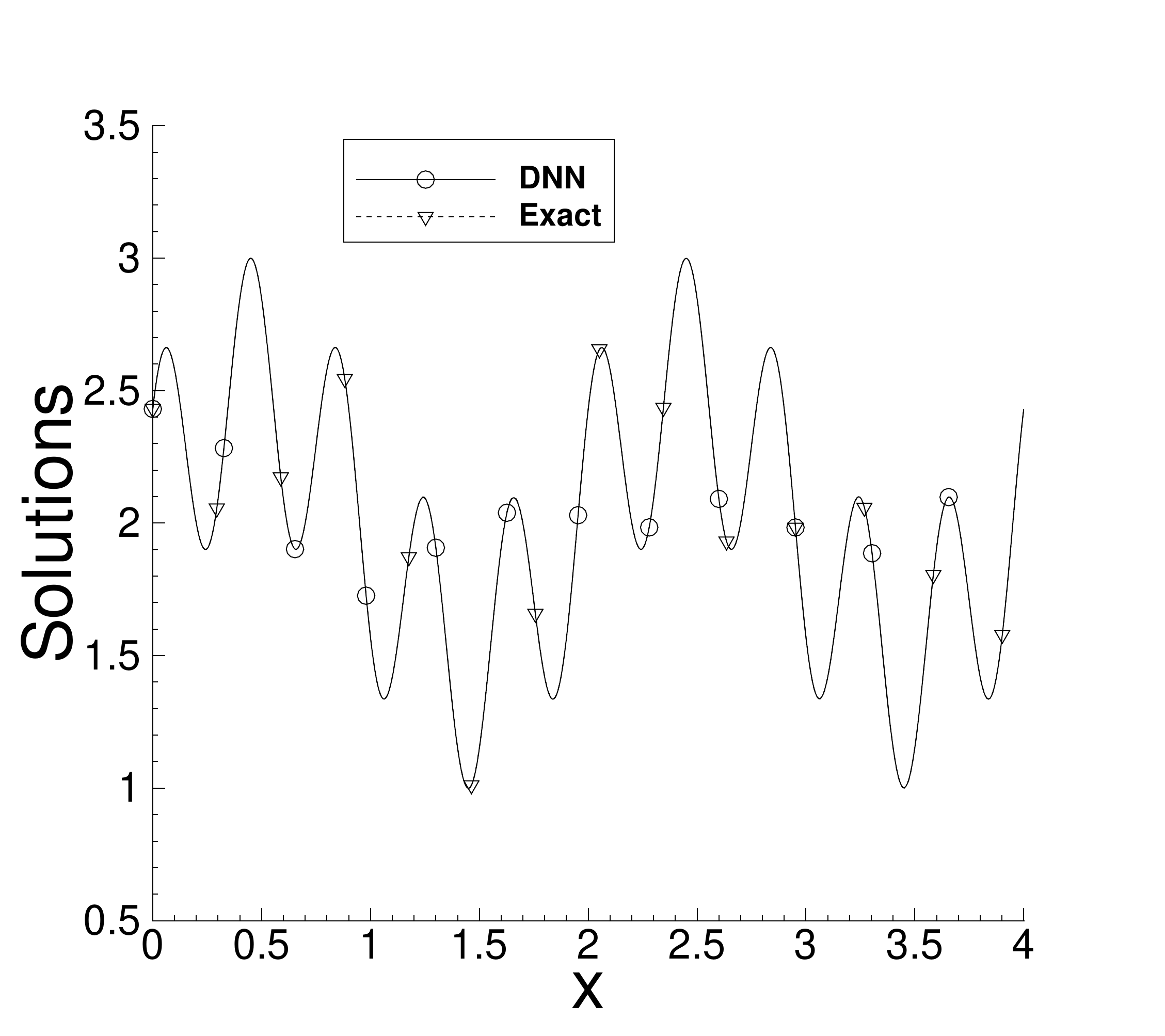}(a)
    \includegraphics[width=3in]{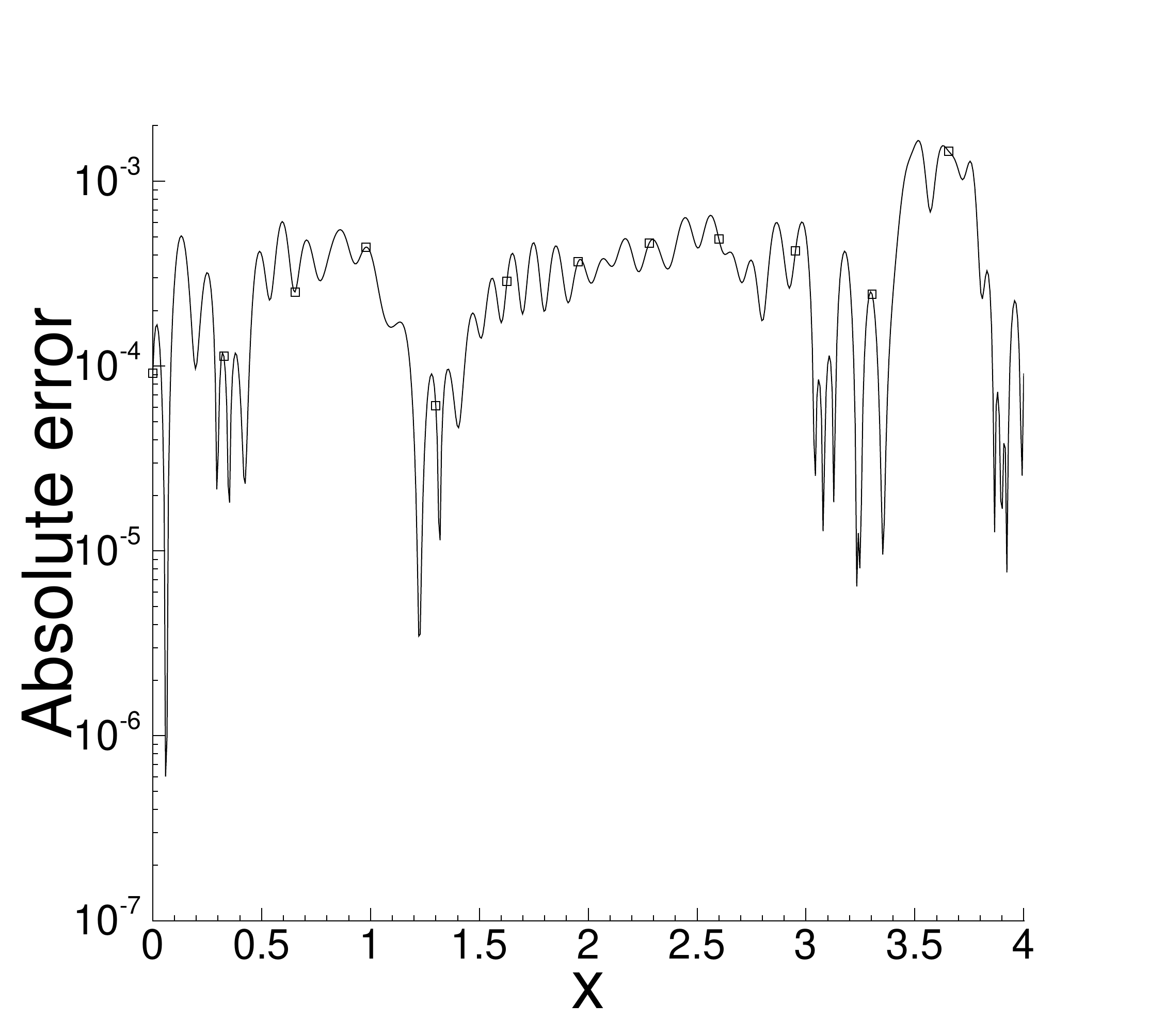}(b)
  }
  \centerline{
    \includegraphics[width=3in]{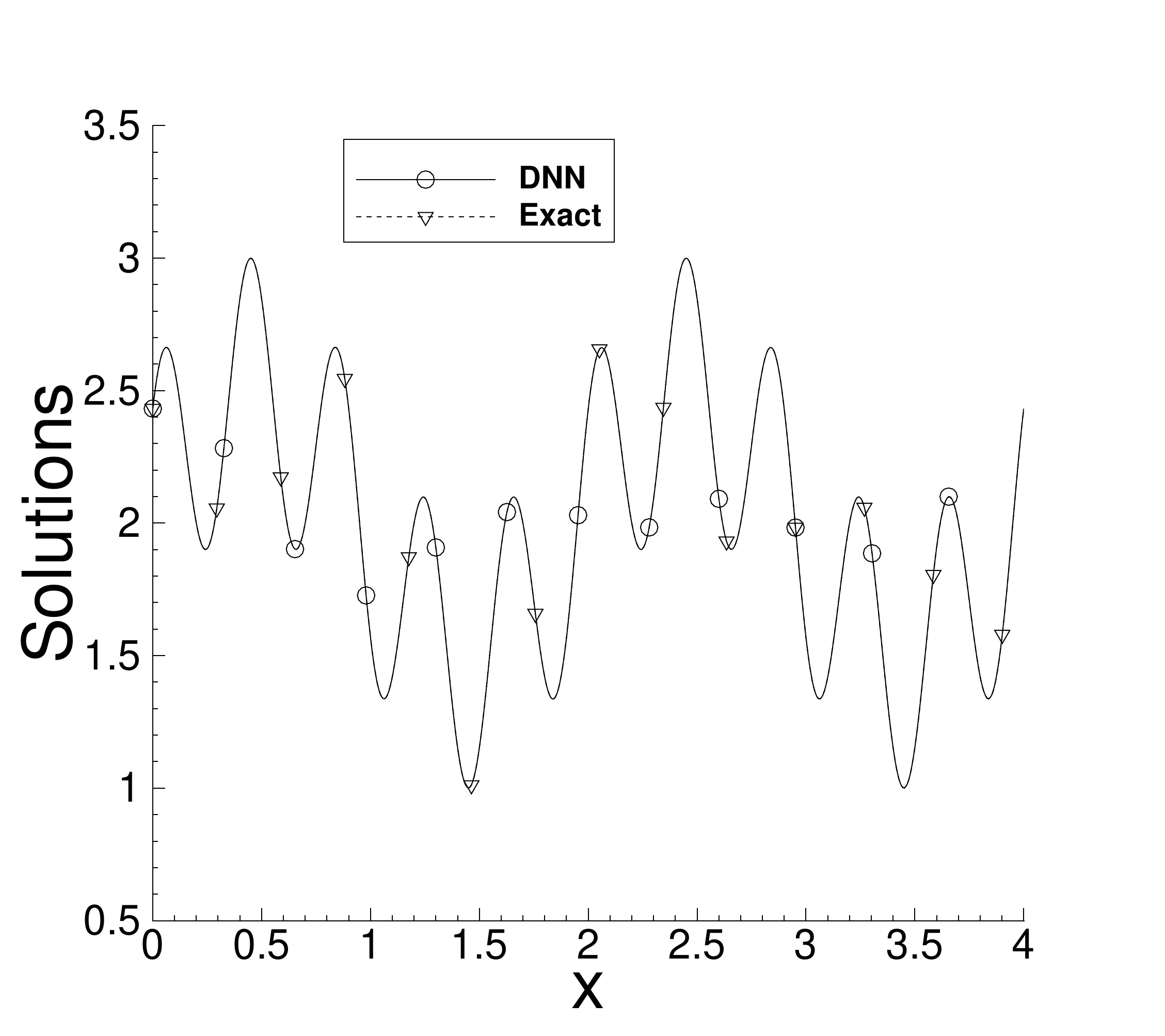}(c)
    \includegraphics[width=3in]{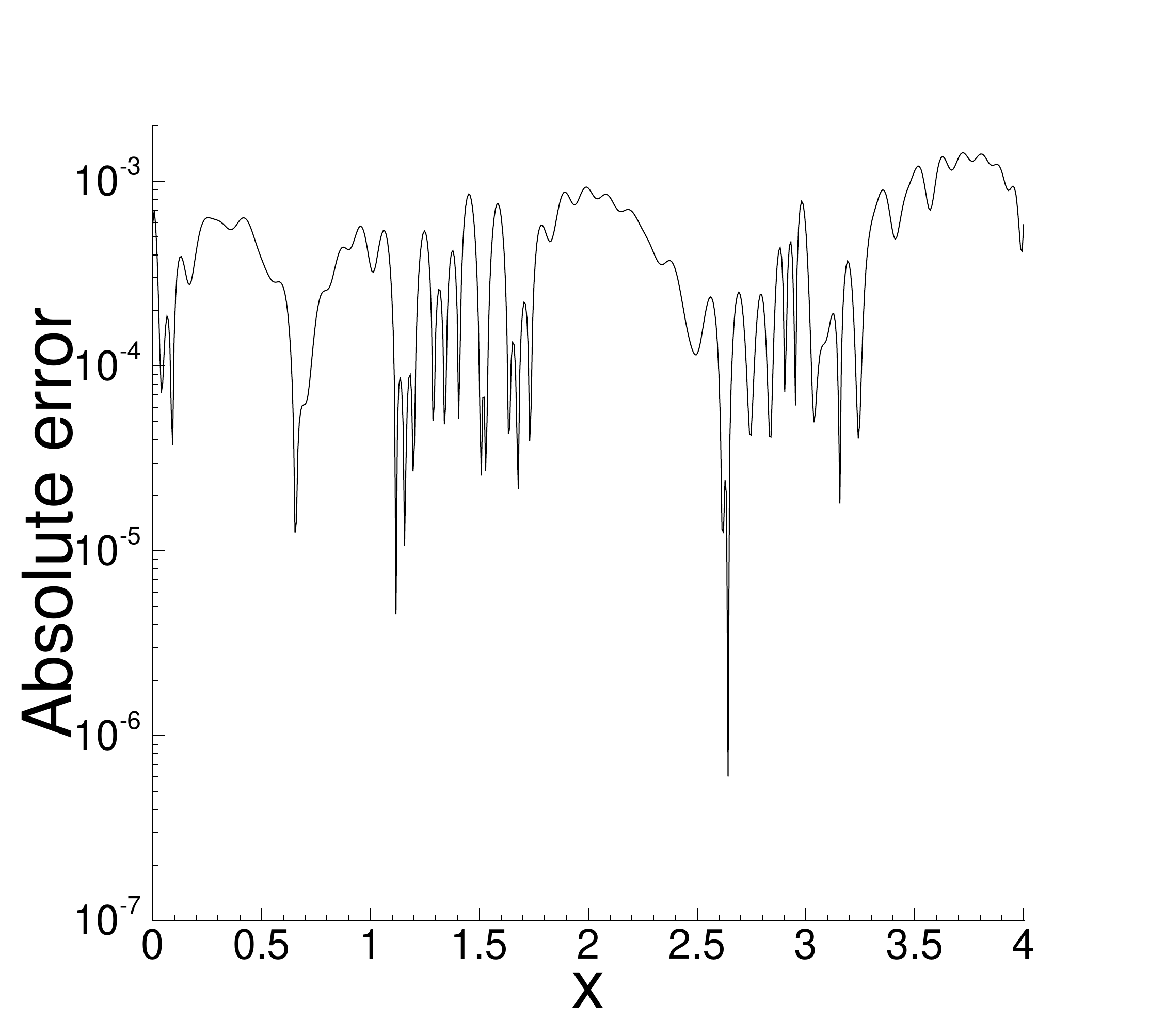}(d)
  }
  \caption{
    1D Helmholtz equation:
    (a,c), comparison between the exact and DNN solutions.
    (b,d), errors of  the DNN solutions against the exact
    solution.
    The results in (a) and (b) are obtained with the $C^{\infty}$ periodic BCs,
    an those in (c) and (d) are obtained with the $C^1$ periodic BCs.
    Periodic BCs are enforced using the method from Section \ref{sec:method}.
  }
  \label{fig:helm1d_soln}
\end{figure}


For the numerical experiments reported below,
we have partitioned the domain into three elements ($N_e=3$),
with these elements being $\Omega_0= [0, 1.3]$,
$\Omega_1=[1.3, 2.6]$ and $\Omega_2=[2.6, 4.0]$.
We employ $40$ quadrature points ($Q=40$) within each element.
The activation functions for all the hidden layers
are the hyperbolic tangent function (``tanh''), and no
activation function is applied to the output layer.
The DNN is trained using the Adam optimizer
for $5000$ epochs with a learning rate $10^{-3}$
for the $C^{\infty}$ periodic BC, and for $15000$ epochs
with a learning rate $5\times 10^{-4}$ for
the $C^1$ periodic BC. The options of ``early stopping''
and ``restore to best weight'' have been used in Tensorflow/Keras
when training the DNN.
The training histories of the loss function
for the $C^{\infty}$ and $C^1$ periodic BCs
are shown in Figure \ref{fig:helm1d_loss}.
We  observe characteristics in the loss histories
similar to those observed in Section \ref{sec:peri_func},
such as the varied loss reduction rates at different stages
and the fluctuations in the loss value with a fixed learning rate.

\begin{figure}
  \centerline{
    \includegraphics[width=3in]{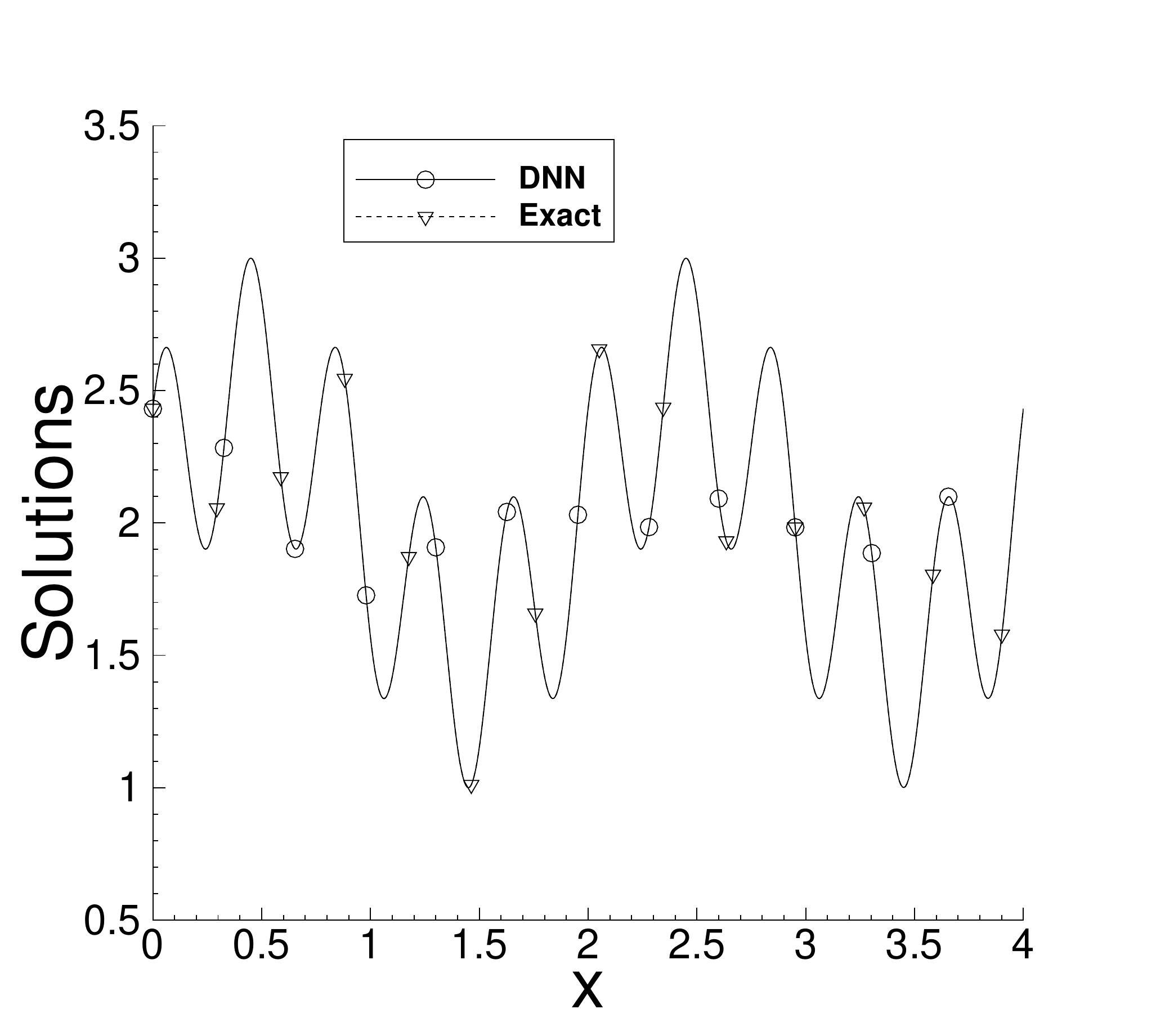}
  }
  \caption{
    1D Helmholtz equation: DNN solution 
    obtained
    with the $C^1$ periodic BC enforced using
    the penalty method.
  }
  \label{fig:helm1d_soln_penalty}
\end{figure}

\begin{table}
  \small
  \centering
  \begin{tabular}{lllll}
    \hline
    & DNN $C^{\infty}$ PBC (Current) & DNN $C^1$ PBC (Current) & DNN $C^1$ PBC (Penalty)
    & Exact solution \\
    $u(0)$ & 2.4318620484799e+00 & 2.4323606718327e+00 & \fbox{2.4314502844043e+00}  & 2.4317706231133e+00  \\
    $u(4)$ & 2.4318620484799e+00 & 2.4323606718327e+00 & \fbox{2.4317127865026e+00}  & 2.4317706231133e+00  \\
    $u'(0)$ & 7.1133416801319e+00 & 7.1449674055594e+00 & \fbox{7.1058987676886e+00}  & 7.1050608901229e+00  \\
    $u'(4)$ & 7.1133416801319e+00 & 7.1449674055594e+00 & \fbox{7.1114177007820e+00}  & 7.1050608901229e+00  \\
    $u''(0)$ & -8.8371916646277e+01 & \fbox{-9.6875622546165e+01} & \fbox{-8.6577693358885e+01}  & -8.8007775637807e+01  \\
    $u''(4)$ & -8.8371916646277e+01 & \fbox{-8.3340776157218e+01} & \fbox{-8.3307846875545e+01}  &  -8.8007775637806e+01  \\
    \hline
  \end{tabular}
  \caption{
    1D Helmholtz equation: Values of the solution and its first and second derivatives on
    the left/right domain boundaries from the exact solution
    and from the DNN solutions with $C^{\infty}$ and $C^1$
    periodic BCs enforced using the current method,
    and with the $C^1$ periodic BCs enforced using
    the penalty method.
  }
  \label{tab:helm1d}
\end{table}

Figure \ref{fig:helm1d_soln} shows a comparison between
the exact solution and the DNN
solutions obtained with the $C^{\infty}$ and $C^1$ periodic BCs enforced
using the current method from Section \ref{sec:method} (left column),
as well as the errors of these DNN solutions against the exact solution (right column).
The DNN solutions are observed to agree with the exact solution
very well. The DNN solution curves almost exactly overlap
with the exact-solution curve, and the maximum errors in the domain
is on the order of $10^{-3}$ with both the $C^{\infty}$ and $C^1$
periodic boundary conditions.

For comparison, we have also computed this problem with the
$C^1$ periodic BCs in another way, by enforcing the $C^1$ periodic BCs
based on the penalty method.
Figure \ref{fig:helm1d_soln_penalty} shows the
DNN solution 
computed  using the penalty method.
Here the DNN has the same parameters ($4$ hidden layers, with $20$ nodes
in each layer). The $C^1$ periodic BCs are enforced by including
a penalty term in the loss function as follows,
\begin{equation}\label{equ:loss_penalty}
  \text{Loss} = \frac{1}{L}\int_{\Omega}\left[
    \frac{d^2u}{dx^2} - \lambda u - f(x)
    \right]^2 dx
  + \theta_{bc}\left(
  \left[u(a)-u(b) \right]^2
  + \left[\left.\frac{du}{dx}\right|_{x=a} - \left.\frac{du}{dx}\right|_{x=b}  \right]^2
  \right),
\end{equation}
where $\theta_{bc}=10$ is the penalty coefficient in front
of the boundary residual terms.
The DNN has been trained with the Adam optimizer for $15000$ epochs.
We observe that the DNN solution resulting from the penalty method
also agrees well with the exact solution.

In Table \ref{tab:helm1d} we list the values of the DNN solution and its
first and second derivatives, with $14$ significant digits shown,
on the left and right domain boundaries obtained with the $C^{\infty}$
and $C^1$ periodic boundary conditions enforced using the current method, together with those
obtained with the $C^1$ periodic BCs enforced using the penalty method.
The boundary values from the exact
solution \eqref{equ:helm1d_anal} are also included in
the table for comparison.
We observe that the current method enforces exactly, to the machine accuracy,
the periodicity for the solution and its derivatives (up to order 2 shown here) with
the $C^{\infty}$ periodic BCs. With the $C^1$ periodic BCs, the current method
enforces exactly the periodicity for the solution and its first derivative, but
not for the second derivative.
In contrast, the penalty method enforces the periodic condition
for none of these quantities exactly.

\subsection{Two-Dimensional Helmholtz Equation with Periodic BCs}

We next test the performance of the proposed methods using the
the Helmholtz equation in two dimensions (2D),
\begin{equation}\label{equ:helm2d}
  \frac{\partial^2 u}{\partial x^2} + \frac{\partial^2 u}{\partial y^2}
  - \lambda u = f(x,y),
\end{equation}
on a rectangular domain
$
\Omega = \{ (x,y) | a_1\leqslant x\leqslant b_1, \
a_2\leqslant y\leqslant b_2 \}.
$
Here $\lambda$, $a_1$, $a_2$, $b_1$ and $b_2$ are given constants,
$u(x,y)$ is the unknown field function to be solved for,
and $f(x,y)$ is a prescribed source term.
We impose periodic boundary conditions in both the $x$ and $y$ directions.


Specifically, we consider $C^1$ and $C^{\infty}$ periodic boundary conditions in 2D.
The 2D $C^1$ periodic BC imposes the relations:
\begin{equation}\label{equ:helm2d_c1}
  \left\{
  \begin{split}
    &
    u(a_1, y) = u(b_1,y), \quad
    \frac{\partial }{\partial x}u(a_1,y) = \frac{\partial }{\partial x}u(b_1,y), \quad
    \forall y\in [a_2,b_2]; \\
    &
    u(x,a_2) = u(x,b_2), \quad
    \frac{\partial}{\partial y}u(x,a_2) = \frac{\partial}{\partial y}u(x,b_2), \quad
    \forall x \in [a_1,b_1].
  \end{split}
  \right.
\end{equation}
The 2D $C^{\infty}$ periodic BC imposes the relations:
\begin{equation}\label{equ:helm2d_cinf}
  \left\{
  \begin{split}
    &
    u(a_1, y) = u(b_1,y), \
    \frac{\partial }{\partial x}u(a_1,y) = \frac{\partial }{\partial x}u(b_1,y), \
    \frac{\partial^2 }{\partial x^2}u(a_1,y) = \frac{\partial^2 }{\partial x^2}u(b_1,y), \
    \dots, \
    \forall y\in [a_2,b_2]; \\
    &
    u(x,a_2) = u(x,b_2), \
    \frac{\partial}{\partial y}u(x,a_2) = \frac{\partial}{\partial y}u(x,b_2), \
    \frac{\partial^2}{\partial y^2}u(x,a_2) = \frac{\partial^2}{\partial y^2}u(x,b_2), \
    \dots, \
    \forall x \in [a_1,b_1].
  \end{split}
  \right.
\end{equation}
With the $C^{\infty}$ periodic BC, we are effectively seeking a smooth
periodic function $u(x,y)$ satisfying
\begin{equation}
  u(x+L_1, y) = u(x,y), \quad
  u(x,y+L_2) = u(x,y), \quad
  \forall x,y \in (-\infty,\infty),
\end{equation}
where $L_1 = b_1-a_1$ and $L_2 = b_2-a_2$.

We specifically consider the following parameter values for
the numerical tests in this section:
\begin{equation}
  \lambda = 10, \quad
  a_1 = a_2 = 0, \quad b_1 = b_2 = 4, \quad
  L_1 = L_2 = 4.
\end{equation}
We choose the source term $f(x,y)$ such that the 2D Helmholtz equation \eqref{equ:helm2d}
has the solution given by,
\begin{equation}\label{equ:helm2d_anal}
  u(x,y) = -\left[
    1.5\cos(\pi x + 0.4\pi)+2\cos(2\pi x-0.2\pi)
    \right]\left[
    1.5\cos(\pi y + 0.4\pi)+2\cos(2\pi y-0.2\pi)
    \right].
\end{equation}
This analytic solution satisfies the periodic boundary
conditions~\eqref{equ:helm2d_c1} and~\eqref{equ:helm2d_cinf}.

\begin{figure}
  \centerline{
    \includegraphics[width=3in]{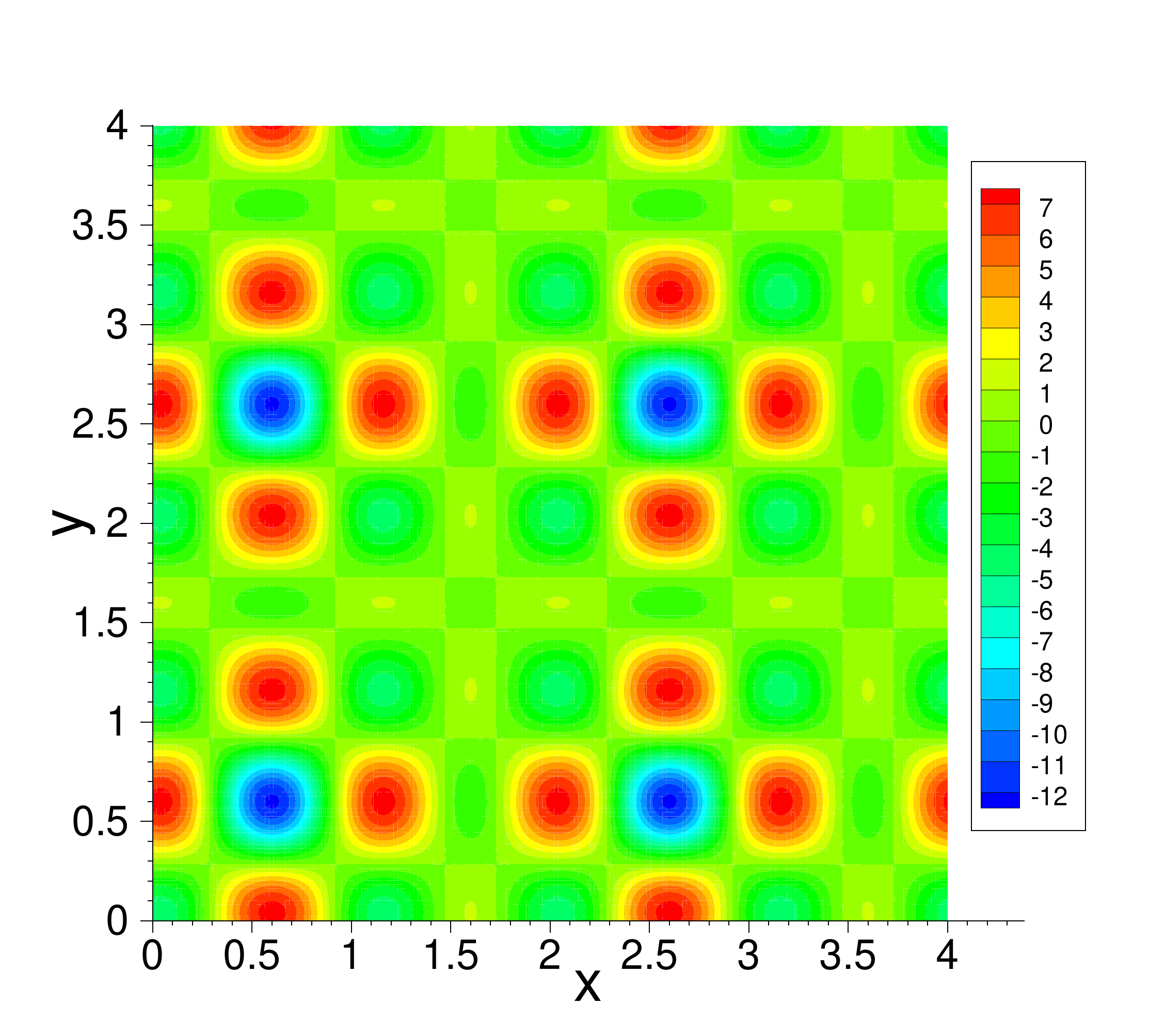}(a)
    \includegraphics[width=3in]{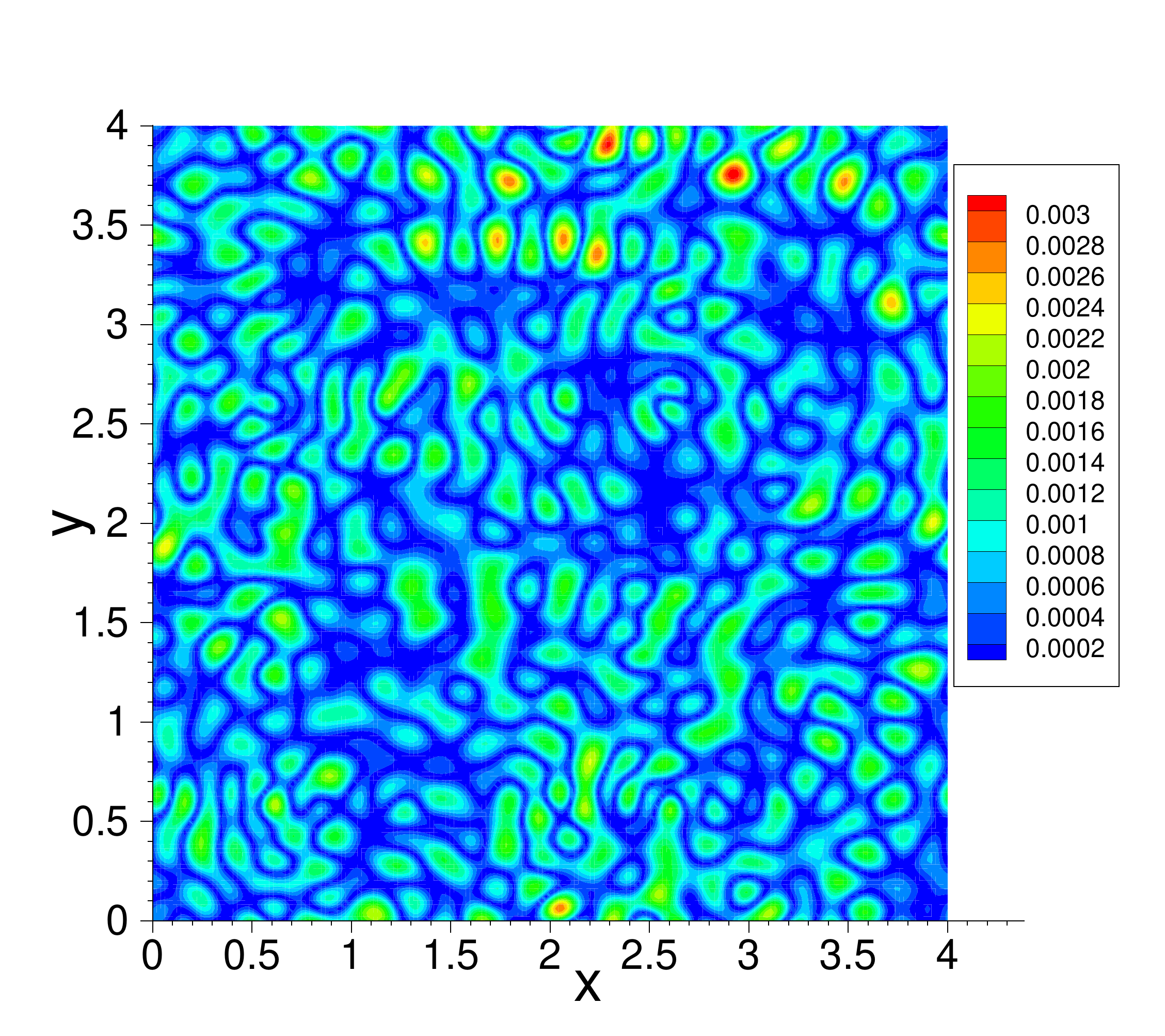}(b)
  }
  \centerline{
    \includegraphics[width=3in]{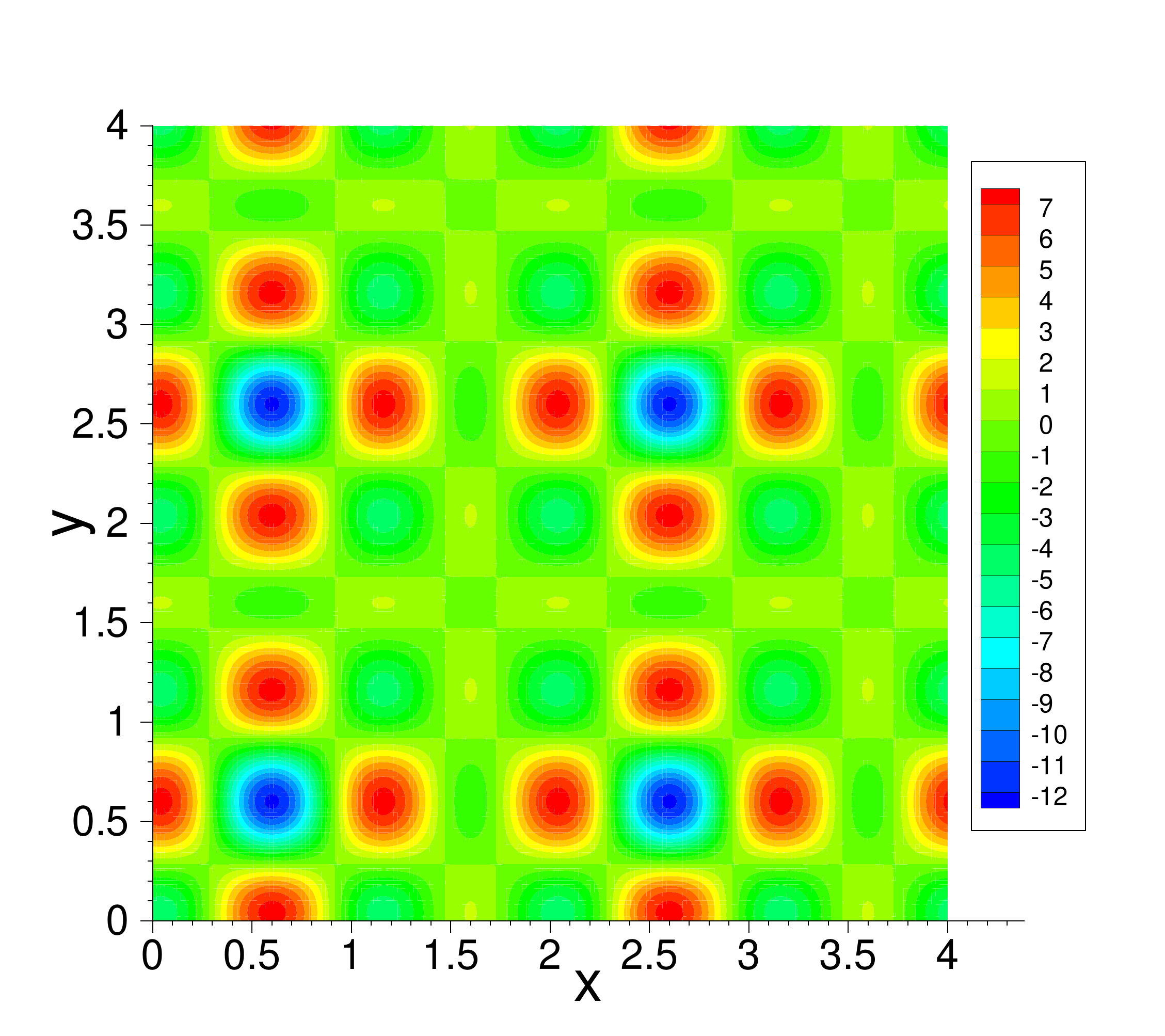}(c)
    \includegraphics[width=3in]{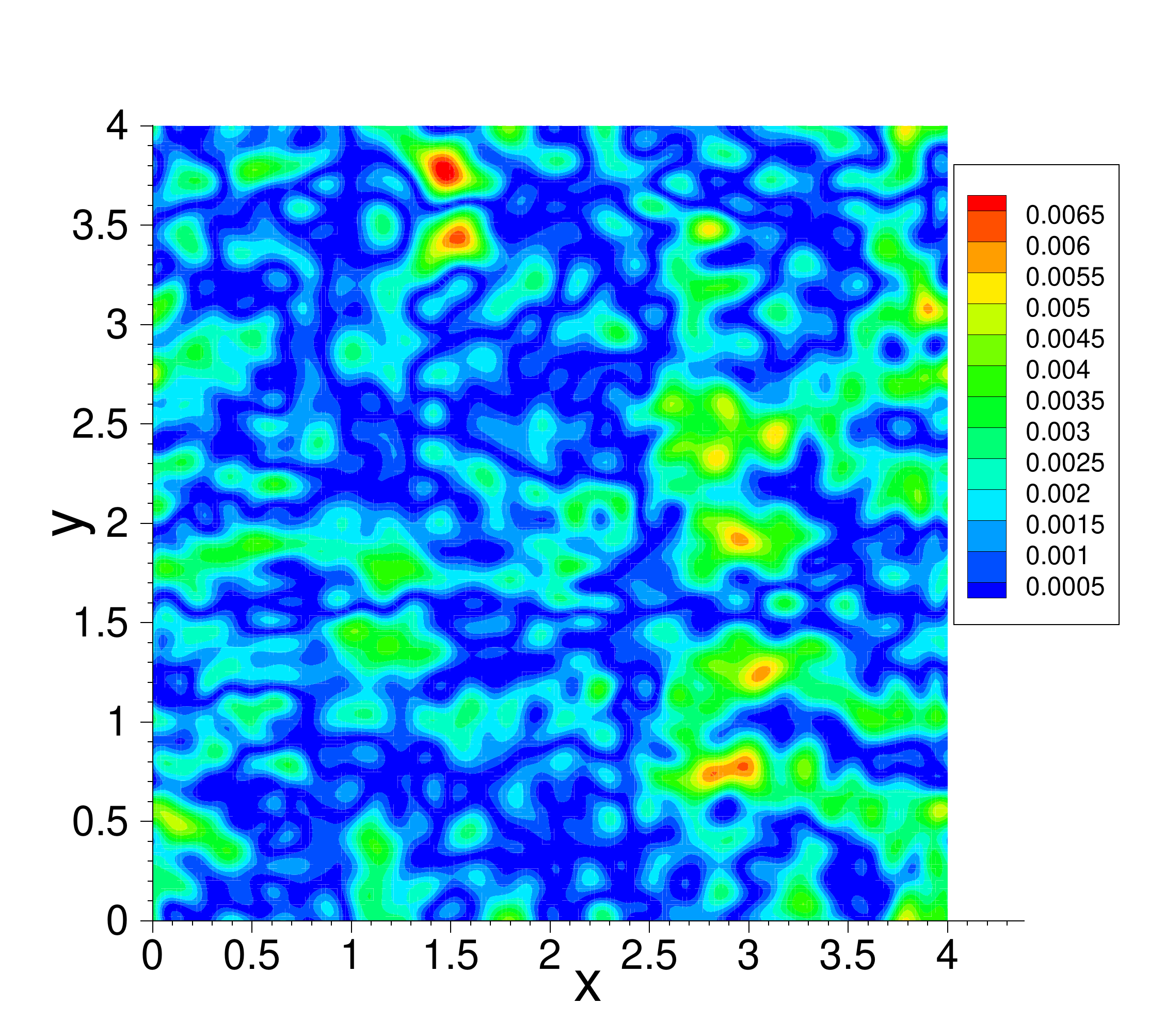}(d)
  }
  \caption{
    2D Helmholtz equation: Contours of DNN Solutions (left column) and their errors
    against the exact solution
    (right column), obtained with $C^{\infty}$ (top row) and $C^1$ (bottom row)
    periodic boundary conditions.
  }
  \label{fig:helm2d}
\end{figure}


To simulate this problem, we employ a feed-forward DNN with 2 nodes
in the input layer, one node in the output layer, and
$4$ hidden layers in between.
The input layer consists of the coordinates $x$ and $y$, and the output
layer is the solution to the 2D Helmholtz equation $u$.
Each of the four hidden layers contains $20$ nodes (neurons) in its
output. For the $C^{\infty}$ periodic BCs,
the second layer of this DNN (i.e.~the first hidden layer)
is set to be a 2D $C^{\infty}$ periodic layer $\mathcal{L}_p^{2D}(m,n)$
with $m=12$ and $n=20$ (see equation \eqref{equ:2d_pbc}),
in which the constants $\omega_1$ and $\omega_2$ are set to
\begin{equation}
  \omega_1 = \frac{2\pi}{L_1} = \frac{\pi}{2}, \quad
  \omega_2 = \frac{2\pi}{L_2} = \frac{\pi}{2}.
\end{equation}
For the $C^1$ periodic BCs, the second layer of this DNN
is set to be a 2D $C^1$ periodic layer $\mathcal{L}_{C^1}^{2D}(m,n)$
with $m=12$ and $n=20$ (see equation \eqref{equ:2d_ckpbc}).

\begin{figure}
  \centerline{
    \includegraphics[width=2.in]{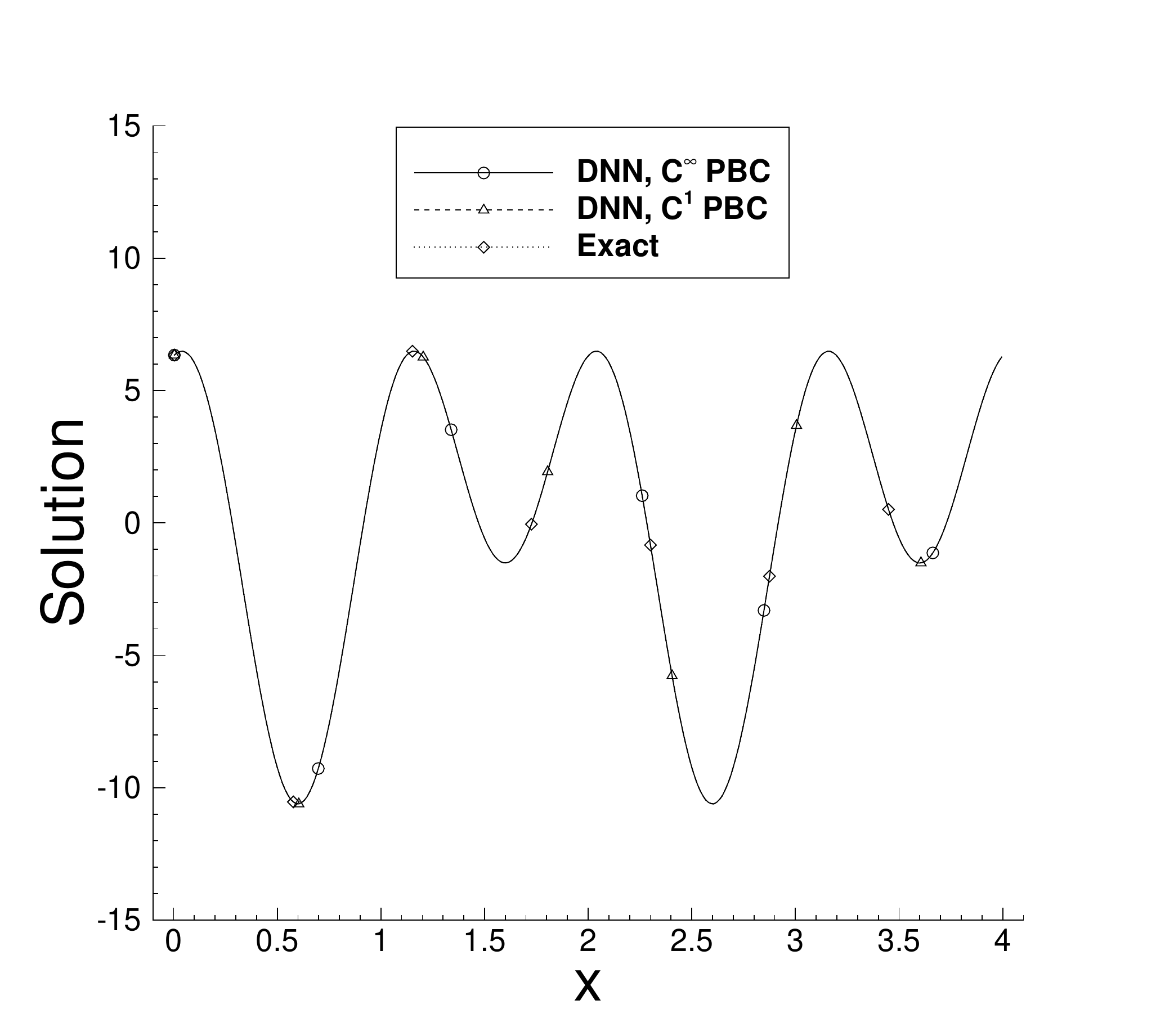}(a)
    \includegraphics[width=2.in]{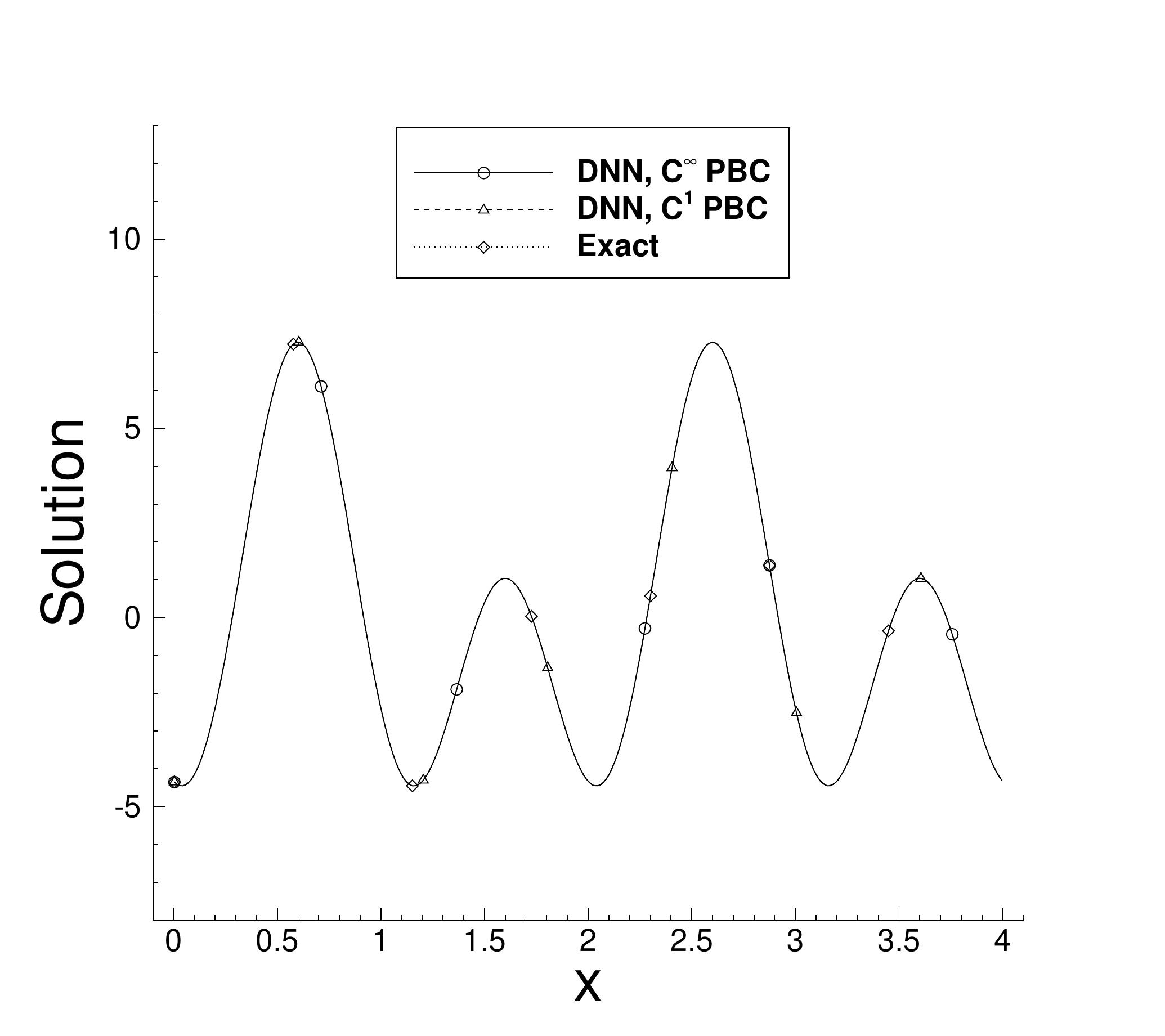}(b)
    \includegraphics[width=2.in]{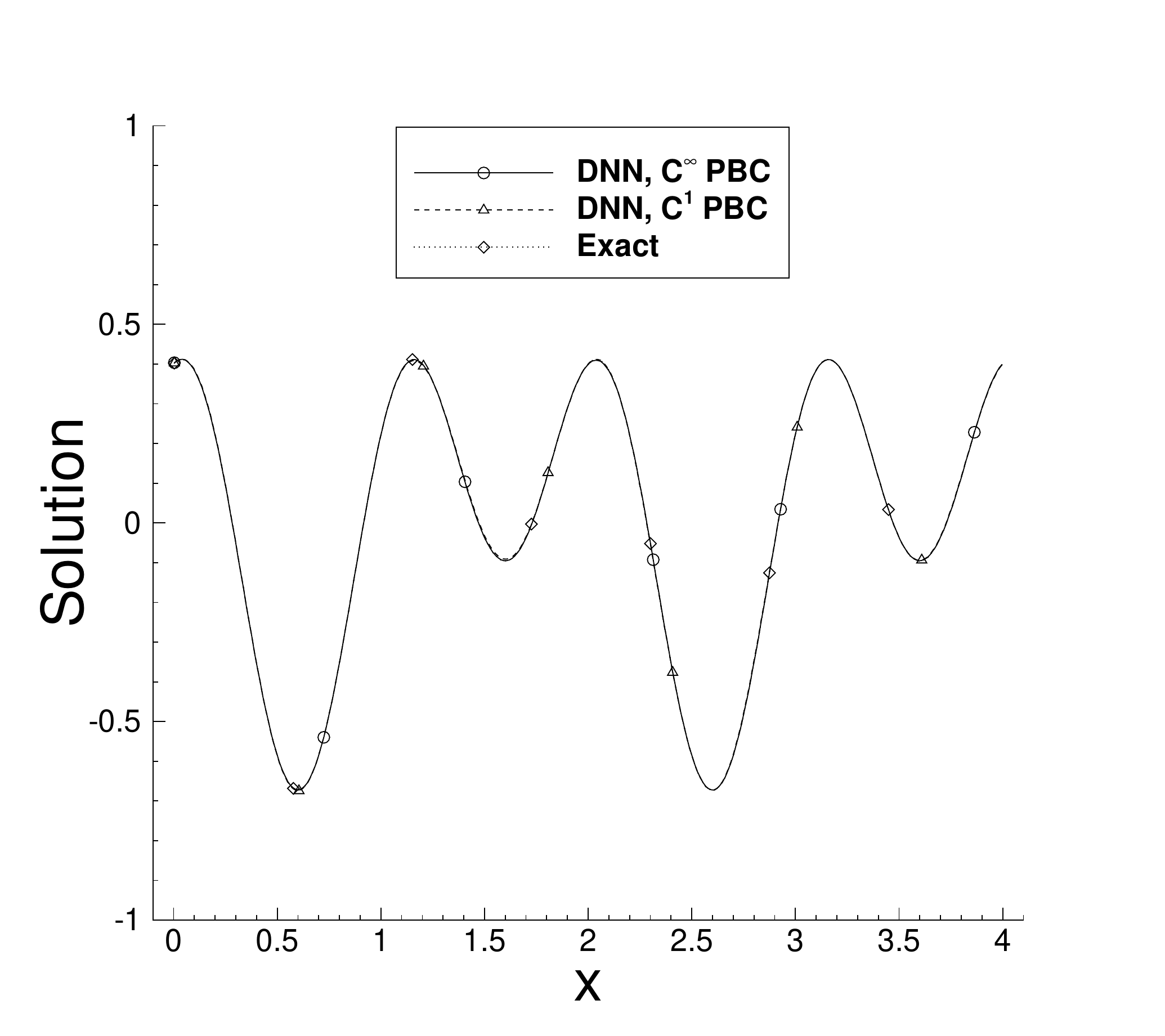}(c)
  }
  \centerline{
    \includegraphics[width=2.in]{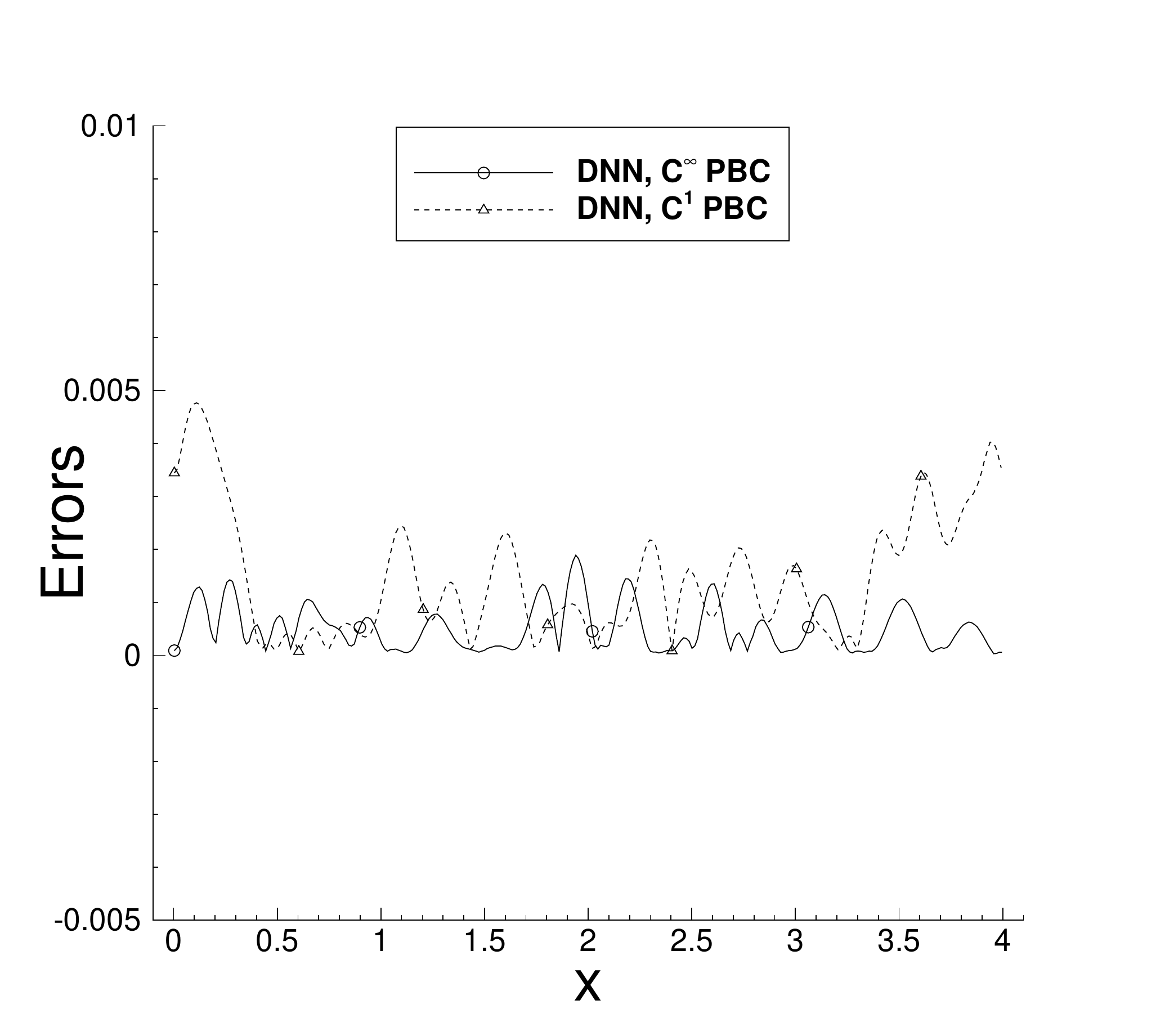}(d)
    \includegraphics[width=2.in]{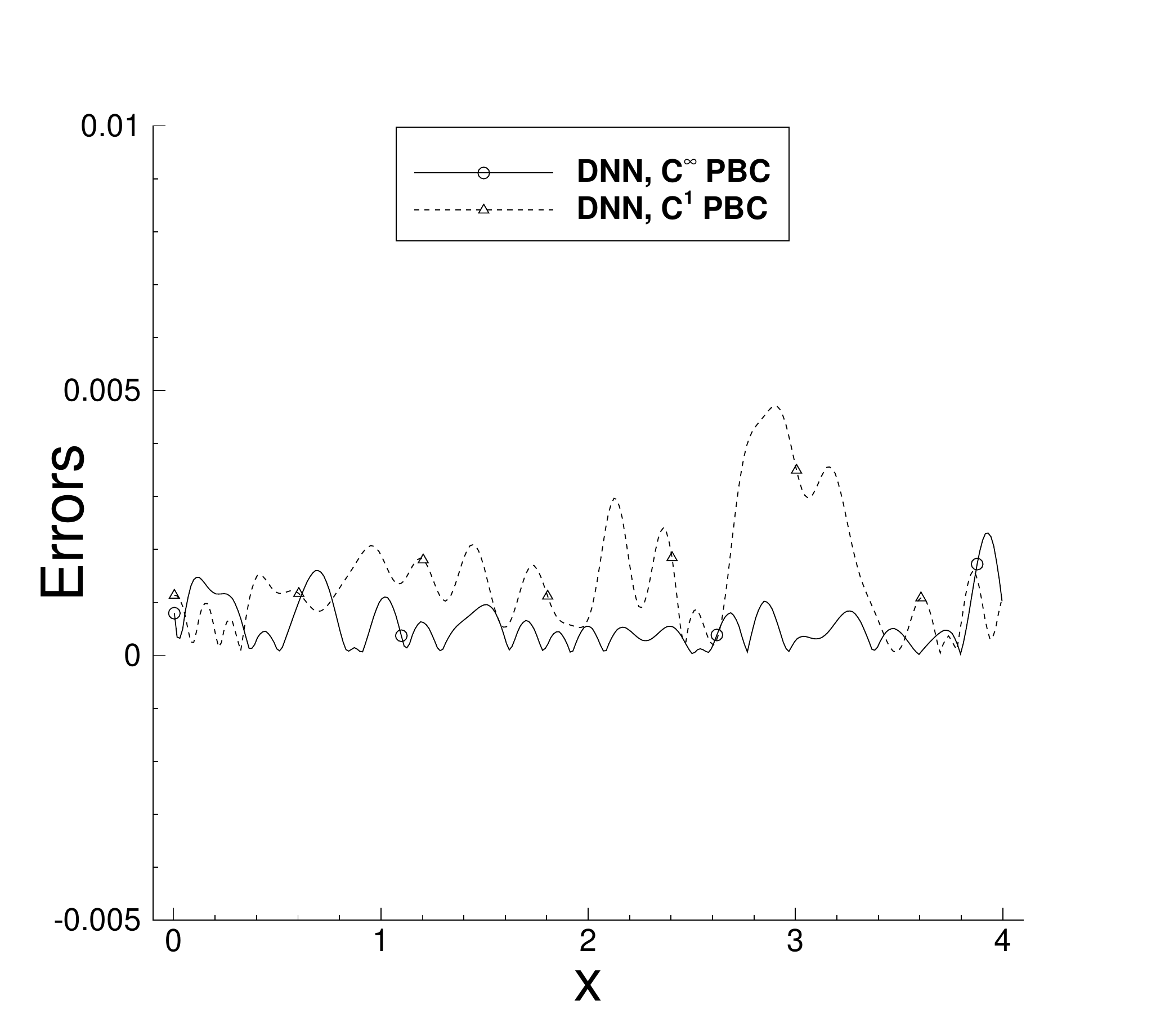}(e)
    \includegraphics[width=2.in]{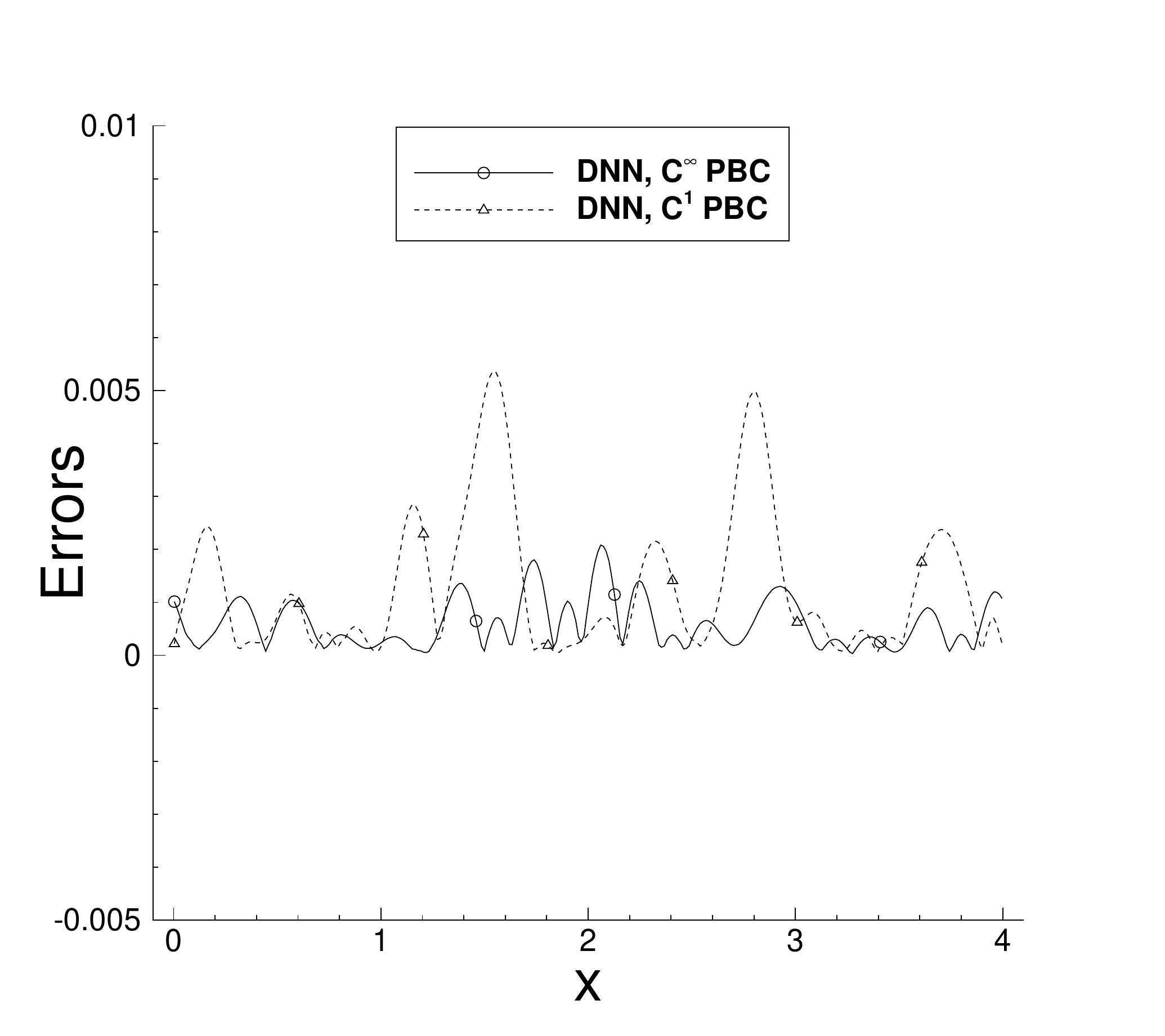}(f)
  }
  \caption{
    2D Helmholtz equation: comparison of profiles
    of the solution (top row) and its error against the exact solution
    (bottom row) along
    several horizontal lines located at (a,d) $y=0.5$,
    (b,e) $y=2.0$, and (c,f) $y=3.5$, from the DNN solutions with
    $C^{\infty}$ and $C^1$ periodic boundary conditions.
    The profiles of the exact solution are also included for comparison.
  }
  \label{fig:helm2d_error}
\end{figure}


We minimize the following loss function,
\begin{equation}
  \begin{split}
  \text{Loss} &= \frac{1}{V_{\Omega}}\int_{\Omega}\left[
    \frac{\partial^2u}{\partial x^2} + \frac{\partial^2u}{\partial y^2}
    - \lambda u - f(x,y)
    \right]^2 d\Omega \\
  &= \frac{1}{V_{\Omega}}\sum_{e=0}^{N_e-1}\sum_{i,j=0}^{Q-1}\left[
    \left.\frac{\partial^2u}{\partial x^2}\right|_{(x_i^e,y_j^e)}
    + \left.\frac{\partial^2u}{\partial y^2}\right|_{(x_i^e,y_j^e)}
    - \lambda u(x_i^e,y_j^e) - f(x_i^e,y_j^e)
    \right]^2 J^{e} w_{ij},
  \end{split}
\end{equation}
where $V_{\Omega}=\int_{\Omega}d\Omega=16$ is the area of the domain,
$N_e$ is the number of elements (sub-domains) we have partitioned the domain $\Omega$ into
for computing the integral,
$Q$ is the number of quadrature points in the $x$ and $y$ directions
within each element, $J^e$ is the Jacobian of the element $e$
($0\leqslant e\leqslant N_e-1$),
$(x_i^e,y_j^e)$ ($0\leqslant i,j\leqslant Q-1$) are the Gauss-Lobatto-Legendre
quadrature points within the element $e$ ($0\leqslant e\leqslant N_e-1$),
and $w_{ij}$ ($0\leqslant i,j\leqslant Q-1$) are the
Gauss-Lobatto-Legendre quadrature weights associated
with $(x_i^e,y_j^e)$.
The input data to the DNN consist of $(x_i^e,y_j^e)$ ($0\leqslant i,j\leqslant Q-1$,
$0\leqslant e\leqslant N_e-1$), and
$f(x_i^e,y_j^e)$ ($0\leqslant i,j\leqslant Q-1$,
$0\leqslant e\leqslant N_e-1$) are passed to the DNN as the label data.
In the loss expression, $u(x_i^e,y_j^e)$ can be obtained
from the output of the DNN, and the derivatives
$\frac{\partial^2u}{\partial x^2}$ and $\frac{\partial^2u}{\partial y^2}$
on $(x_i^e,y_j^e)$ can be computed by auto-differentiation.


In the numerical tests below, we partition the domain into
$4$ elements ($N_e=4$), with $2$ uniform elements in both
the $x$ and $y$ directions.
We use $30$ quadrature points ($Q=30$) in each direction within each
element. We employ the hyperbolic tangent (``tanh'') function
as the activation function for each of the hidden layers,
and no activation is applied to the output layer.
The DNN is trained using the L-BFGS optimizer for $10500$ iterations
with the $C^{\infty}$ periodic BCs, and for $6500$ iterations with
the $C^1$ periodic BCs.


\begin{table}
  \centering
  \begin{tabular}{llll}
    \hline
    & DNN $C^{\infty}$ PBC & DNN $C^1$ PBC & Exact solution \\
    $u(0,0.5)$ & 6.3375720647119e+00 & 6.3341020999642e+00 &  6.3375550504603e+00  \\
    $u(4,0.5)$ & 6.3375720647119e+00 & 6.3341020999642e+00 &  6.3375550504603e+00  \\
    $u_x(0,0.5)$ & 8.8479796805350e+00 &  8.8535226659745e+00 & 8.8433359504634e+00  \\
    $u_x(4,0.5)$ & 8.8479796805351e+00 &  8.8535226659745e+00 & 8.8433359504634e+00 \\
    $u_{xx}(0,0.5)$ &  -2.0815263069180e+02 &  \fbox{-2.1045671256754e+02} & -2.0841095826411e+02  \\
    $u_{xx}(4,0.5)$ &  -2.0815263069180e+02 &  \fbox{-2.0946606085833e+02} & -2.0841095826411e+02  \\
    \hline
    $u(0,3.5)$ & 3.9958327765965e-01 & 3.9859455580613e-01 & 3.9851292703942e-01   \\
    $u(4,3.5)$ & 3.9958327765962e-01 & 3.9859455580613e-01 & 3.9851292703942e-01   \\
    $u_x(0,3.5)$ & 5.4769599613000e-01 & 5.8968112421753e-01 & 5.5607938177296e-01  \\
    $u_x(4,3.5)$ & 5.4769599613003e-01 & 5.8968112421753e-01 & 5.5607938177296e-01  \\
    $u_{xx}(0,3.5)$ & -1.3286736511194e+01 & \fbox{-1.4294530675460e+01} & -1.3105126558055e+01  \\
    $u_{xx}(4,3.5)$ & -1.3286736511194e+01 & \fbox{-1.2645789026374e+01} & -1.3105126558055e+01  \\
    \hline
    $u(1,0)$ & -2.4020925534027e+00 & -2.4038799915243e+00 & -2.4031781074217e+00  \\
    $u(1,4)$ & -2.40209255340274e+00 & -2.4038799915243e+00 & -2.40317810742171e+00   \\
    $u_y(1,0)$ & -3.3463550988308e+00 & -3.3630705573155e+00 & -3.3533612226666e+00   \\
    $u_y(1,4)$ & -3.3463550988308e+00 & -3.3630705573155e+00 & -3.3533612226666e+00   \\
    $u_{yy}(1,0)$ & 7.8621899432667e+01 & \fbox{7.9457755849148e+01} & 7.9028686655862e+01   \\
    $u_{yy}(1,4)$ & 7.8621899432667e+01 & \fbox{7.9025311915153e+01} & 7.9028686655862e+01   \\
    \hline
    $u(3,0)$ & -2.4018445486575e+00 & -2.4022228345948e+00 & -2.4031781074217e+00  \\
    $u(3,4)$ & -2.4018445486575e+00 & -2.4022228345948e+00 & -2.4031781074217e+00   \\
    $u_y(3,0)$ & -3.3590252565867e+00 & -3.3499280675496e+00 & -3.3533612226666e+00   \\
    $u_y(3,4)$ & -3.3590252565867e+00 & -3.3499280675496e+00 & -3.3533612226666e+00  \\
    $u_{yy}(3,0)$ & 7.8580508305656e+01 & \fbox{7.8911124269480e+01} & 7.9028686655862e+01  \\
    $u_{yy}(3,4)$ & 7.85805083056566e+01 & \fbox{7.8910907593259e+01} & 7.9028686655862e+01  \\
    \hline
  \end{tabular}
  \caption{
    2D Helmholtz equation: Values of the solution and its derivatives on
    selected corresponding points of the left/right and top/bottom
    boundaries, obtained from the DNN solutions
    with $C^{\infty}$ and $C^1$ periodic BCs and from the exact solution.
    $u_x=\frac{\partial u}{\partial x}$, $u_y = \frac{\partial u}{\partial y}$,
    $u_{xx}=\frac{\partial^2u}{\partial x^2}$, and $u_{yy}=\frac{\partial^2u}{\partial y^2}$.
  }
  \label{tab:helm2d_bv}
\end{table}


Figure \ref{fig:helm2d} shows contours of the DNN solutions (left column) and their
errors (right column) against the exact solution \eqref{equ:helm2d_anal}, computed with
the $C^{\infty}$ periodic boundary conditions (top row) and the $C^1$
periodic boundary conditions.
The distributions of the DNN solutions are qualitatively the same as
that of the exact solution, and no difference can be discerned visually.
The maximum absolute error of the DNN solution in the domain
is less
than $5\times 10^{-3}$ with the $C^{\infty}$ periodic BC and less than
$10^{-2}$ with the $C^1$ periodic BC.
Figure \ref{fig:helm2d_error} provides a quantitative comparison between the
DNN solutions and the exact solution. It shows the profiles of
the DNN solutions obtained using $C^{\infty}$ and $C^1$ periodic BCs,
as well as the exact solution, 
along several horizontal lines across the domain located at
$y=0.5$, $2$, and $3.5$.
The error profiles of the DNN solutions along these lines are also shown
in this figure.
We observe that the DNN solutions with both the $C^{\infty}$ and
$C^1$ periodic BCs obtained using the current method agree very well with
the exact solution.

To examine how well the current methods enforce the
periodic boundary conditions for the 2D Helmholtz equation,
we have extracted the values of the DNN solution and its
partial derivatives (up to second order) on several corresponding
points on the left and right boundaries, and on the top and bottom
boundaries. Note that these derivatives are computed by
auto-differentiation, and they are the exact derivatives corresponding
to the given DNN representation of the field.
Table \ref{tab:helm2d_bv} lists the values of the DNN solutions
and their partial derivatives on several corresponding points
on the left/right boundaries and top/bottom boundaries.
The second-order mixed derivatives and some first derivatives
are not listed in the table, such as $\frac{\partial^2 u}{\partial x\partial y}$,
$\frac{\partial u}{\partial y}$ on the left/right boundaries, and
$\frac{\partial u}{\partial x}$ on the top/bottom boundaries.
These unlisted values are exactly the same on the corresponding boundary
points with both the $C^{\infty}$ and $C^1$ periodic BCs.
The boxed values in this table highlight the difference in the second partial
derivatives on the corresponding boundary points of the DNN solution
obtained with $C^1$ periodic BCs.
As expected, the current methods have enforced the
periodic boundary conditions exactly for the solution and
the corresponding higher-order derivatives.

\subsection{Diffusion Equation with Periodic BCs}

The next test problem is the unsteady diffusion equation:
\begin{equation}\label{equ:diffu}
\frac{\partial u}{\partial t} - \nu\frac{\partial^2u}{\partial x^2} = f(x,t),
\end{equation}
where the constant $\nu>0$ is the diffusion coefficient,
$u(x,t)$ is the unknown field function to be solved for,
$f(x,t)$ is a prescribed source term,
$x$ is the spatial coordinate, and $t$ is time.
We consider the spatial-temporal domain
$\Omega = \{(x,t) | a\leqslant x\leqslant b, \ 0\leqslant t\leqslant T  \}$,
where $a$, $b$, $T$ are prescribed constants whose values
are specified below.
This equation is supplemented by the initial condition,
\begin{equation}\label{equ:ic}
  u(x,0) = u_{in}(x),
\end{equation}
where $u_{in}$ denotes the initial distribution.


We impose periodic boundary conditions on the spatial  boundaries
$x=a$ and $b$. We specifically consider the $C^{\infty}$ and $C^1$
periodic BCs. The $C^{\infty}$ periodic BC requires,
\begin{equation}\label{equ:diffu_cinf}
  u(a,t) = u(b,t), \quad
  \frac{\partial }{\partial x}u(a,t) = \frac{\partial}{\partial x}u(b,t), \quad
  \frac{\partial^2}{\partial x^2}u(a,t) = \frac{\partial^2}{\partial x^2}u(b,t), \quad
  \dots, \quad
  \forall t\in [0,T].
\end{equation}
The $C^1$ periodic BC requires,
\begin{equation}\label{equ:diffu_c1}
  u(a,t) = u(b,t), \quad
  \frac{\partial }{\partial x}u(a,t) = \frac{\partial}{\partial x}u(b,t), \quad
  \forall t\in [0,T].
\end{equation}

\begin{figure}
  \centerline{
    \includegraphics[width=2.5in]{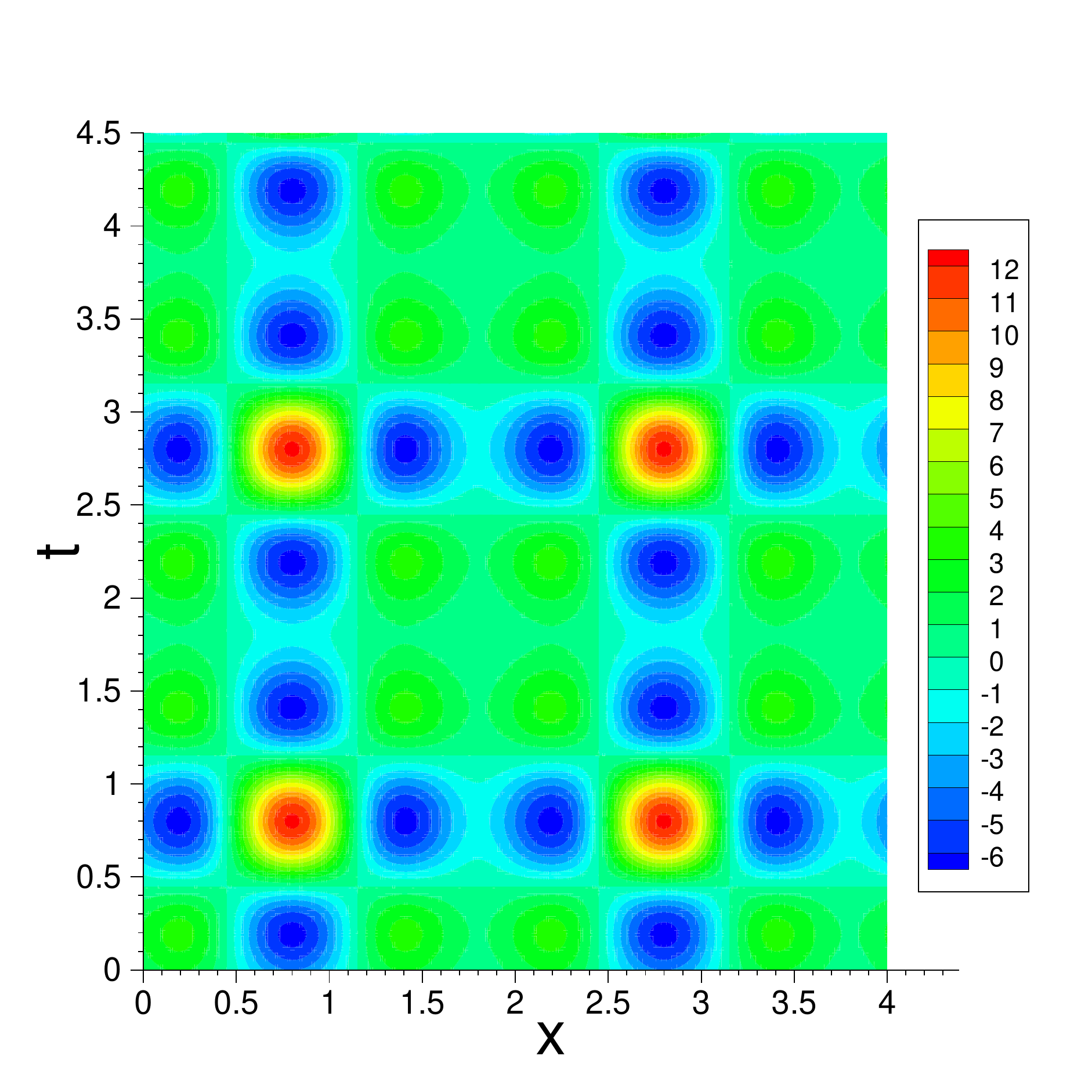}(a)
    \includegraphics[width=2.5in]{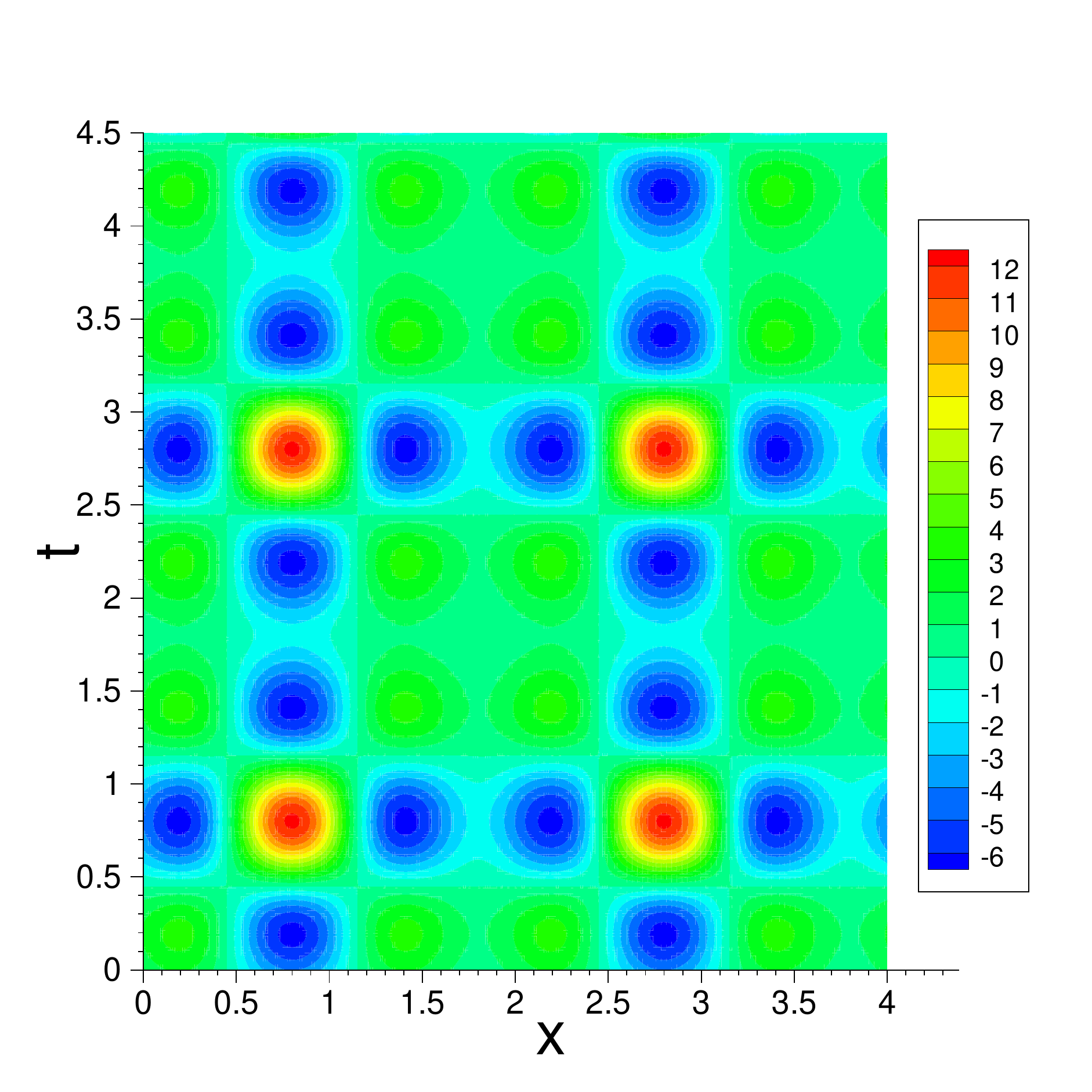}(b)
  }
  \caption{
    Diffusion equation: DNN solutions
    obtained with $C^{\infty}$ (a) and $C^1$ (b)
    periodic boundary conditions in the $x$ direction.
  }
  \label{fig:diffu_soln}
\end{figure}


For the numerical tests in this subsection
we employ the following parameter values,
\begin{equation}
  \nu = 0.01, \quad
  a = 0, \quad
  b = 4, \quad
  T = 4.5, \quad
  L = b-a = 4.
\end{equation}
We choose the source term $f(x,t)$ such that
the function,
\begin{equation}\label{equ:diffu_anal}
  u(x,t) = \left(
    2\cos(\pi x + 0.2\pi)+1.5\cos(2\pi x - 0.6\pi)
    \right)
    \left(
    2\cos(\pi t + 0.2\pi)+1.5\cos(2\pi t - 0.6\pi)
    \right),
\end{equation}
is a solution to the equation \eqref{equ:diffu}.
We choose the initial distribution $u_{in}(x)$ by
using the analytic expression \eqref{equ:diffu_anal}
and setting $t=0$.
Note that the expression \eqref{equ:diffu_anal} satisfies
the boundary conditions~\eqref{equ:diffu_c1} and \eqref{equ:diffu_cinf}.
So under the initial condition \eqref{equ:ic}
and the periodic boundary conditions, the exact solution to
the diffusion equation is given by \eqref{equ:diffu_anal}.

We solve this initial/boundary value problem using
DNN together with the method from Section \ref{sec:method}
for enforcing the periodic BCs in $x$.
We employ a DNN with two nodes in the input layer, which
represent the spatial coordinate $x$ and the time $t$,
and one node in the output layer, which represents
the unknown function $u(x,t)$ to be solved for.
This DNN contains $3$ hidden layers between the input
and the output layers. Each of the hidden layers
has an output consisting of $30$ nodes.
Note that periodic BCs are imposed only in the $x$ direction,
not in time.
For the $C^{\infty}$ periodic BCs, the second layer of
this DNN (or the first hidden layer) is a set to be
a modified 2D $C^{\infty}$ periodic layer as discussed in the Remark \ref{lem:lem_a}.
For the periodic direction $x$, this modified 2D $C^{\infty}$ periodic layer
corresponds to a 1D $C^{\infty}$ periodic layer $\mathcal{L}_p(m,n)$
with $m=12$ and $n=30$ (see equation \eqref{equ:def_zi}),
in which the constant $\omega$ is set to
\begin{equation}
  \omega = \frac{2\pi}{L} = \frac{\pi}{2}.
\end{equation}
For the $C^1$ periodic BCs, the second layer of this DNN
is set to be a modified 2D $C^1$ periodic layer as discussed in Remark \ref{lem:lem_b}.
For the periodic direction $x$, this modified 2D $C^{1}$ periodic layer
corresponds to a 1D $C^1$ periodic layer $\mathcal{L}_{C^1}(m,n)$
with $m=12$ and $n=30$ (see equation \eqref{equ:ckp_layer}).


We minimize the following loss function,
\begin{equation}\label{equ:diffu_loss}
  \begin{split}
  \text{Loss} &= \theta_{eq}\frac{1}{V_{\Omega}}\int_{\Omega}\left[
    \frac{\partial u}{\partial t} - \nu\frac{\partial^2u}{\partial x^2} -f(x,t)
    \right]^2 d\Omega + \theta_{ic}\frac{1}{b-a}\int_{a}^{b}\left[
    u(x,0) - u_{in}(x)
    \right]^2 dx \\
  &= \theta_{eq}\frac{1}{V_{\Omega}}\sum_{e=0}^{N_{el}-1}\int_{\Omega_e}\left[
    \frac{\partial u}{\partial t} - \nu\frac{\partial^2u}{\partial x^2} -f(x,t)
    \right]^2 d\Omega
  + \theta_{ic}\frac{1}{b-a}\sum_{e=0}^{N_{el}^x-1}\int_{a_e}^{b_e}\left[
    u(x,0) - u_{in}(x)
    \right]^2 dx
  \\
  &= \frac{\theta_{eq}}{(b-a)T}\sum_{e=0}^{N_{el}-1}\sum_{i,j}^{Q-1}\left[
    \left.\frac{\partial u}{\partial t} \right|_{(x_i^e,t_j^e)}
    - \left.\nu\frac{\partial^2u}{\partial x^2}\right|_{(x_i^e,t_j^e)}
    - f(x_i^e,t_j^e)
    \right]^2 J^e w_{ij} \\
  &\quad
  + \frac{\theta_{ic}}{b-a}\sum_{e=0}^{N_{el}^x-1}\sum_{i=0}^{Q-1}\left[
    u(x_i^e,0) - u_{in}(x_i^e)
    \right]^2 J^{ex} w_i,
  \end{split}
\end{equation}
where $V_{\Omega} = \int_{\Omega}d\Omega=(b-a)T$ is the
volume of the spatial-temporal domain $\Omega$,
$N_{el}$ is the number of spatial-temporal elements we have partitioned
$\Omega$ into in order to compute the integral,
$\Omega_e$ denotes the spatial-temporal element $e$ for
$0\leqslant e\leqslant N_{el}-1$, and $Q$ is the number of
Gauss-Lobatto-Legendre quadrature points in both
the $x$ and $t$ directions within each spatial-temporal element.
We use $N_{el}^x$ to denote the number of elements in the spatial direction,
and $N_{el}^t$ to denote the number of elements in time,
and then $N_{el}=N_{el}^xN_{el}^t$.
The sub-interval $[a_e,b_e]$ 
denotes the spatial element $e$ for $0\leqslant e\leqslant N_{el}^x-1$.
The constants $\theta_{eq}$ and $\theta_{ic}$ are the penalty
coefficients for the equation residual term and the initial condition
residual term in \eqref{equ:diffu_loss}.
The Gauss-Lobatto-Legendre quadrature points within
the spatial-temporal element $\Omega_e$ are denoted by $(x_i^e,t_j^e)$,
with the associated quadrature weights $w_{ij}$.
$J^e$ is the Jacobian associated with the element $\Omega_e$
for $0\leqslant e\leqslant N_{el}-1$.
$J^{ex}$ is the Jacobian of the spatial element $[a_e,b_e]$
for $0\leqslant e\leqslant N_{el}^x-1$.
$w_i$ ($0\leqslant i\leqslant Q-1$) denote the quadrature weights
associated with Gauss-Lobatto-Legendre quadrature points $x_i^e$
in the spacial direction.
The input data to the DNN consist of all the quadrature points
within the domain, $(x_i^e,t_j^e)$, for $0\leqslant i,j\leqslant Q-1$
and $0\leqslant e\leqslant N_{el}-1$.
The values of the source term on the quadrature points, $f(x_i^e,t_j^e)$,
are passed to the DNN as the label data.
The terms $\left.\frac{\partial u}{\partial t}\right|_{(x_i^e,t_j^e)}$
and $\left.\frac{\partial^2 u}{\partial x^2}\right|_{(x_i^e,t_j^e)}$ in
\eqref{equ:diffu_loss} are computed based on auto-differentiation.
It can be observed that here we are enforcing the initial condition
by the penalty method.

In the numerical tests reported below,
the domain $\Omega$ is partitioned into $9$ spatial-temporal elements ($N_{el}=9$),
with $3$ uniform elements in time ($N_{el}^t=3$) and also $3$ elements
in the $x$ direction ($N_{el}^x=3$). Along the $x$ direction, the two interior boundaries
of the elements are located at $x=1.3$ and $2.6$.
We employ $20$ quadrature points ($Q=20$) in space and time within
each spatial-temporal element.
The penalty coefficients are set to be $\theta_{eq}=0.9$ and $\theta_{ic}=1-\theta_{eq}=0.1$.
The ``tanh'' activation function has been employed for each hidden layer,
and no activation function is applied to the output layer of the DNN.
The L-BFGS optimizer has been employed to train the DNN
for $13000$ iterations with the $C^{\infty}$ periodic BCs,
and for $14500$ iterations with the $C^1$ periodic BCs.


\begin{figure}
  \centerline{
    \includegraphics[width=2.in]{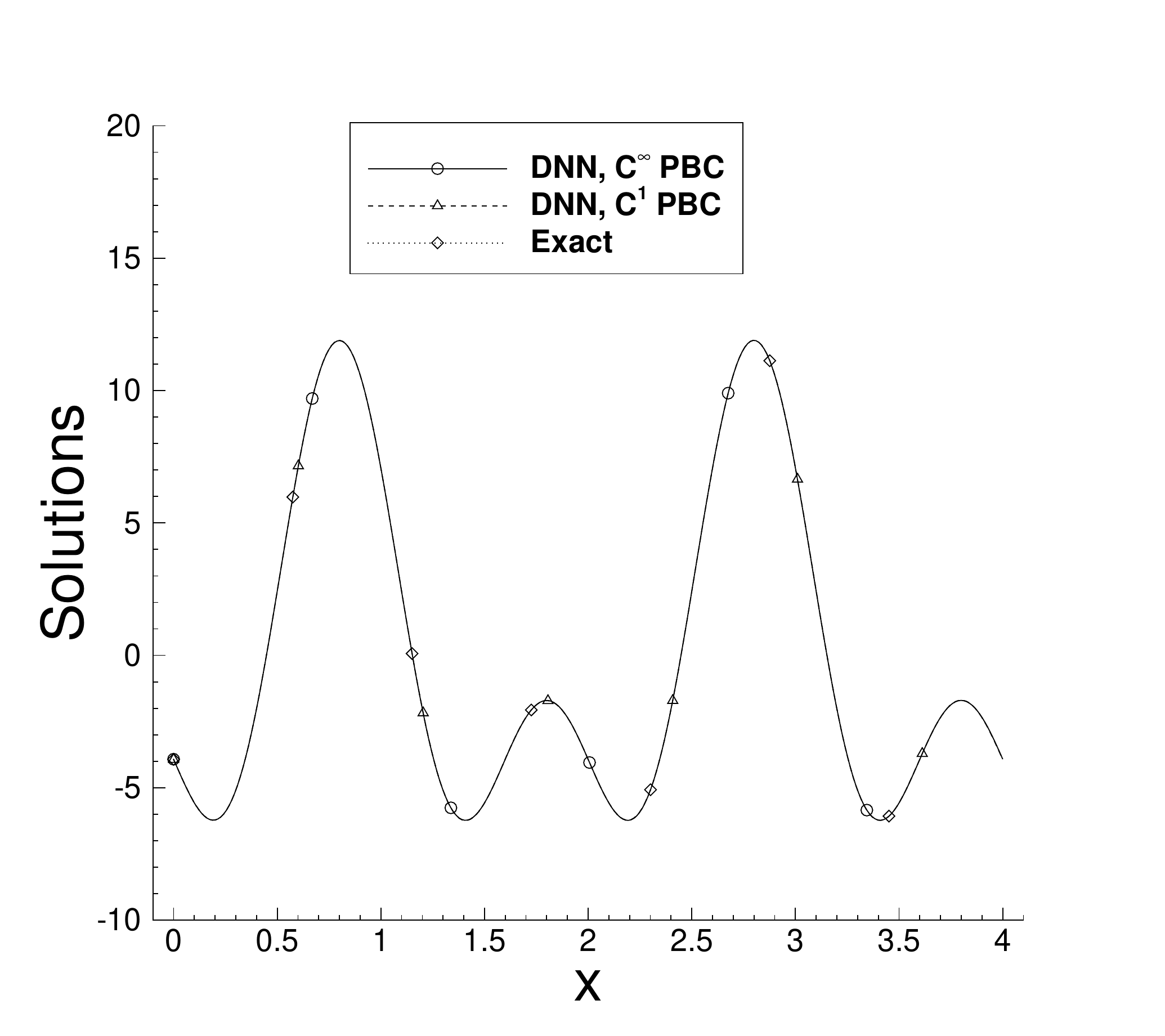}(a)
    \includegraphics[width=2.in]{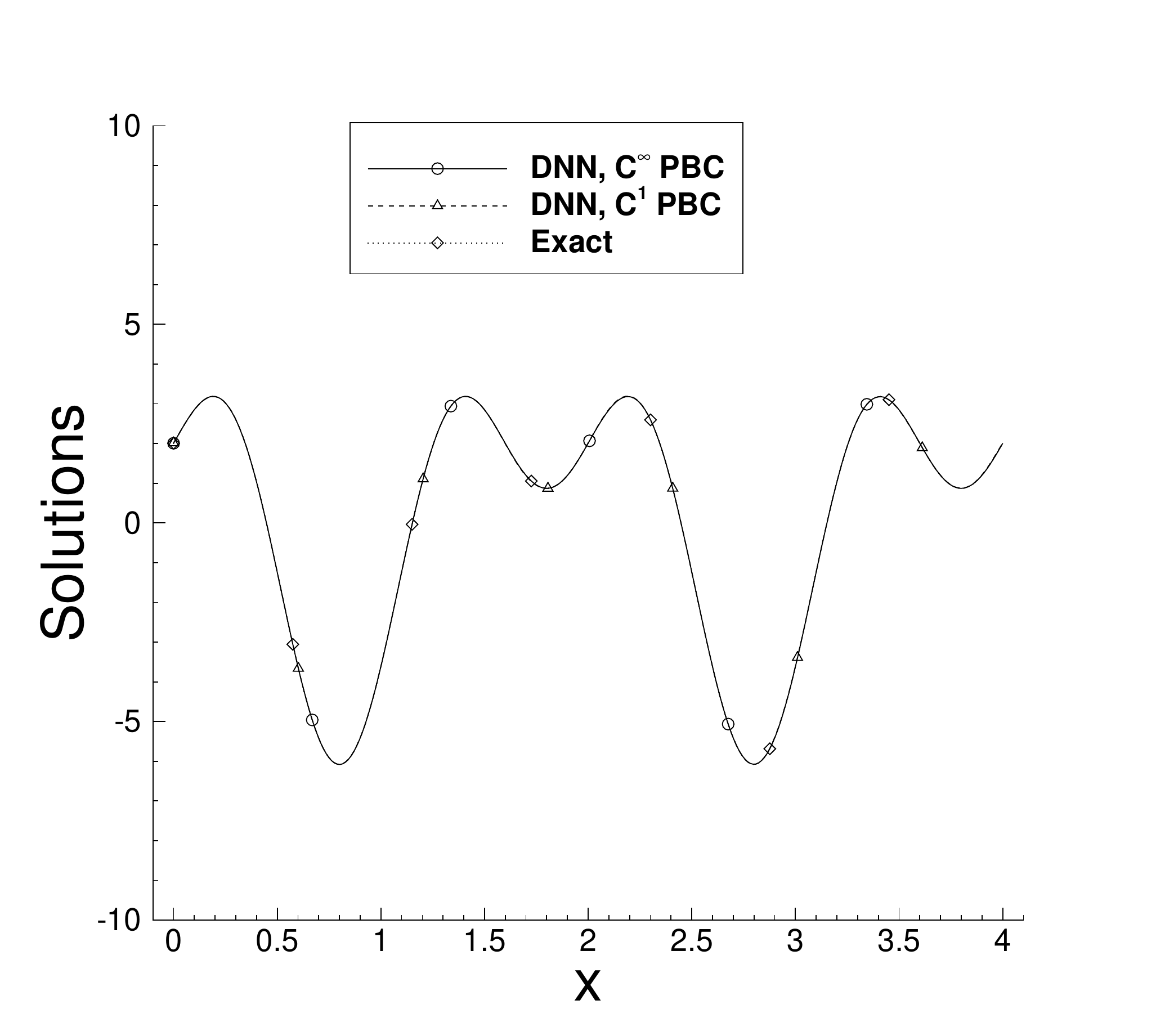}(b)
    \includegraphics[width=2.in]{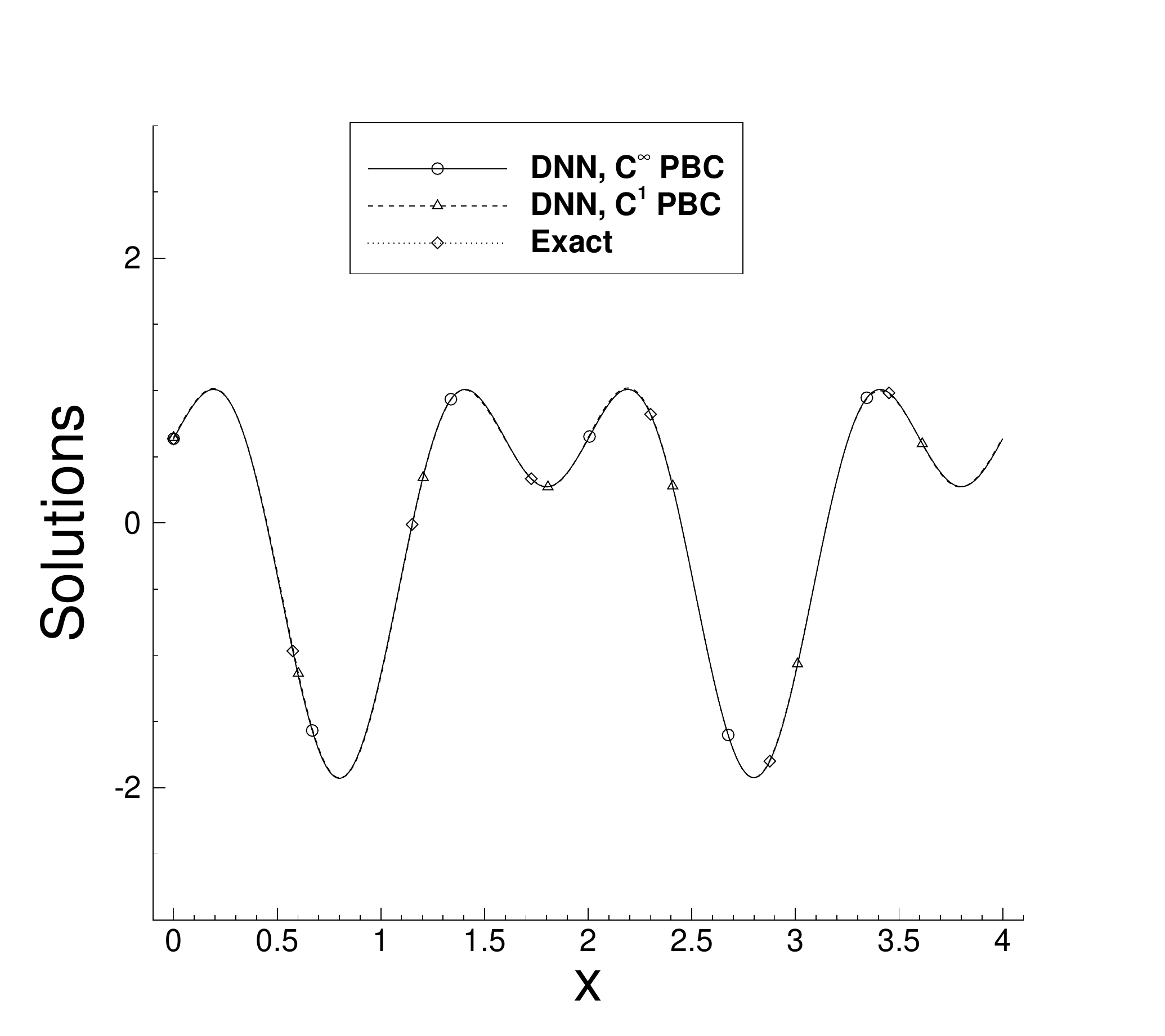}(c)
  }
  \centerline{
    \includegraphics[width=2.in]{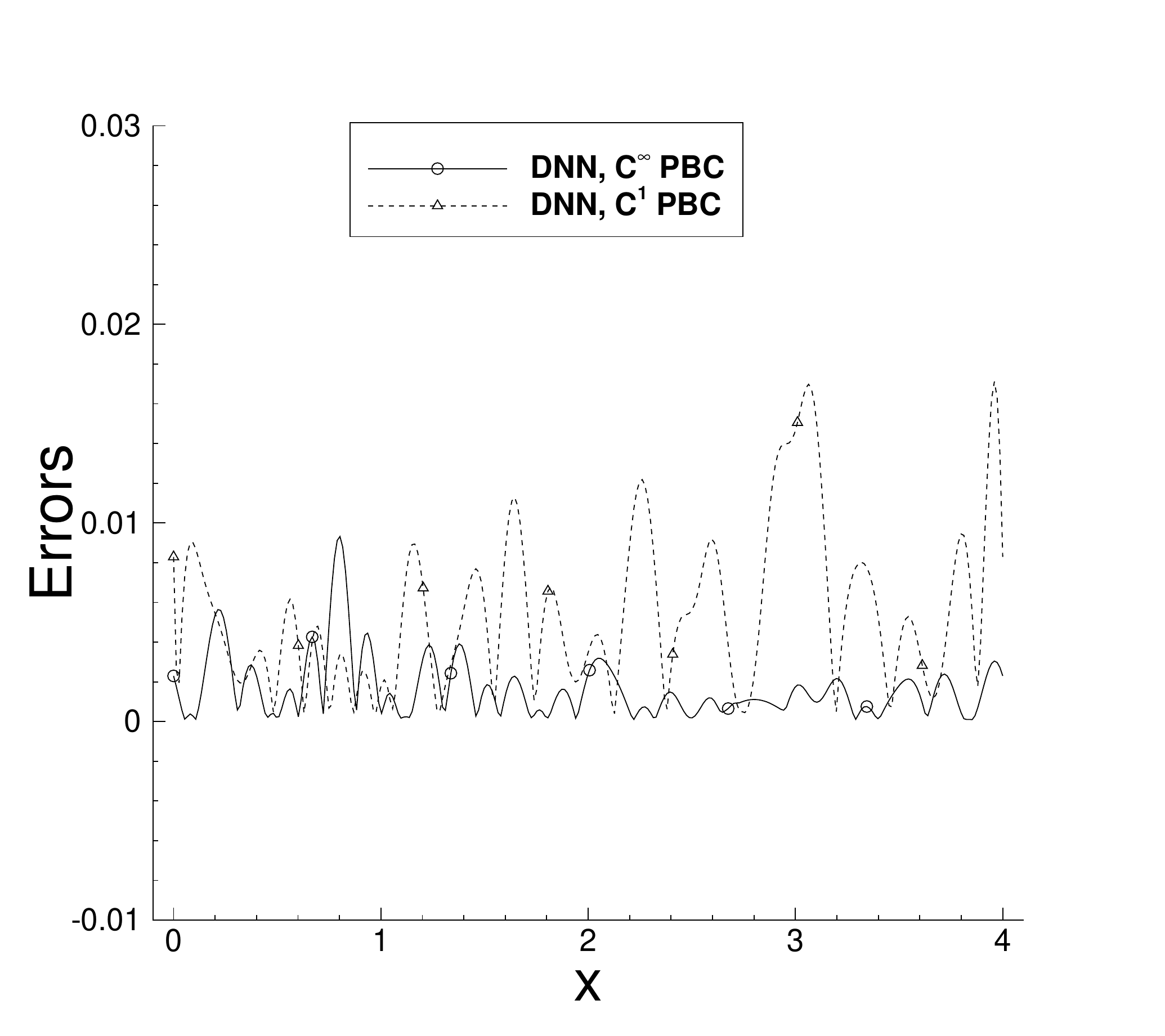}(d)
    \includegraphics[width=2.in]{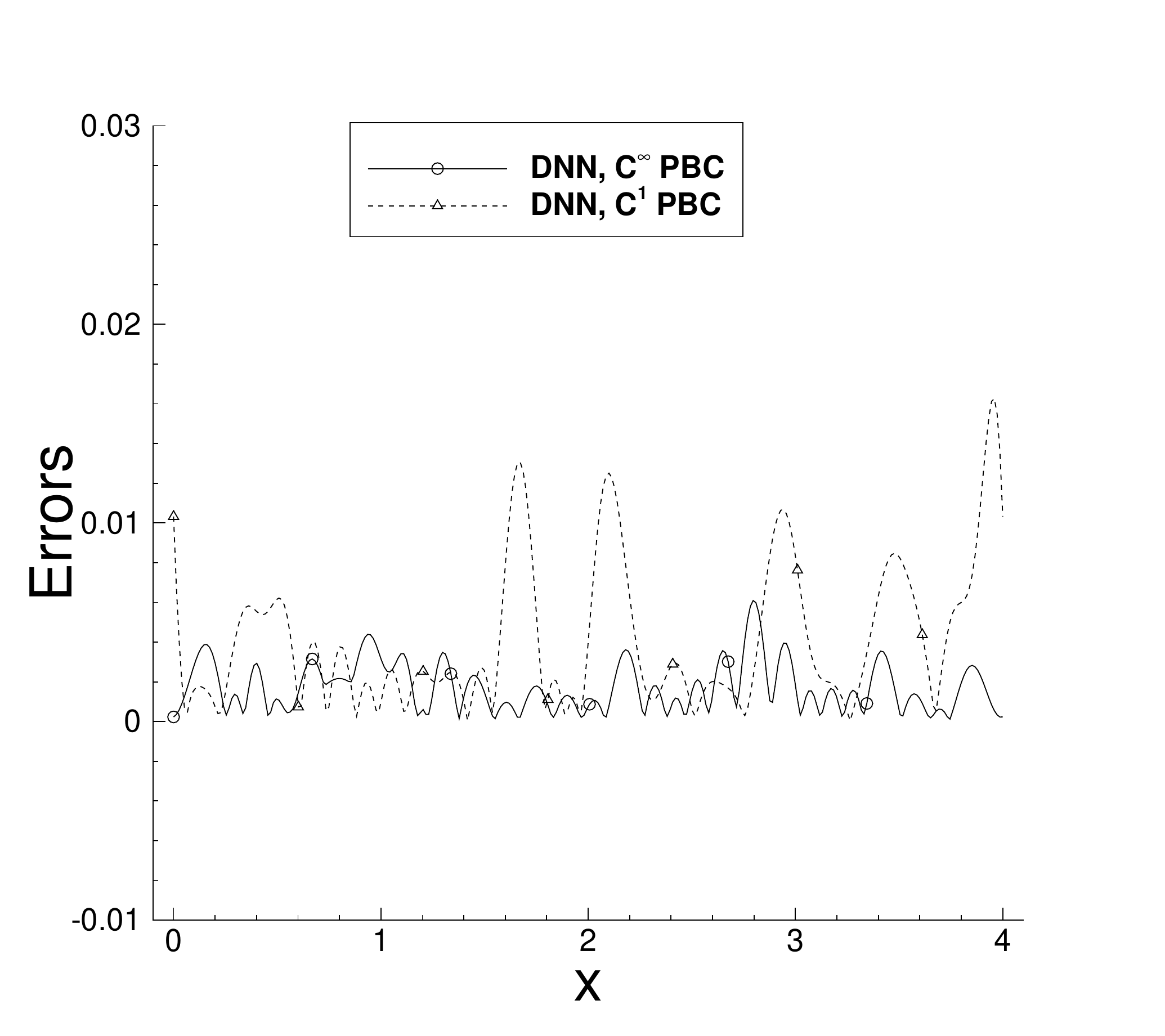}(e)
    \includegraphics[width=2.in]{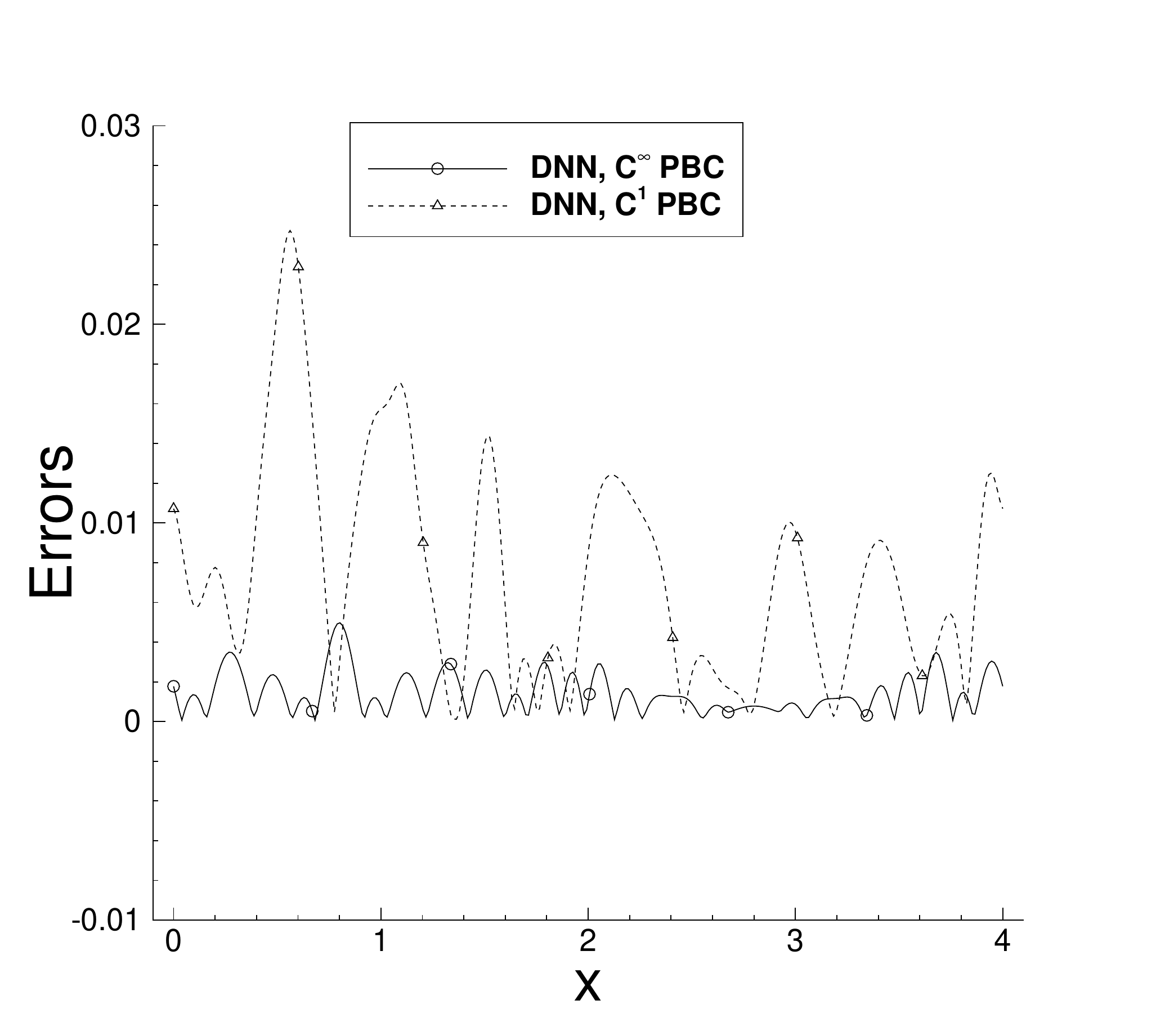}(f)
  }
  \caption{
    Diffusion equation: comparison of profiles
    of the solution (top row) and the absolute error (bottom row) at time instants
    (a,d) $t=0.75$, (b,e) $t=2.25$, (c,f) $t=3.75$
    from the DNN solutions with $C^{\infty}$ and $C^1$ periodic boundary
    conditions and from the exact solution.
  }
  \label{fig:diffu_prof}
\end{figure}


Figure \ref{fig:diffu_soln} shows contours in the spatial-temporal
($x-t$) plane of the DNN solutions (left column)
and their errors against the exact solution (right column), obtained using the
$C^{\infty}$ periodic BCs (top row) and the $C^1$ periodic BCs (bottom row).
Qualitatively, no difference can be discerned between distributions of the DNN solutions and
the exact solution given by \eqref{equ:diffu_anal}.
Figure \ref{fig:diffu_prof} provides a quantitative  comparison between the DNN
solutions and the exact solution.
It shows profiles of the exact solution and the DNN solutions
corresponding to $C^{\infty}$ and $C^1$ periodic BCs at three
time instants $t=0.75$, $2.25$ and $3.75$.
The profiles of the absolute errors of the DNN solutions are shown in
this figure as well.
The DNN solution profiles corresponding to both the $C^{\infty}$ and $C^1$
periodic BCs are observed to overlap with those of the exact solution
at different time instants.
It can be observed from the error profiles that the DNN 
with the $C^{\infty}$ periodic BCs appears to result in generally smaller errors
than that with the $C^1$ periodic BCs for this problem.


\begin{table}
  \centering
  \begin{tabular}{llll}
    \hline
    & DNN $C^{\infty}$ PBC & DNN $C^1$ PBC & Exact solution \\
    $u(0,0.6)$ & -2.4017654769459e+00 & -2.4116317389834e+00 & -2.4031781074217e+00  \\
    $u(4,0.6)$ & -2.4017654769459e+00 & -2.4116317389834e+00 & -2.4031781074217e+00  \\
    $u_x(0,0.6)$ & -1.0998296722163e+01 & -1.0634488873369e+01 & -1.0970511273439e+01  \\
    $u_x(4,0.6)$ & -1.0998296722163e+01 & -1.0634488873369e+01 & -1.0970511273439e+01  \\
    $u_{xx}(0,0.6)$ & -4.6844809230852e+00 & \fbox{-9.0735064404064e+00} & -4.8498203327098e+00  \\
    $u_{xx}(4,0.6)$ & -4.6844809230855e+00 & \fbox{3.4743060900784e+00} & -4.8498203327099e+00  \\
    \hline
    $u(0,1.92)$ & 8.8799799206032e-01 & 8.8941798966940e-01 & 8.8446899507452e-01  \\
    $u(4,1.92)$ & 8.8799799206032e-01 & 8.8941798966940e-01 & 8.8446899507452e-01  \\
    $u_x(0,1.92)$ & 4.0457999629891e+00 & 3.8734589665624e+00 & 4.0376021450541e+00  \\
    $u_x(4,1.92)$ & 4.0457999629891e+00 & 3.8734589665624e+00 & 4.0376021450541e+00  \\
    $u_{xx}(0,1.92)$ & 1.1406604203965e+00 & \fbox{5.2000604027409e+00} & 1.7849345842143e+00  \\
    $u_{xx}(4,1.92)$ & 1.1406604203965e+00 & \fbox{7.0346713619307e-01} & 1.7849345842144e+00  \\
    \hline
    $u(0,3.3)$ & 1.7326144730460e+00 & 1.7391187048218e+00 & 1.7317627457812e+00  \\
    $u(4,3.3)$ & 1.7326144730459e+00 & 1.7391187048218e+00 & 1.7317627457812e+00  \\
    $u_x(0,3.3)$ & 7.9202902481403e+00 & 7.7216199212532e+00 & 7.9054992498656e+00  \\
    $u_x(4,3.3)$ & 7.9202902481403e+00 & 7.7216199212532e+00 & 7.9054992498656e+00  \\
    $u_{xx}(0,3.3)$ & 3.2478290437743e+00 & \fbox{6.6397883264174e+00} & 3.4948463245322e+00  \\
    $u_{xx}(4,3.3)$ & 3.2478290437744e+00 & \fbox{2.3875667090972e-01} & 3.4948463245323e+00  \\
    \hline
  \end{tabular}
  \caption{
    Diffusion equation: values of the solution and its derivatives
    on the spatial boundaries ($x=0$ and $4$) at several time instants
    ($14$ significant digits shown),
    obtained from the DNN solutions
    with $C^{\infty}$ and $C^1$ periodic BCs and from the exact solution.
    The boxes highlight the differences in the obtained values
    for the second derivatives.
  }
  \label{tab:diffu_bv}
\end{table}

To assess how well the current method enforces the periodic boundary conditions,
we have extracted the values of the solution and its partial derivatives (up to
order two) on the spatial boundaries ($x=0$ and $4$) at several time
instants from the exact solution and the DNN solutions obtained
using the $C^{\infty}$ and $C^1$ periodic boundary conditions.
Table \ref{tab:diffu_bv} lists these boundary values for the solution and its
derivatives, with $14$ significant digits shown.
As expected, the current method has enforced exactly the periodicity
for the solution and all its derivatives extracted here with the $C^{\infty}$ periodic BCs,
while with the $C^{1}$ periodic BCs it enforces exactly the periodicity
only for the solution and its first derivative.

\subsection{Wave Equation with Periodic BCs}

In the last numerical example, we test the proposed method
using the wave equation,
\begin{equation}\label{equ:wave}
  \frac{\partial u}{\partial t} - c\frac{\partial u}{\partial x} = 0,
\end{equation}
where the prescribed constant $c$ represents the wave speed, $u(x,t)$ is the unknown
field function to be solved for, $x$ is the spatial coordinate and
$t$ is the time.
We consider the spatial-temporal domain
$\Omega = \{(x,t)| a\leqslant x\leqslant b,\ 0\leqslant t\leqslant T \}$
for this problem,
where $a$, $b$ and $T$ are prescribed constants whose values are
to be specified below.
We consider the following initial condition,
\begin{equation}\label{equ:wave_ic}
  u(x,0) = u_{in}(x) =  2\sech \frac{3(x-x_0)}{\delta},
  \quad \forall x\in[a,b],
\end{equation}
where $x_0\in[a,b]$ and $\delta$ are prescribed constants whose values are specified below.

We impose the periodic boundary condition on the spatial boundaries
of the domain, $x=a$ and $b$.
Specifically, we consider $C^{\infty}$, $C^0$, $C^1$ and $C^2$ periodic
boundary conditions in this test.
The $C^k$ ($k=0,1,2$) periodic BCs involve the conditions:
\begin{align}
  &
  u(a,t)=u(b,t), \quad \forall t\in[0,T];
  \label{equ:wave_c0} \\
  &
  \frac{\partial}{\partial x}u(a,t) = \frac{\partial}{\partial x}u(b,t), 
  \quad \forall t\in[0,T]; \label{equ:wave_c1} \\
  &
  \frac{\partial^2}{\partial x^2}u(a,t) = \frac{\partial^2}{\partial x^2}u(b,t), 
  \quad \forall t\in[0,T]. \label{equ:wave_c2} 
\end{align}
The $C^0$ periodic BC imposes the condition \eqref{equ:wave_c0}.
The $C^1$ periodic BC imposes the conditions \eqref{equ:wave_c0} and \eqref{equ:wave_c1}.
The $C^2$ periodic BC imposes the conditions \eqref{equ:wave_c0}--\eqref{equ:wave_c2}.
The $C^{\infty}$ periodic BC imposes the conditions:
\begin{equation}\label{equ:wave_cinf}
  u(a,t)=u(b,t), \quad
  \frac{\partial}{\partial x}u(a,t) = \frac{\partial}{\partial x}u(b,t), \quad
  \dots, \quad
  \frac{\partial^{m}}{\partial x^{m}}u(a,t) = \frac{\partial^{m}}{\partial x^{m}}u(b,t),
  \quad
  \dots, \quad \forall t\in[0,T].
\end{equation}

This initial/boundary value problem has the solution,
\begin{equation}\label{equ:wave_exact}
  u(x,t) =\left\{
  \begin{array}{ll}
    u_{in}(x+ct) =  2\sech \frac{3(x-x_0+ct)}{\delta}, & \text{if} \ (x+ct)\in[a,b], \\
    u(x\pm L, t), & \text{otherwise},
  \end{array}
  \quad \forall (x,t)\in \Omega,
  \right.
\end{equation}
where $L=b-a$.
In the numerical tests reported below we have employed the following
values for the parameters:
\begin{equation}
  c = 2, \quad
  T = 4, \quad
  a = 0, \quad
  b = 4, \quad
  L = b-a = 4, \quad
  \delta = 1, \quad
  x_0 = 2.
\end{equation}


To solve this initial/boundary value problem, we employ a
feed-forward DNN together with the method from Section \ref{sec:method}
for enforcing the periodic boundary conditions.
The input layer of the DNN consists of two nodes, which represent
the spatial coordinate $x$ and the time $t$.
The output layer of the DNN consists of one node, which represents
the field function $u$ to be solved for.
We employ $3$ hidden layers between the input and the output layers.
Each hidden layer has an output with $30$ nodes.
Since the periodic BC is only imposed in the $x$ direction,
we employ the modified 2D periodic layers to enforce
periodic boundary conditions; see the Remarks \ref{lem:lem_a}
and \ref{lem:lem_b}.
For the $C^{\infty}$ periodic BCs, the second layer of
this DNN (or the first hidden layer) is set to be
a modified 2D $C^{\infty}$ periodic layer as discussed in Remark \ref{lem:lem_a}.
For the $x$ direction, this modified periodic layer corresponds to
the 1D $C^{\infty}$ periodic layer $\mathcal{L}_p(m,n)$
with $m=12$ and $n=30$ (see equation \eqref{equ:p_layer}),
in which the constant $\omega$ is set to
$\omega=\frac{2\pi}{L}=\frac{\pi}{2}$.
For the $C^k$ ($k=0,1,2$) periodic BCs, the second layer of this DNN
is set to be a modified 2D $C^k$ periodic layer, which for
the $x$ direction corresponds to the 1D $C^k$ periodic layer $\mathcal{L}_{C^k}(m,n)$
with $m=12$ and $n=30$ (see equation \eqref{equ:ckp_layer}).

We employ the following loss function for this DNN,
\begin{equation}\label{equ:wave_loss}
  \begin{split}
    \text{Loss} &= \theta_{eq}\frac{1}{V_{\Omega}}\int_{\Omega}\left(
    \frac{\partial u}{\partial t} - c\frac{\partial u}{\partial x}
    \right)^2d\Omega
    + \theta_{ic}\frac{1}{b-a}\int_{a}^{b}\left[
      u(x,0) - u_{in}(x)
      \right]^2dx \\
    &=  \frac{\theta_{eq}}{V_{\Omega}}\sum_{e=0}^{N_{el}-1}\int_{\Omega_e}\left(
    \frac{\partial u}{\partial t} - c\frac{\partial u}{\partial x}
    \right)^2d\Omega
    + \frac{\theta_{ic}}{b-a}\sum_{e=0}^{N_{el}^x-1}\int_{a_e}^{b_e}\left[
      u(x,0) - u_{in}(x)
      \right]^2dx \\
    &= \frac{\theta_{eq}}{(b-a)T}\sum_{e=0}^{N_{el}-1}\sum_{i,j=0}^{Q-1}\left[
      \frac{\partial}{\partial t}u(x_i^e,t_j^e) - c\frac{\partial}{\partial x}u(x_i^e,t_j^e)
      \right]^2 J^e w_{ij} \\
    &\quad
    + \frac{\theta_{ic}}{b-a}\sum_{e=0}^{N_{el}^x-1}\sum_{i=0}^{Q-1}\left[
      u(x_i^e,0) - u_{in}(x_i^e)
      \right]^2 J^{ex} w_i.
  \end{split}
\end{equation}
In the above expression, $V_{\Omega}=\int_{\Omega}d\Omega=(b-a)T$
is the volume of the spatial-temporal domain $\Omega$, and
the constants $\theta_{eq}$ and $\theta_{ic}$ are the penalty coefficients
for the loss terms associated with the equation and the initial condition,
respectively. In order to compute the integrals, we have partitioned
the domain $\Omega$ into $N_{el}$ spatial-temporal elements,
with $N_{el}^x$ elements in the spatial direction
and $N_{el}^t$ elements in time,
leading to the relation $N_{el}=N_{el}^xN_{el}^t$.
$\Omega_e$ denotes the region occupied by the spatial-temporal element
$e$ for $0\leqslant e\leqslant N_{el}-1$. The interval $[a_e,b_e]$
denotes the region of the spatial element $e$ for $0\leqslant e\leqslant N_{el}^x-1$.
$Q$ is the number of quadrature points in both the spatial and temporal directions
within each spatial-temporal element.
$(x_i^e,t_j^e)$ ($0\leqslant i,j\leqslant Q-1$) are the Gauss-Lobatto-Legendre
quadrature points within the spatial-temporal element $\Omega_e$, for
$0\leqslant e\leqslant N_{el}-1$.
$J^e$ is the Jacobian of the spatial-temporal element $\Omega_e$
($0\leqslant e\leqslant N_{el}-1$),
and $J^{ex}$ is the Jacobian of the spatial element $[a_e,b_e]$
($0\leqslant e\leqslant N_{el}^{x}-1$).
$w_{ij}$ ($0\leqslant i,j\leqslant Q-1$) denote the weights associated
with the Gauss-Lobatto-Legendre
quadrature points $(x_i^e,t_j^e)$.
$w_i$ ($0\leqslant i\leqslant Q-1$) denote the weights associated with
the spatial Gauss-Lobatto-Legendre quadrature point $x_i^e$.
The input data to the DNN are the quadrature points
$(x_i^e,t_j^e)$ for $0\leqslant i,j\leqslant Q-1$ and $0\leqslant e\leqslant N_{el}-1$.
In the loss function \eqref{equ:wave_loss}, $u(x_i^e,t_j^e)$
is obtained from the output of the DNN, and the terms
$\left.\frac{\partial u}{\partial t} \right|_{(x_i^e,t_j^e)}$ and
$\left.\frac{\partial u}{\partial x} \right|_{(x_i^e,t_j^e)}$ can be
computed based on auto-differentiation.
It can be observed that the initial condition is enforced
by the penalty method.


For the numerical results reported below, we have partitioned
the domain $\Omega$ into $4$ spatial-temporal elements ($N_{el}=4$),
with $2$ uniform elements along the spatial and temporal
directions ($N_{el}^x=N_{el}^t=2$).
We employ $30$ quadrature points in both space and time ($Q=30$)
within each spatial-temporal element.
The penalty coefficients are set to be $\theta_{eq}=0.9$ and
$\theta_{ic}=1-\theta_{eq}=0.1$.
We use ``tanh'' as
the activation functions for the hidden layers. No activation
is applied to the output layer.
The Adam optimizer has been employed to train the DNN
for $60,000$ epochs with the $C^{\infty}$ and $C^1$ periodic BCs,
for $90,000$ epochs with the $C^0$ periodic BC,
and for $80,000$ epochs with the $C^2$ periodic BCs.
The options for ``early stopping'' and ``restore to best weight''
in Tensorflow/Keras are employed during the training of
the DNNs.

\begin{figure}
  \centerline{
    \includegraphics[width=2.2in]{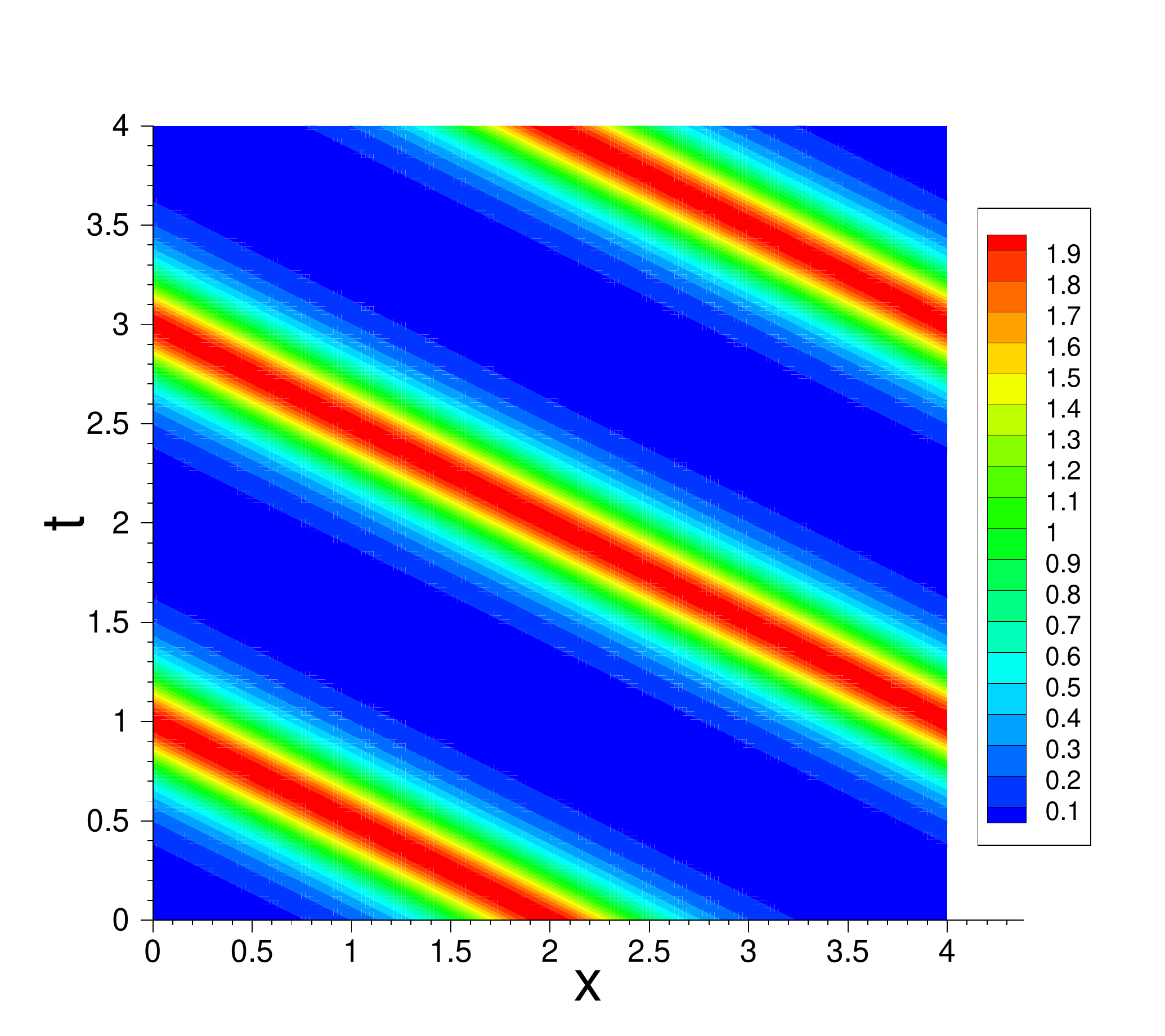}(a)
    \includegraphics[width=2.2in]{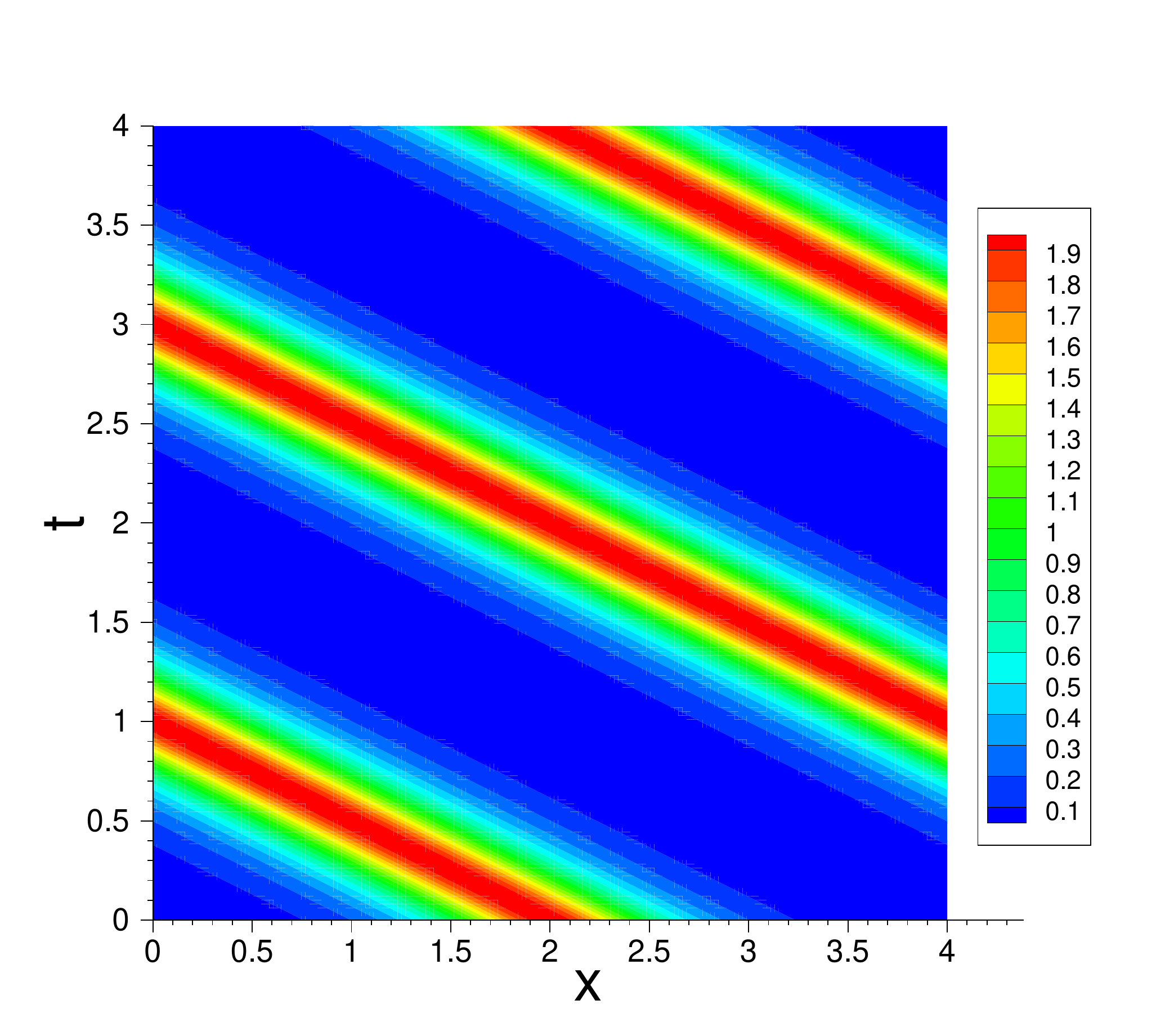}(b)
  }
  \centerline{
    \includegraphics[width=2.2in]{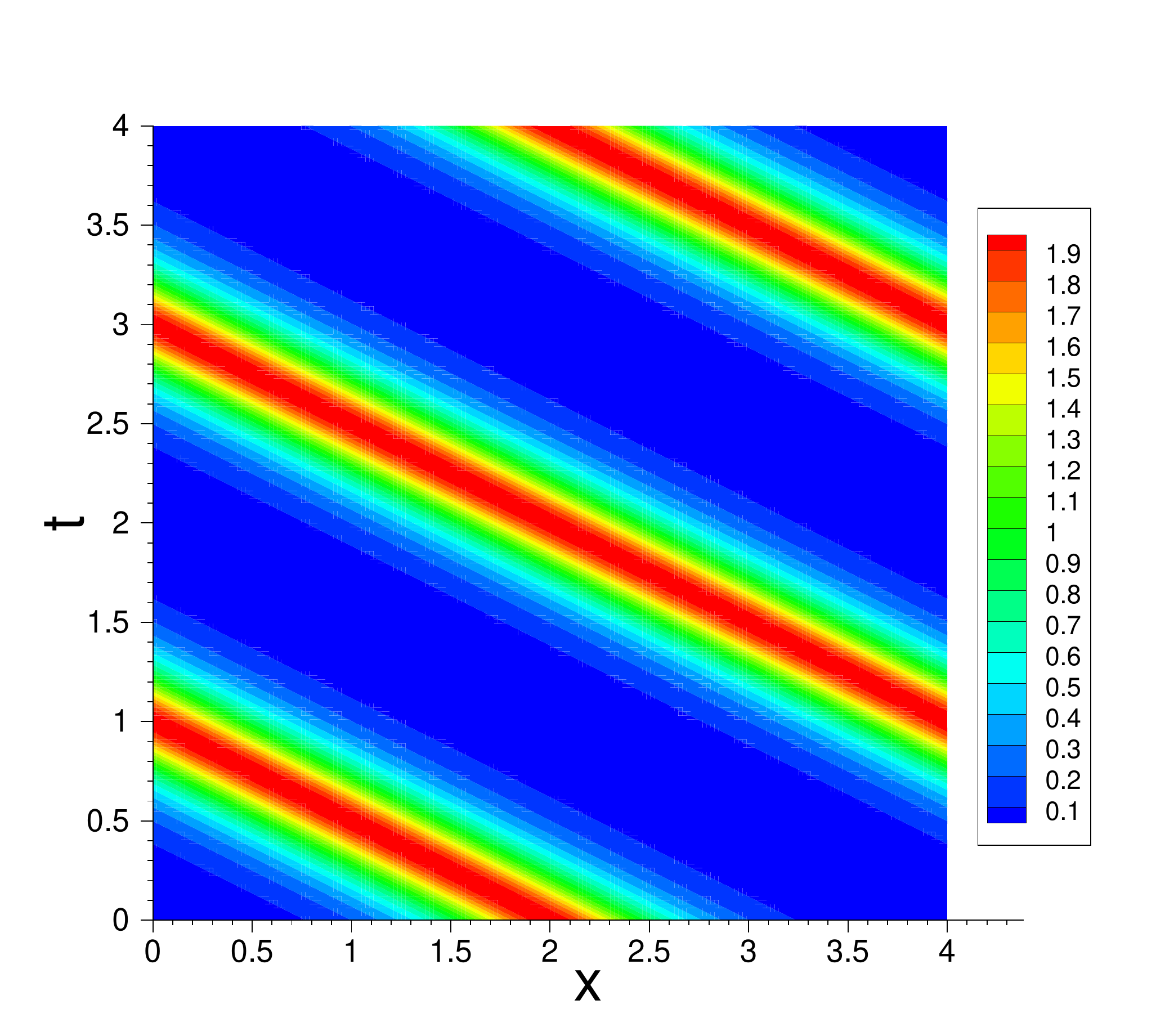}(c)
    \includegraphics[width=2.2in]{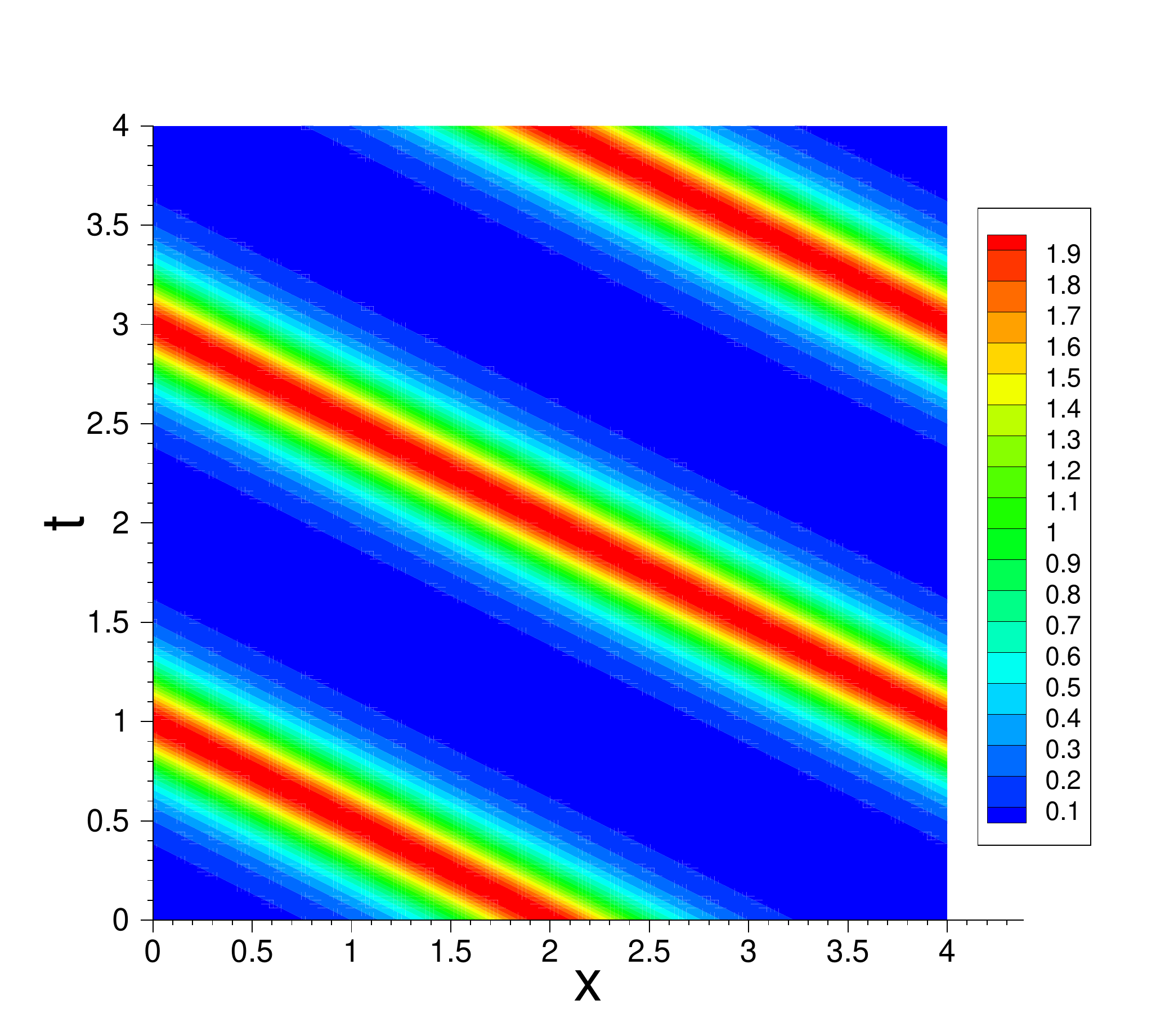}(d)
  }
  \caption{
    Wave equation: Contours of the DNN solutions
    obtained with the $C^{\infty}$ (a),
    $C^0$ (b), $C^1$ (c), $C^2$ (d)
    periodic boundary conditions in the $x$ direction.
  }
  \label{fig:wave_soln}
\end{figure}


Figure \ref{fig:wave_soln} illustrates the distributions of the DNN
solutions and their absolute errors.
The plots in the left column of this figure are contours of
the DNN solutions in the spatial-temporal ($x-t$) plane,
obtained with the $C^{\infty}$ and $C^k$ ($k=0,1,2$) periodic
BCs on the spatial boundaries.
The plots in the right column are contours of
the absolute errors of these solutions against the
exact solution \eqref{equ:wave_exact}.
Qualitatively, no difference can be discerned of the distributions
between the DNN solutions and the exact solution.
The maximum errors in the domain are approximately on
the order of magnitude $10^{-2}$.

\begin{figure}
  \centerline{
    \includegraphics[width=1.5in]{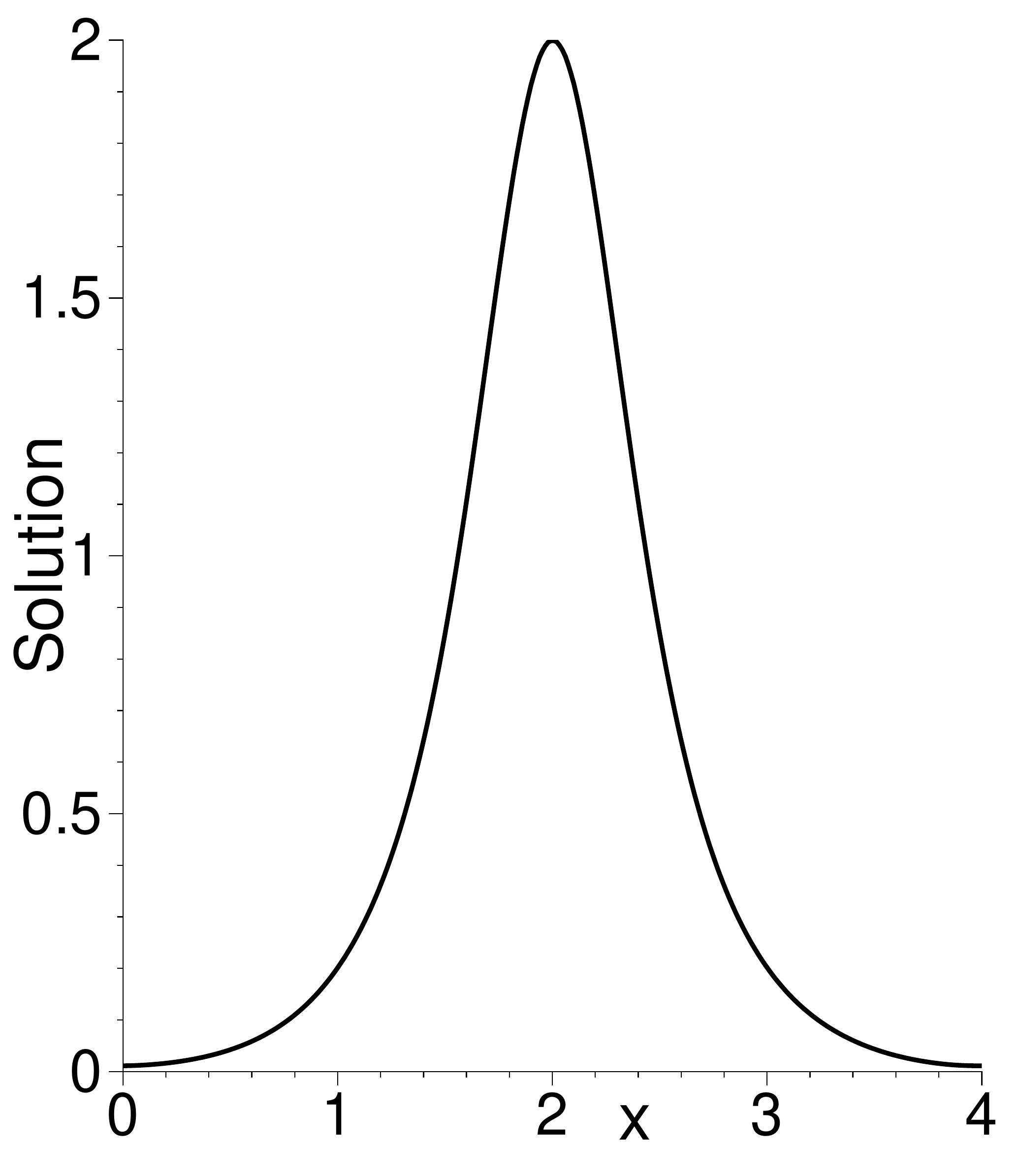}(a)
    \qquad\qquad
    \includegraphics[width=1.5in]{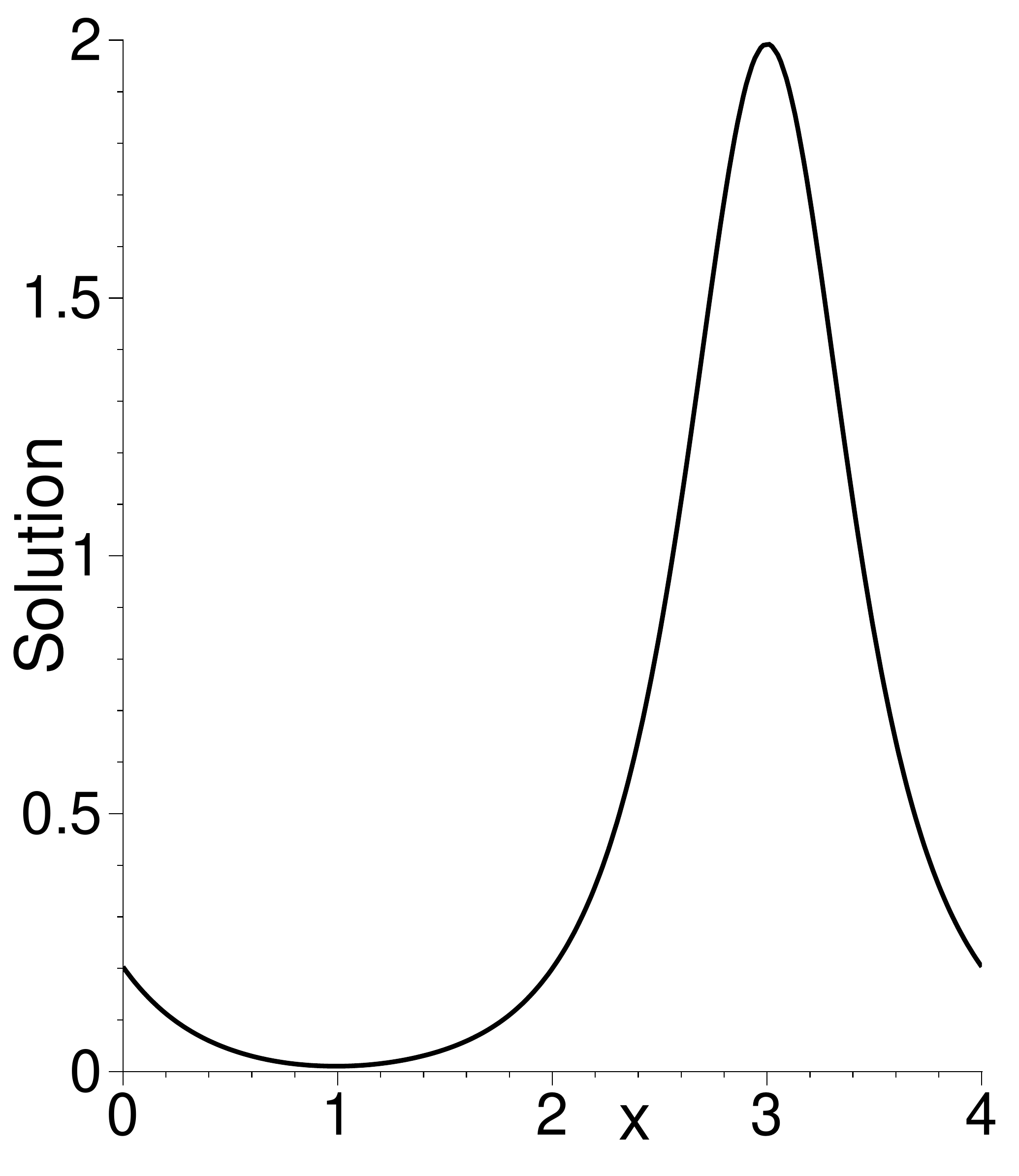}(d)
    \qquad\qquad
    \includegraphics[width=1.5in]{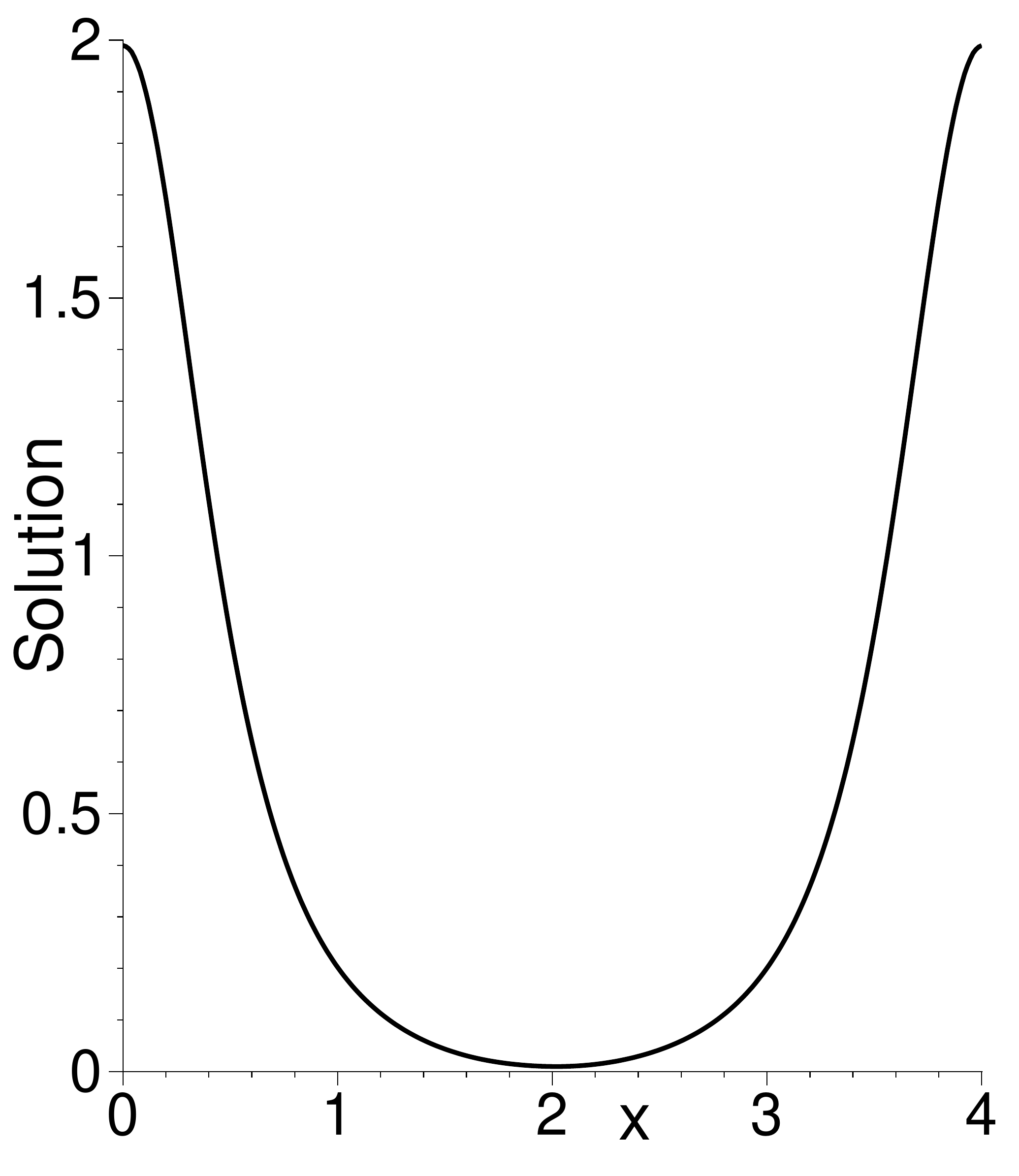}(g)
  }
  \centerline{
    \includegraphics[width=1.5in]{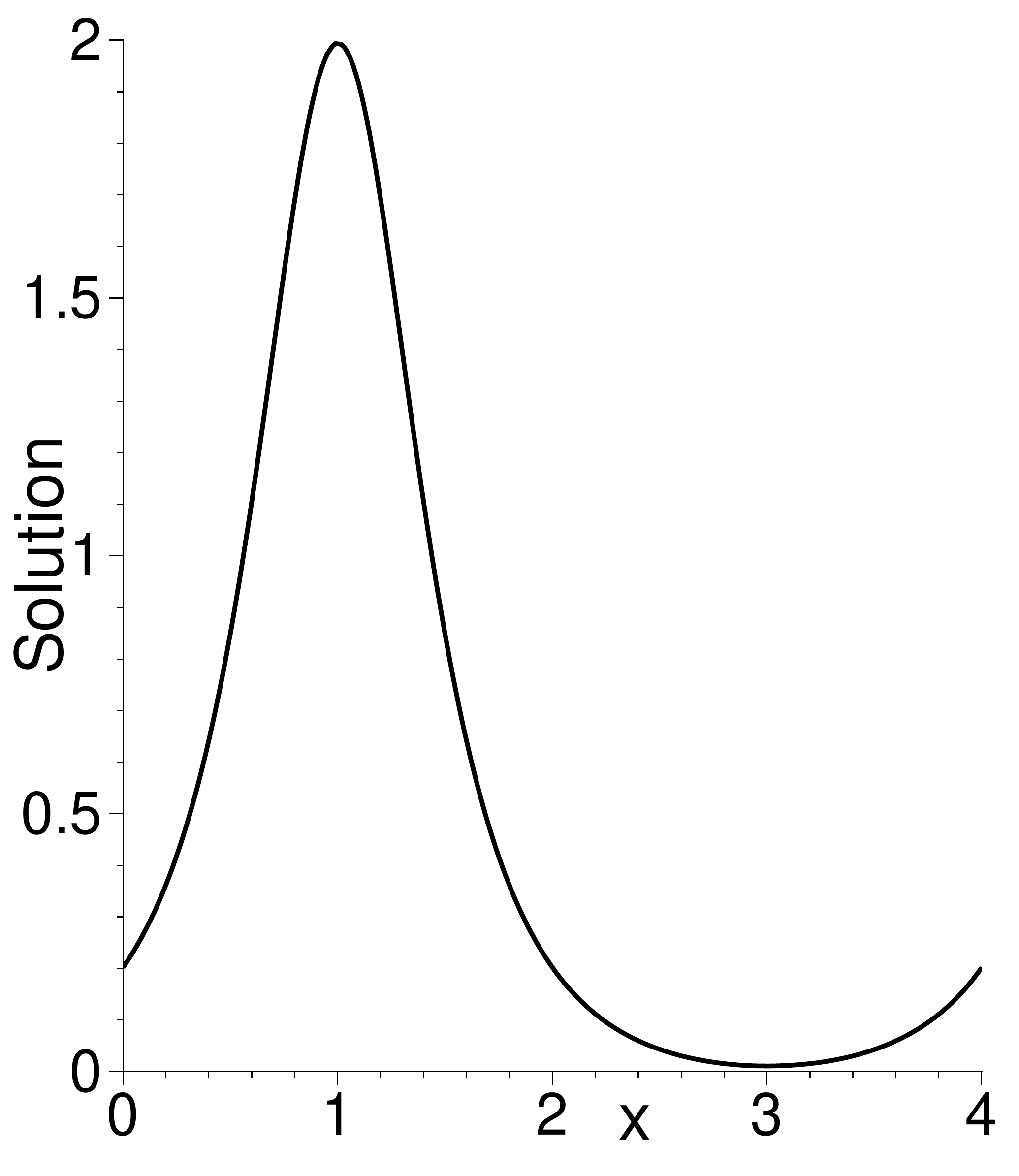}(b)
    \qquad\qquad
    \includegraphics[width=1.5in]{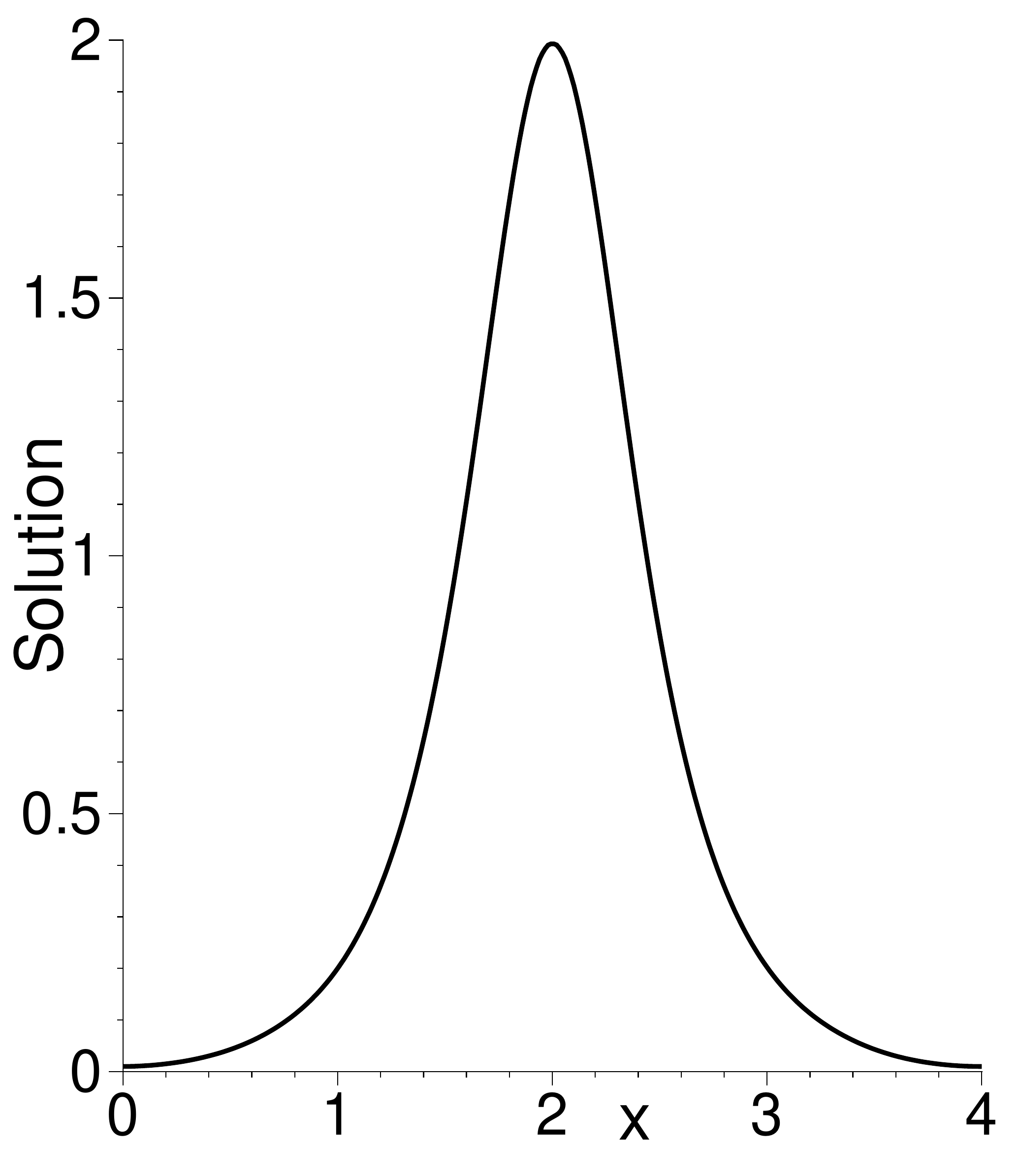}(e)
    \qquad\qquad
    \includegraphics[width=1.5in]{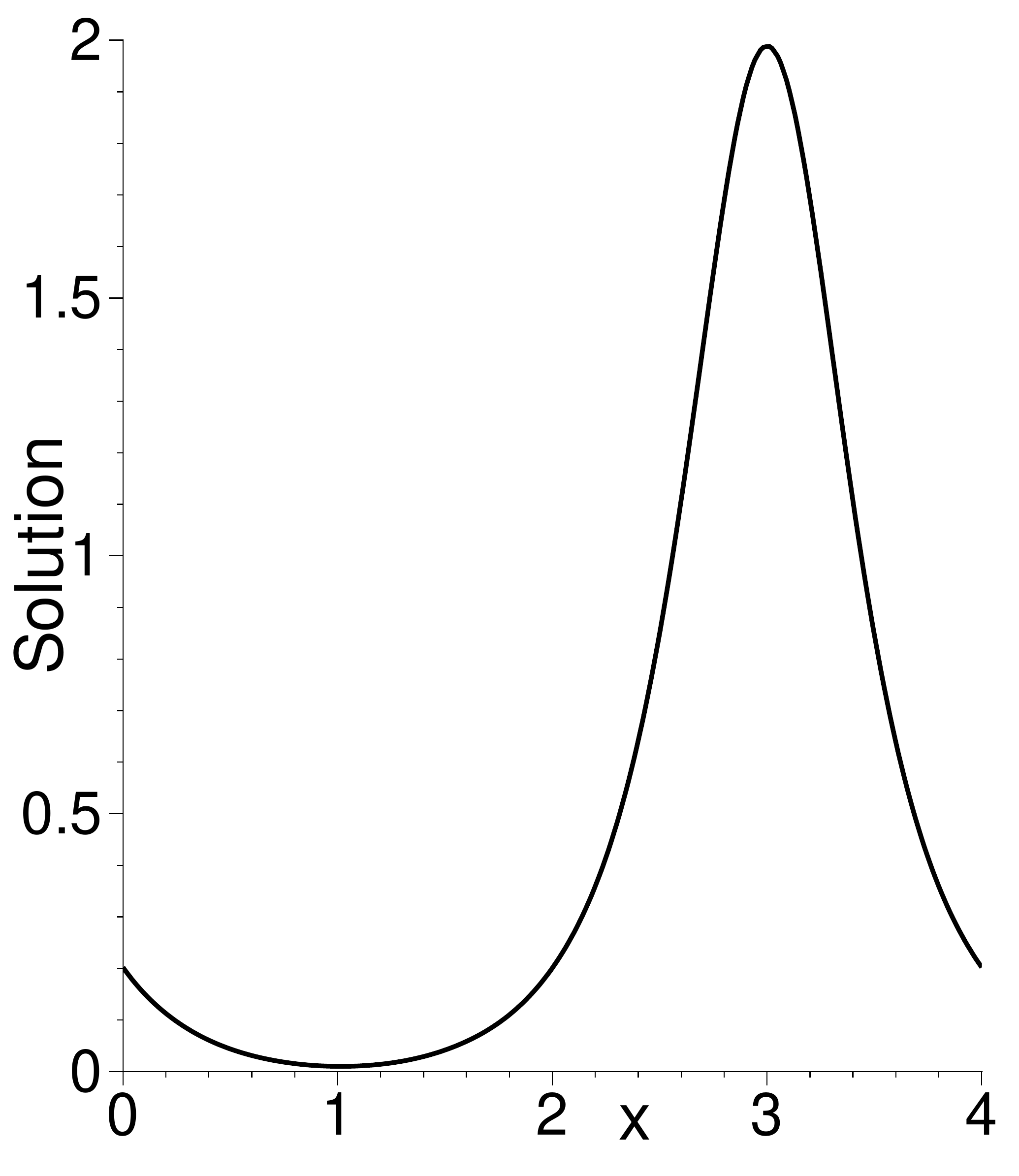}(h)
  }
  \centerline{
    \includegraphics[width=1.5in]{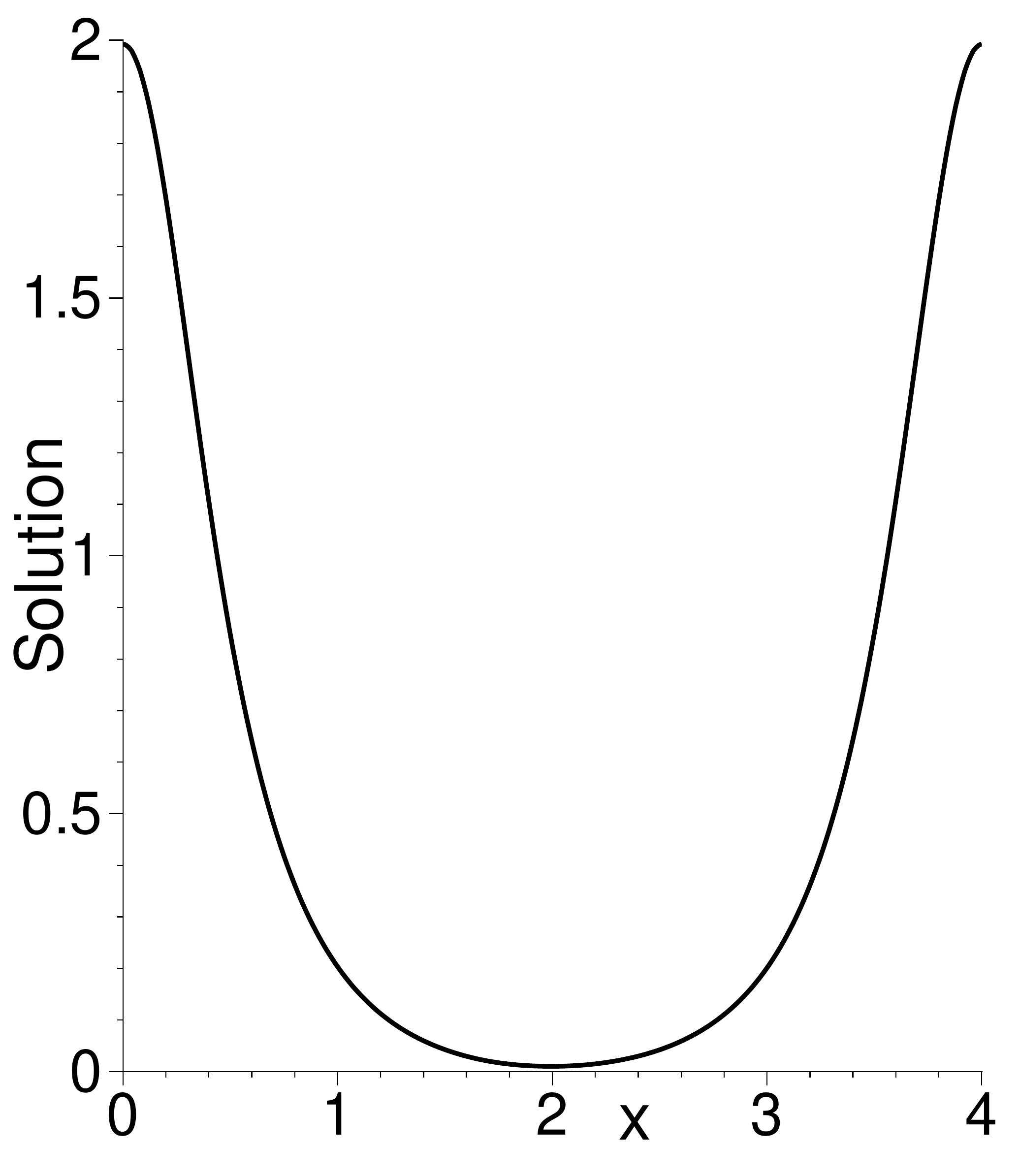}(c)
    \qquad\qquad
    \includegraphics[width=1.5in]{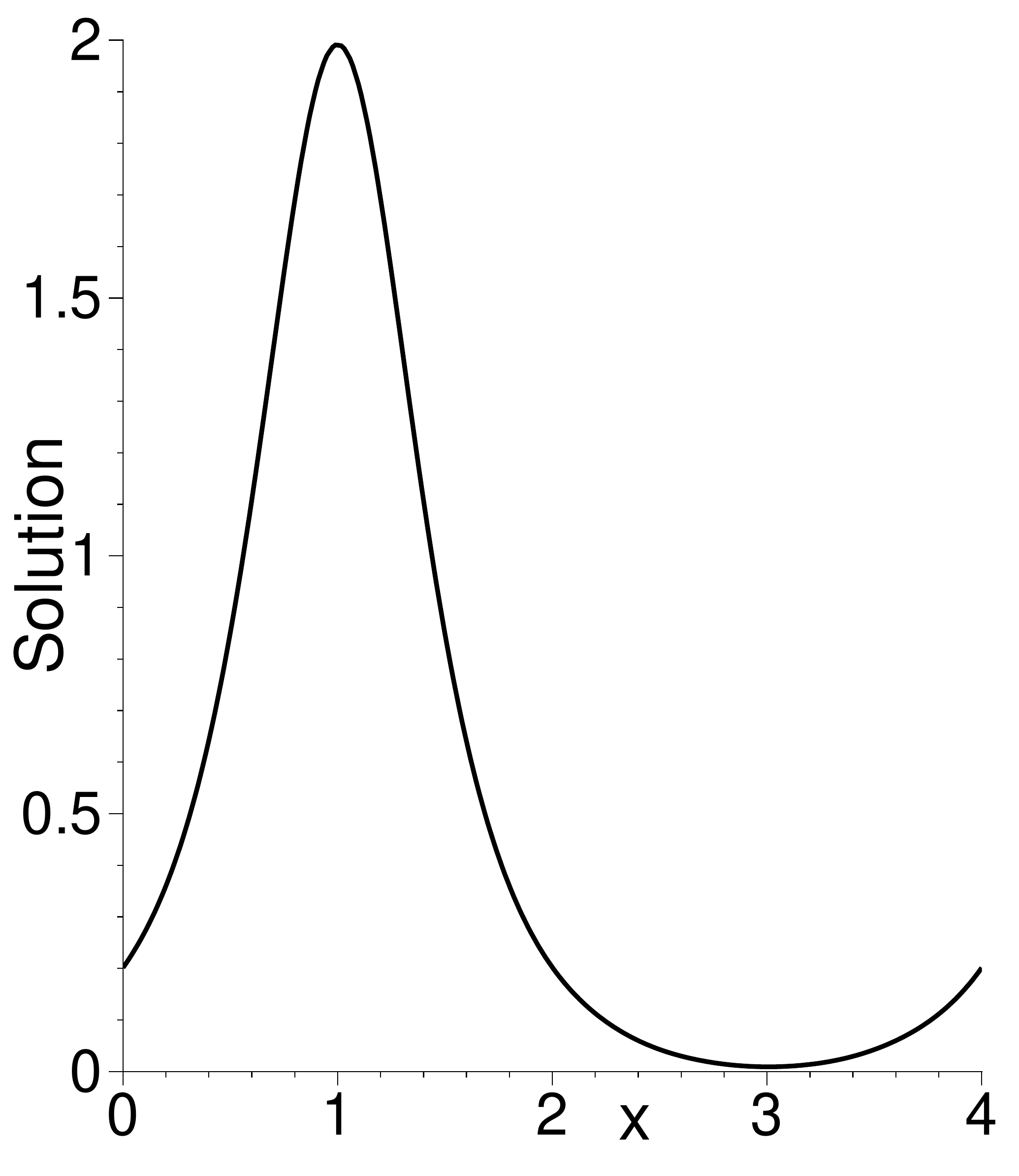}(f)
    \qquad\qquad
    \includegraphics[width=1.5in]{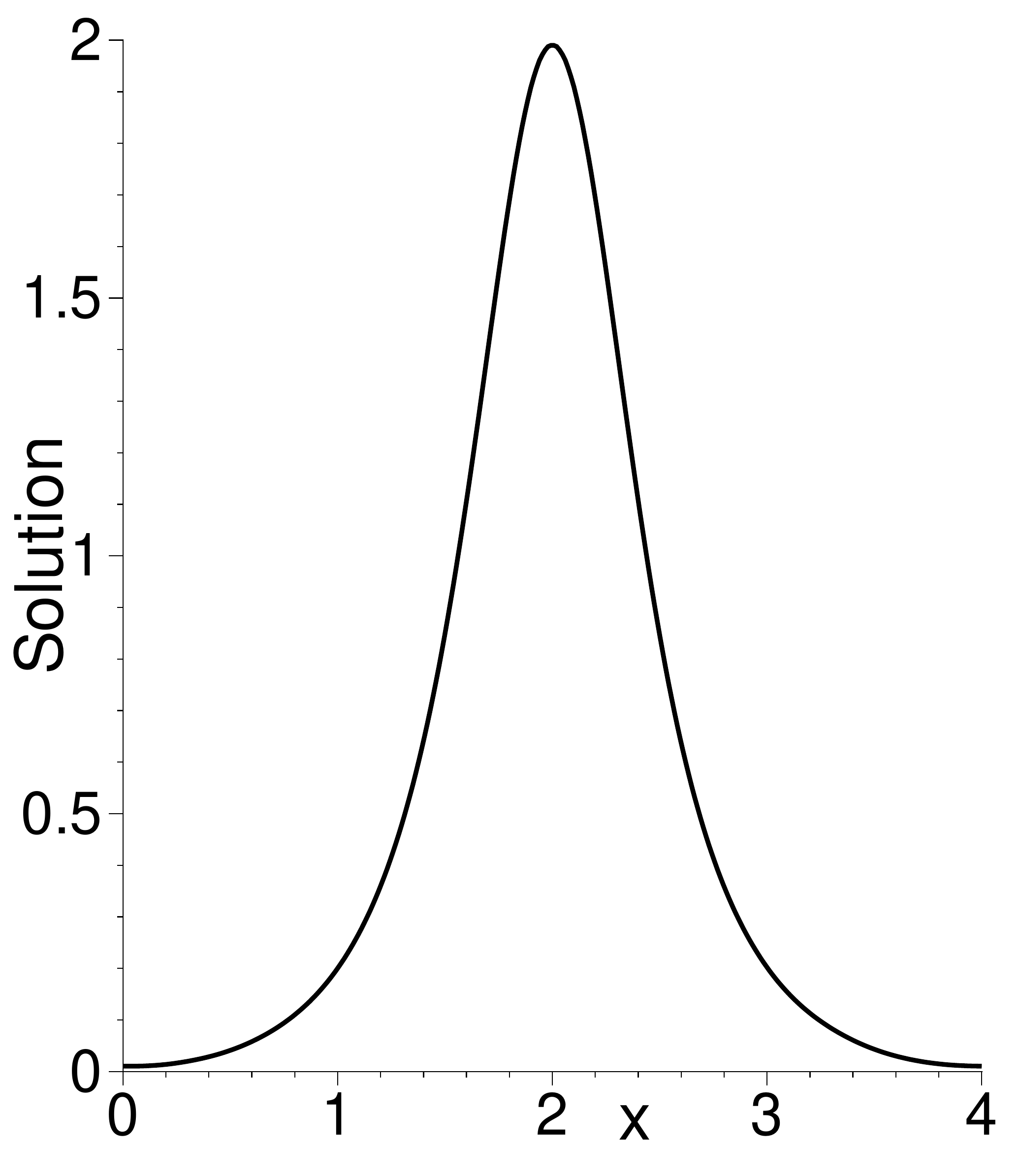}(i)
  }
  \caption{
    Wave equation: Snapshots of the wave profiles obtained with
    the $C^{\infty}$ periodic BC at time instants
    (a) $t=0$,
    (b) $t=0.5$,
    (c) $t=1.0$,
    (d) $t=1.5$,
    (e) $t=2.0$,
    (f) $t=2.5$,
    (g) $t=3.0$,
    (h) $t=3.5$,
    (i) $t=4.0$.
  }
  \label{fig:wave_snap}
\end{figure}

Figure \ref{fig:wave_snap} shows a temporal sequence of snapshots
of the wave form, obtained from the DNN solution with the
$C^{\infty}$ periodic boundary conditions. One can clearly observe
the propagation of the wave form in the $-x$ direction at a constant
speed. Because of the imposed periodic conditions, as soon as
the wave exits the left boundary ($x=0$), it re-enters the domain
through the right boundary ($x=4$) in a seamless and smooth fashion.

\begin{figure}
  \centerline{
    \includegraphics[width=2in]{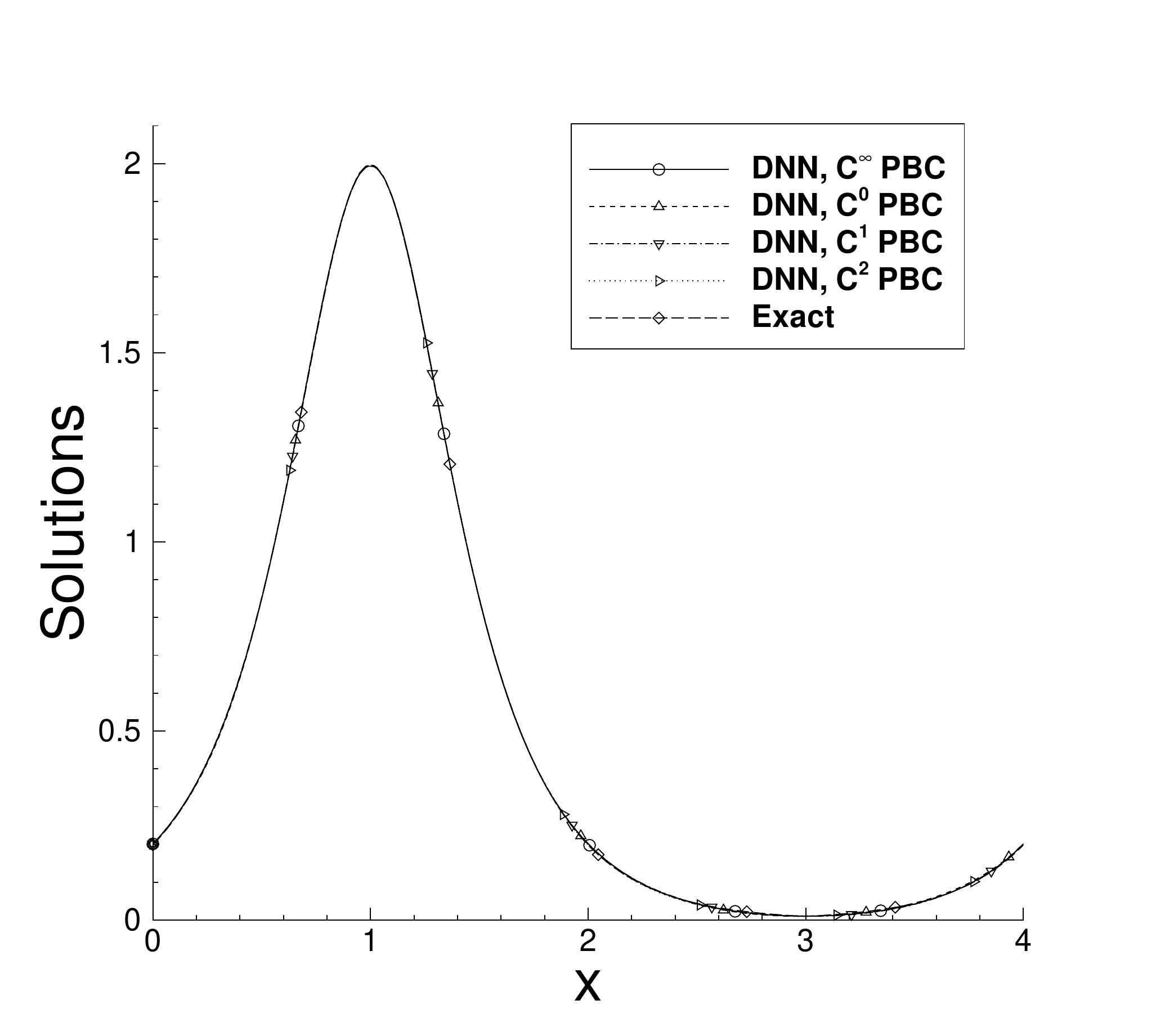}(a)
    \includegraphics[width=2in]{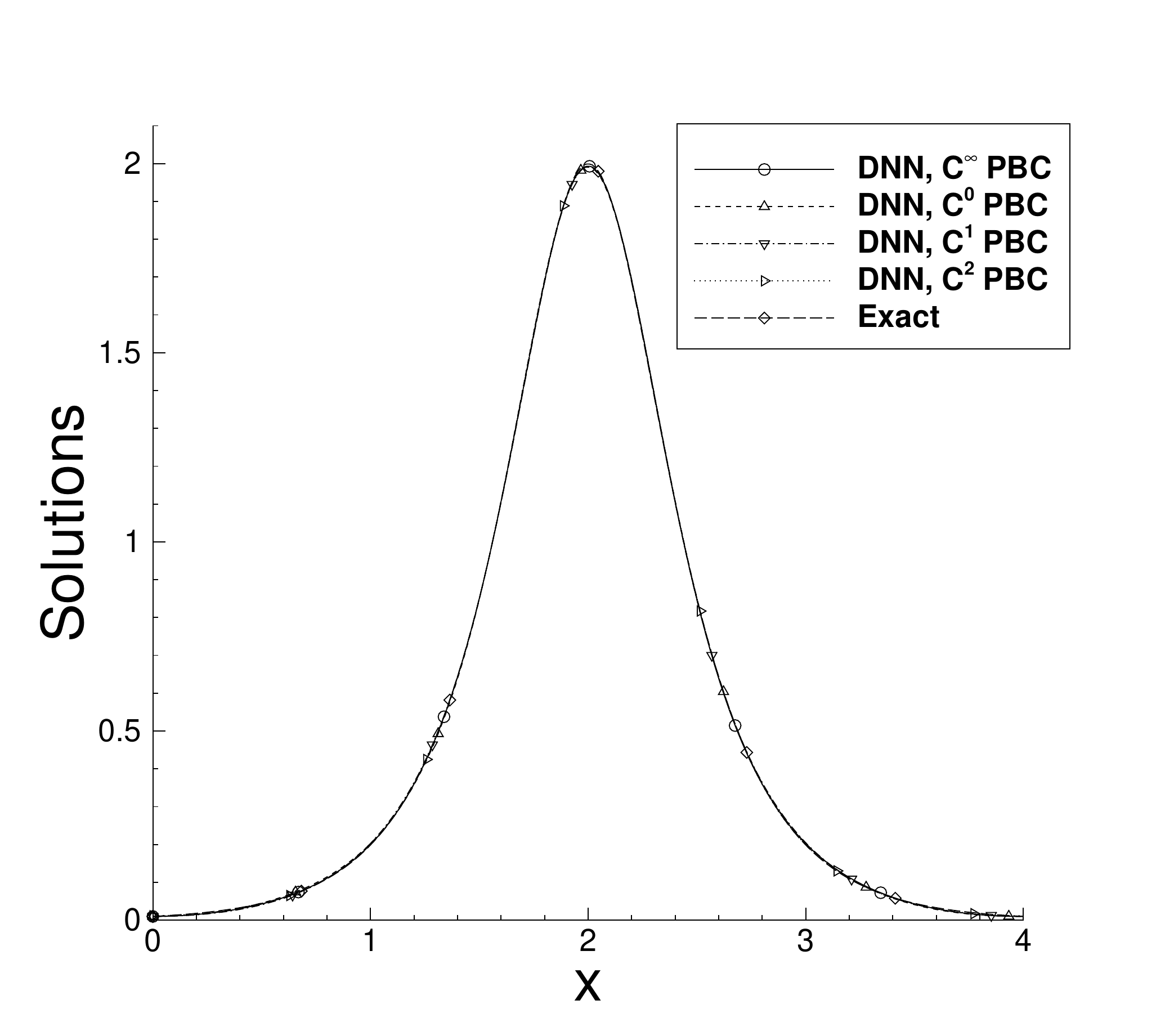}(b)
    \includegraphics[width=2in]{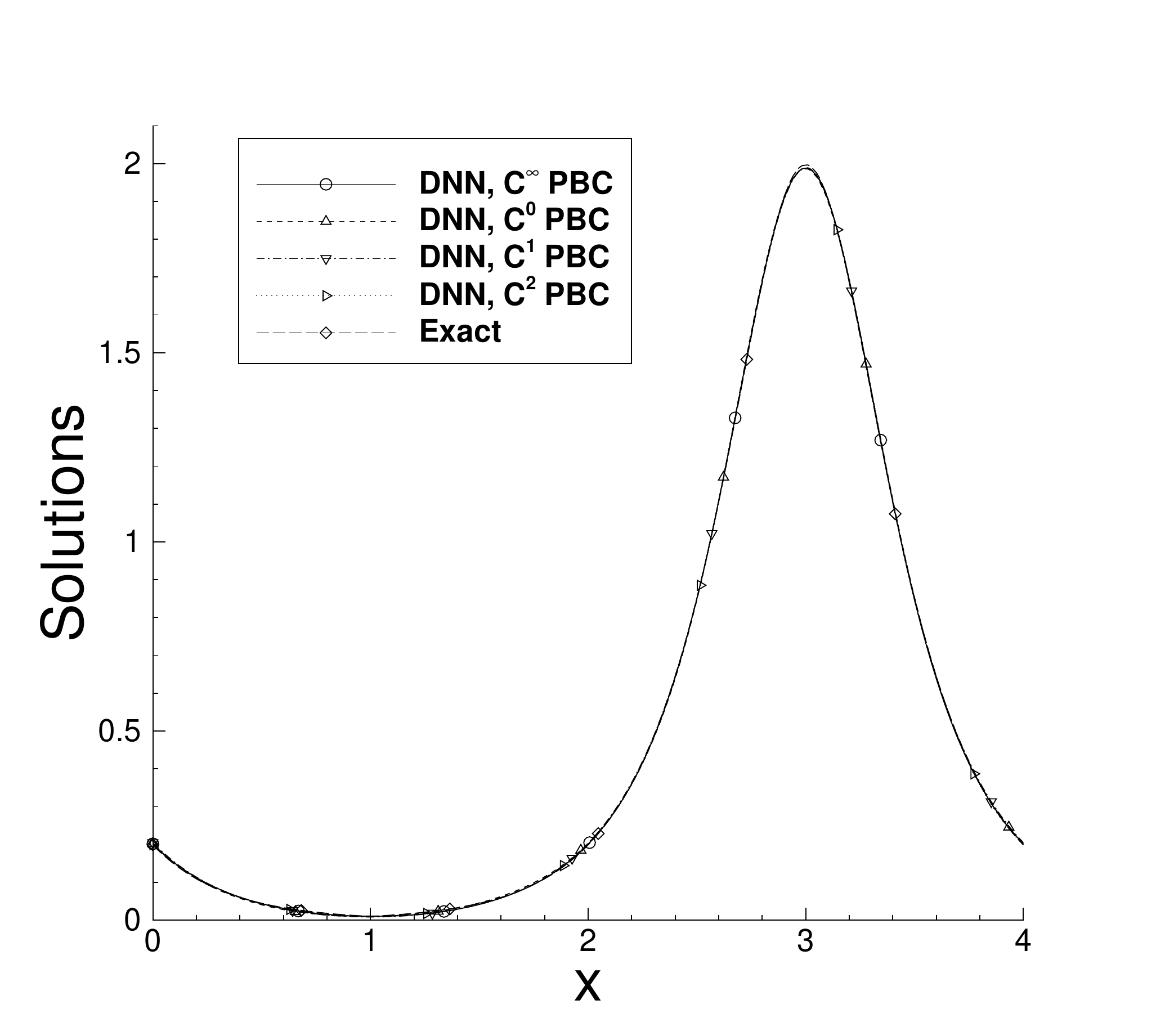}(c)
  }
  \centerline{
    \includegraphics[width=2in]{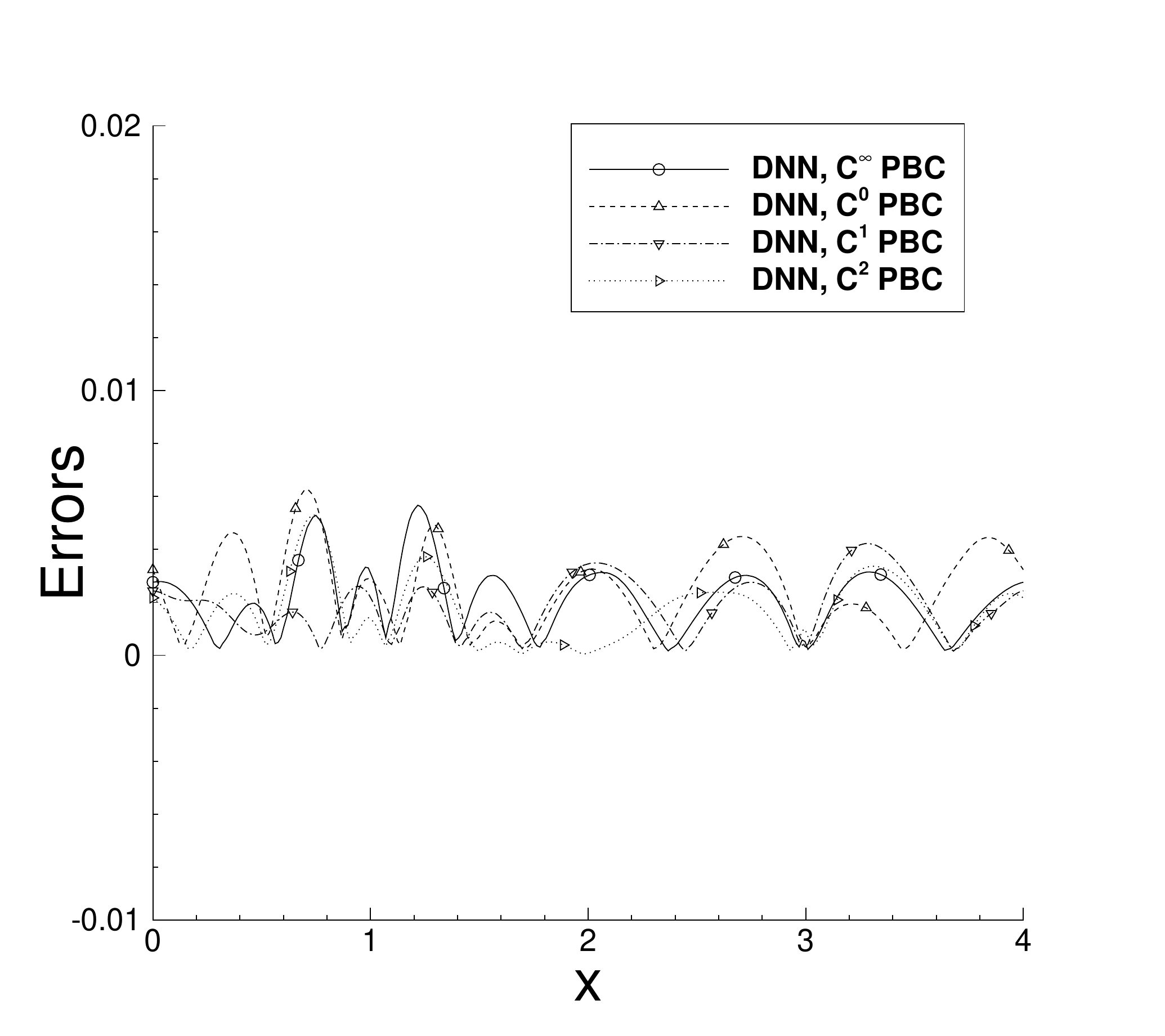}(d)
    \includegraphics[width=2in]{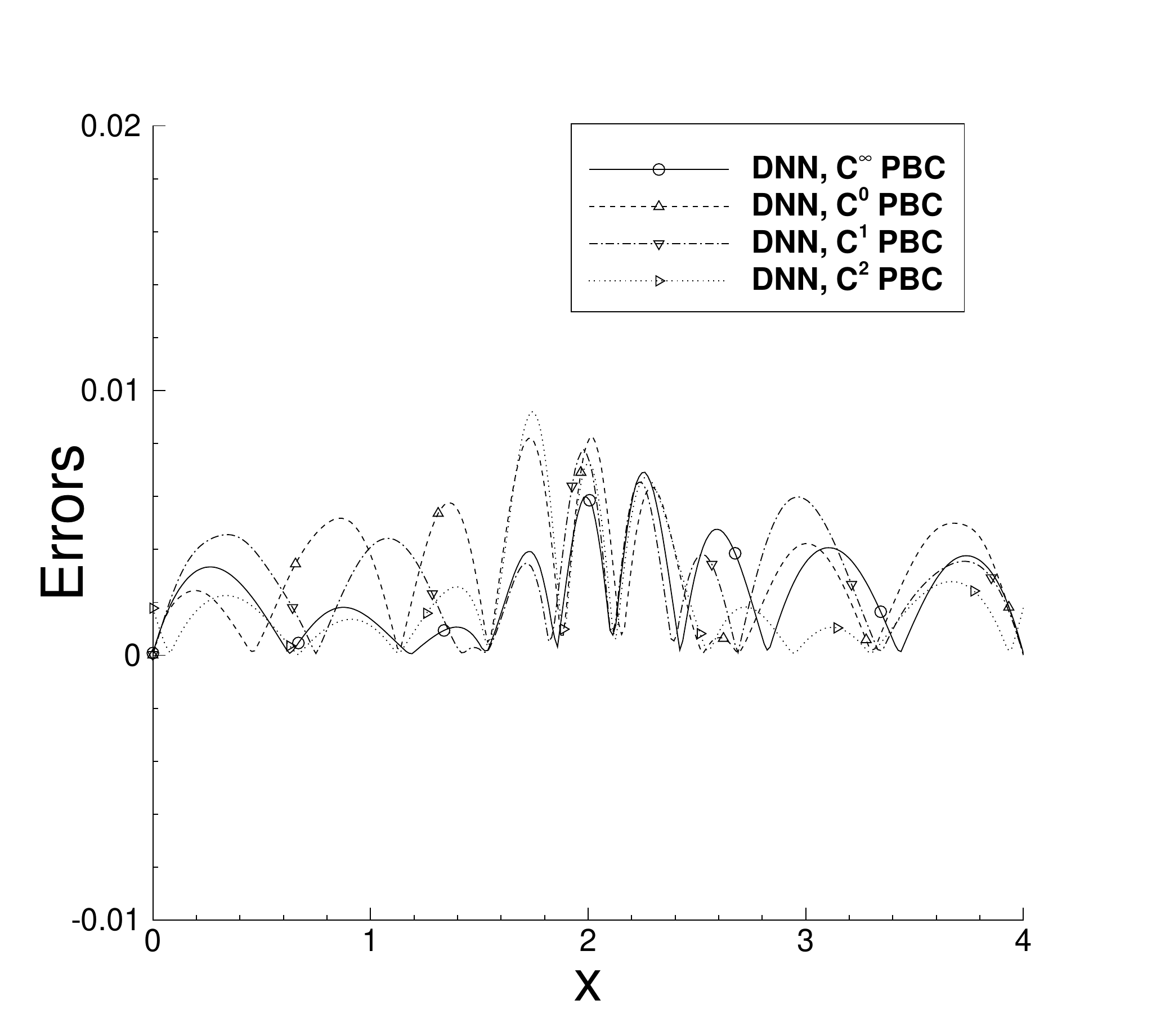}(e)
    \includegraphics[width=2in]{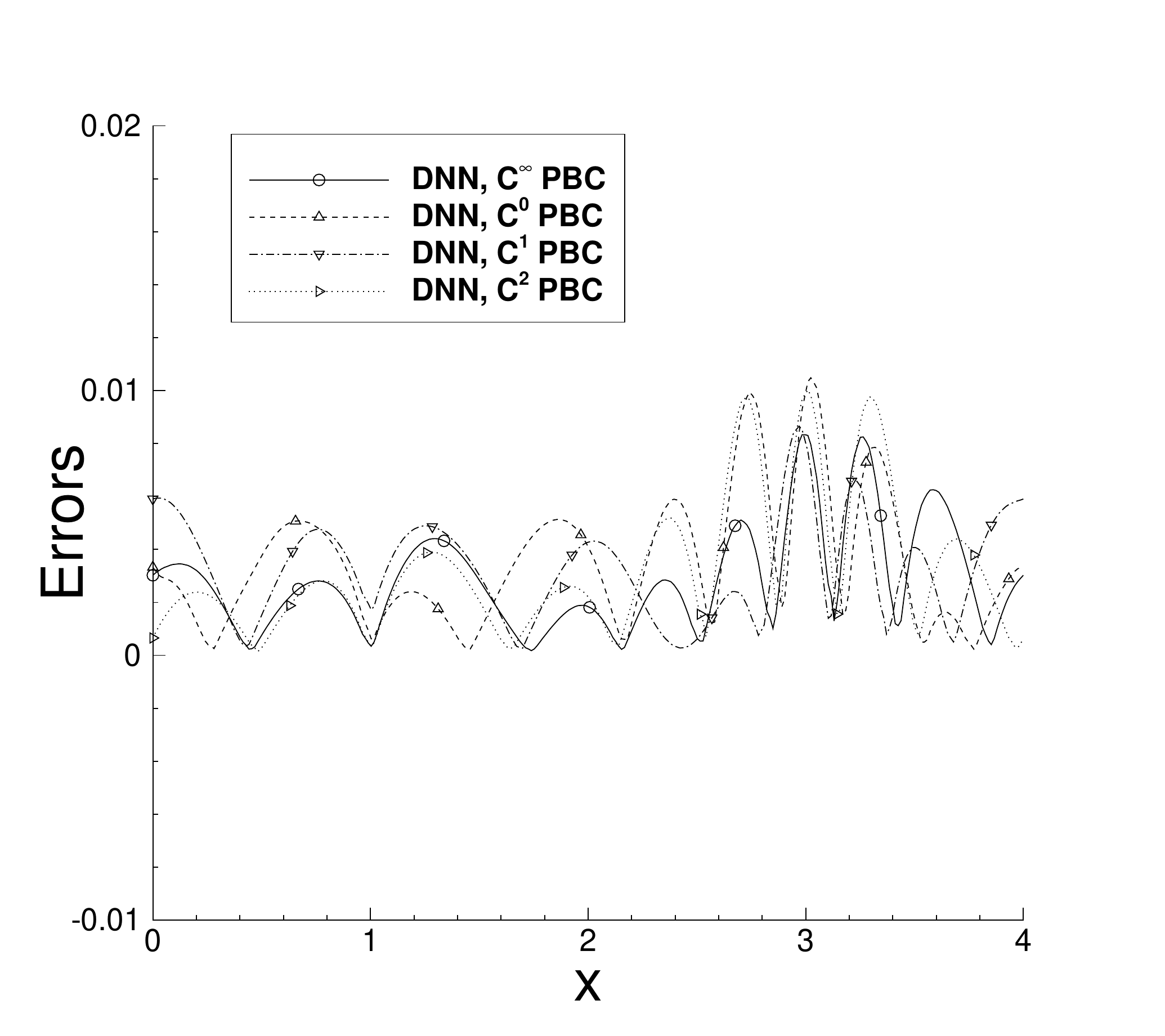}(f)
  }
  \caption{
    Wave equation: comparison of profiles of the solution (top row) and
    the absolute error (bottom row) obtained with the $C^{\infty}$, $C^0$, $C^1$
    and $C^2$ periodic boundary conditions and with the exact solution
    at several time instants:
    (a, d) $t=0.5$, (b, e) $t=2.0$, (c, f) $t=3.5$.
  }
  \label{fig:wave_prof}
\end{figure}

In Figure \ref{fig:wave_prof} we compare the wave profiles
from the exact solution \eqref{equ:wave_exact} and
from the DNN solutions with different types of periodic BCs
at three time instants ($t=0.5$, $2$ and $3.5$).
The error profiles of these DNN solutions are also included
in this figure (bottom row).
It is observed that the wave profiles obtained from the DNN
with various types of periodic BCs are in good agreement with that of
the exact solution, and that
the largest errors appear to coincide with the peak of the wave.

\begin{table}
  \footnotesize
  \centering
  \begin{tabular}{llllll}
    \hline
    & DNN $C^{\infty}$ PBC & DNN $C^0$ PBC & DNN $C^1$ PBC & DNN $C^2$ PBC & Exact solution\\
    $u(0,0.5)$ & 2.0142799261625e-01 & 2.0190756733086e-01 & 2.0109869263559e-01 &
    2.0083996838890e-01 & 1.9865585483886e-01 \\
    $u(4,0.5)$ & 2.0142799261625e-01 & 2.0190756733086e-01 & 2.0109869263559e-01 &
    2.0083996838891e-01 & 1.9865585483886e-01 \\
    $u_x(0,0.5)$ & 5.9482715294468e-01 & \fbox{5.7394660776802e-01} & 5.9323728164392e-01 &
    5.8746219179701e-01 & 5.9302035811534e-01 \\
    $u_x(4,0.5)$ & 5.9482715294468e-01 & \fbox{5.8248575768435e-01} & 5.9323728164392e-01 &
    5.8746219179701e-01 & 5.9302035811534e-01 \\
    $u_{xx}(0,0.5)$ & 1.6980636297410e+00 & \fbox{1.4696979666932e+00} & \fbox{1.5994875423905e+00} &
    1.6679798466138e+00 & 1.7526236647042e+00 \\
    $u_{xx}(4,0.5)$ & 1.6980636297410e+00 & \fbox{1.9395212811018e+00} & \fbox{1.5277304396975e+00} &
    1.6679798466138e+00 & 1.7526236647042e+00  \\
    \hline
    $u(0,2)$ & 9.8284051583224e-03 & 9.9241674367022e-03 & 9.9440310013471e-03 &
    1.1696938690667e-02 & 9.9149477871207e-03 \\
    $u(4,2)$ & 9.8284051583222e-03 & 9.9241674367022e-03 & 9.9440310013471e-03 &
    1.1696938690667e-02 & 9.9149477871207e-03 \\
    $u_x(0,2)$ & 1.5256678688244e-03 & \fbox{-4.8521014846574e-03} & 9.9336770382177e-04 &
    2.0456109624086e-03 & 2.9744477846340e-02 \\
    $u_x(4,2)$ & 1.5256678688247e-03 & \fbox{2.3815674708337e-03} & 9.9336770382177e-04 &
    2.0456109624087e-03 & 2.9744477846340e-02 \\
    $u_{xx}(0,2)$ & 2.4222108105432e-01 & \fbox{4.2985279066538e-01} & \fbox{1.1927388753390e-01} &
    2.1278569005620e-01 & 8.9230143930769e-02 \\
    $u_{xx}(4,2)$ & 2.4222108105432e-01 & \fbox{2.2731343637548e-01} & \fbox{2.6553423293645e-01} &
    2.1278569005620e-01 & 8.9230143930769e-02 \\
    \hline
    $u(0,3.5)$ & 2.0170117943758e-01 & 2.0200326210422e-01 & 2.0458115518013e-01 &
    1.9932351260077e-01 & 1.9865585483886e-01 \\
    $u(4,3.5)$ & 2.0170117943758e-01 & 2.0200326210422e-01 & 2.0458115518013e-01 &
    1.9932351260077e-01 & 1.9865585483886e-01 \\
    $u_x(0,3.5)$ & -5.8538598646821e-01 & \fbox{-6.1113064140580e-01} & -5.8981323269116e-01 &
    -5.7592303278591e-01 & -5.9302035811534e-01 \\
    $u_x(4,3.5)$ & -5.8538598646821e-01 & \fbox{-5.9639323869798e-01} & -5.8981323269116e-01 &
    -5.7592303278591e-01 & -5.9302035811534e-01 \\
    $u_{xx}(0,3.5)$ & 1.6523878185422e+00 & \fbox{2.3592894577992e+00} & \fbox{1.6262664810111e+00} &
    1.6631571855496e+00 & 1.7526236647042e+00 \\
    $u_{xx}(4,3.5)$ & 1.6523878185422e+00 & \fbox{1.2305332426305e+00} & \fbox{1.8162774394782e+00} &
    1.6631571855496e+00 & 1.7526236647042e+00 \\
    \hline
  \end{tabular}
  \caption{
    Wave equation: values of the solution and its derivatives
    on the left/right boundaries at several time instants ($t=0.5$, $2$ and $3.5$),
    obtained from the DNN solutions with $C^{\infty}$, $C^0$, $C^1$ and $C^2$ periodic
    boundary conditions and from the exact solution.
  }
  \label{tab:wave}
\end{table}

Table \ref{tab:wave} provides a verification that the current method
enforces the periodic conditions exactly for the wave equation as expected.
Here we list the boundary values of the DNN solutions obtained
with different types of periodic BCs and the exact solution,
as well as their partial derivatives (up to order two),
at several time instants ($t=0.5$, $2$ and $3.5$).
We have again included $14$ significant digits for each value.
It is evident that, with the $C^{\infty}$ and $C^2$ periodic BCs,
the current method has enforced exactly the periodic conditions
for the solution and its first and second derivatives.
With the $C^0$ periodic BC, the method enforces exactly the periodic condition
only for  the solution. With the $C^1$ periodic BC,
the method enforces exactly the periodic condition
for the solution and its first derivative, but not for its second derivative.

\section{Concluding Remarks}
\label{sec:summary}


In this paper we have presented a method for enforcing exactly
the $C^{\infty}$ and $C^k$ (for any $k\geqslant 0$) periodic
conditions with deep neural networks. The method stems
from some simple properties about function compositions
involving periodic functions.
The method essentially composes an arbitrary DNN-represented function
with a set of independent known periodic functions
with adjustable (training) parameters.
More specifically,
we have defined the operations
that constitute a $C^{\infty}$ periodic layer and
a $C^k$ periodic layer. The DNN with a $C^{\infty}$ periodic layer
incorporated as the second layer of the network (behind the input)
automatically and exactly satisfies the $C^{\infty}$
periodic boundary conditions in its output.
The DNN with a $C^k$ periodic layer incorporated as the second
layer automatically and exactly satisfies the $C^k$
periodic boundary conditions in its output.
The $C^{\infty}$ periodic layer comprises constructions of
a set of independent $C^{\infty}$ periodic functions with a prescribed period,
based on sinusoidal functions, affine mappings, and nonlinear
activation functions.
The $C^k$ periodic layer comprises constructions of a set of independent
$C^{k}$ periodic functions, based on the generalized Hermite interpolation polynomials,
affine mappings, and nonlinear activation functions.
We have tested the method in extensive numerical experiments
with ordinary and partial differential equations involving
$C^{\infty}$ and $C^k$ periodic boundary conditions.
The numerical results demonstrate
that the proposed method indeed enforces exactly,
to the machine accuracy, the periodicity  for
the solution and its derivatives. 


The proposed method can be implemented in a straightforward way.
The $C^{\infty}$ and $C^k$ periodic layers defined herein can be
implemented as user-defined Keras layers, and used in the same way
as the built-in core Keras layers. All the numerical examples
in the current work are implemented based on Tensorflow and Keras.

Periodic functions and periodic boundary conditions have widespread
applications in computational science of various disciplines.
The proposed method provides an effective tool, based on deep neural
networks, for
representing periodic functions and enforcing exactly the periodic
boundary conditions. We anticipate that this method will be
instrumental in expanding DNN-based techniques to new classes of
applications that are unexplored or scarcely explored so far.


An outstanding question concerning the method developed herein
is the following:
Can an arbitrary periodic function of a certain regularity
be represented by 
the current periodic DNNs to arbitrary accuracy?
This is an important question and it is open at this point.
Our extensive
numerical experiments seem to suggest that the answer to this question
is positive. The periodic DNNs from
the current method are essentially compositions of
an arbitrary DNN-represented function 
with a set of independent periodic functions with adjustable parameters. 
Can the universal approximation power of the original DNN carry over to the
resultant periodic DNN as the set of independent periodic functions
becomes sufficiently large?
Can theoretical analysis establish an analogous universal approximation
property for such periodic DNNs with respect to periodic functions?
These are interesting questions that call for future research
and should be pursued by the community.


\section*{Acknowledgement}
This work was partially supported by
NSF (DMS-1522537). 

\bibliographystyle{plain}
\bibliography{mypub,dnn,sem,obc}

\end{document}